\documentstyle[12pt]{article}

\newcommand \beq{\begin{eqnarray}}
\newcommand \eeq{\end{eqnarray}}
\begin{document}
\input epsf


\def\bfgamma{\mbox{\boldmath$\gamma$}}
\def\bfalpha{\mbox{\boldmath$\alpha$}}
\def\bftau{\mbox{\boldmath$\tau$}}
\def\bfgrad{\mbox{\boldmath$\nabla$}}
\def\bfsigma{\mbox{\boldmath$\sigma$}}
\def\BN{\hbox{Bloch-Nordsiek}}
\def\vp{\mbox{$\bf v\cdot p$}}
\def\vq{\mbox{$\bf v\cdot q$}}
\def\vpq{\mbox{$\bf v\cdot(p+ q)$}}
\def\tilA{\mbox{$v\cdot A$}}
\def\tilQ{\mbox{v\cdot q}}
\def\tilQ1{\mbox{$v\cdot q_1$}}
\def\tilQ2{\mbox{$v\cdot q_2$}}
\def\bfp{\mbox{\boldmath$p$}}

\def\square{\hbox{{$\sqcup$}\llap{$\sqcap$}}}   
\def\grad{\nabla}                               
\def\del{\partial}                              

\def\frac#1#2{{#1 \over #2}}
\def\smallfrac#1#2{{\scriptstyle {#1 \over #2}}}
\def\half{\ifinner {\scriptstyle {1 \over 2}}
   \else {1 \over 2} \fi}

\def\bra#1{\langle#1\vert}              
\def\ket#1{\vert#1\rangle}              

\def\simge{\mathrel{%
   \rlap{\raise 0.511ex \hbox{$>$}}{\lower 0.511ex \hbox{$\sim$}}}}
\def\simle{\mathrel{
   \rlap{\raise 0.511ex \hbox{$<$}}{\lower 0.511ex \hbox{$\sim$}}}}


\def\parenbar#1{{\null\!                        
   \mathop#1\limits^{\hbox{\fiverm (--)}}       
   \!\null}}                                    
\def\nunubar{\parenbar{\nu}}
\def\ppbar{\parenbar{p}}


\def\buildchar#1#2#3{{\null\!                   
   \mathop#1\limits^{#2}_{#3}                   
   \!\null}}                                    
\def\overcirc#1{\buildchar{#1}{\circ}{}}


\def\slashchar#1{\setbox0=\hbox{$#1$}           
   \dimen0=\wd0                                 
   \setbox1=\hbox{/} \dimen1=\wd1               
   \ifdim\dimen0>\dimen1                        
      \rlap{\hbox to \dimen0{\hfil/\hfil}}      
      #1                                        
   \else                                        
      \rlap{\hbox to \dimen1{\hfil$#1$\hfil}}   
      /                                         
   \fi}                                         %


\def\subrightarrow#1{
  \setbox0=\hbox{
    $\displaystyle\mathop{}
    \limits_{#1}$}
  \dimen0=\wd0
  \advance \dimen0 by .5em
  \mathrel{
    \mathop{\hbox to \dimen0{\rightarrowfill}}
       \limits_{#1}}}                           

%
\def\journal#1#2#3#4{\ {#1}{\bf #2} ({#3})\  {#4}}

\def\AdvPhys{\journal{Adv.\ Phys.}}
\def\AnnPhys{\journal{Ann.\ Phys. }}
\def\EurophysLett{\journal{Europhys.\ Lett.}}
\def\JApplPhys{\journal{J.\ Appl.\ Phys.}}
\def\JMathPhys{\journal{J.\ Math.\ Phys.}}
\def\LettNuovoCimento{\journal{Lett.\ Nuovo Cimento}}
\def\Nature{\journal{Nature}}
\def\NPA{\journal{Nucl.\ Phys.\ {\bf A}}}
\def\NPB{\journal{Nucl.\ Phys.\ {\bf B}}}
\def\NuovoCimento{\journal{Nuovo Cimento}}
\def\Physica{\journal{Physica}}
\def\PLA{\journal{Phys.\ Lett.\ {\bf A}}}
\def\PLB{\journal{Phys.\ Lett.\ {\bf B}}}
\def\PhysRev{\journal{Phys.\ Rev.}}
\def\PR{\journal{Phys.\ Rev.}}
\def\PRC{\journal{Phys.\ Rev.\ {\bf C}}}
\def\PRD{\journal{Phys.\ Rev.\ {\bf D}}}
\def\PRB{\journal{Phys.\ Rev.\ {\bf B}}}
\def\PRL{\journal{Phys.\ Rev.\ Lett. }}
\def\PhysRept{\journal{Phys.\ Repts. }}
\def\ProcNatlAcadSci{\journal{Proc.\ Natl.\ Acad.\ Sci.}}
\def\ProcRoySoc{\journal{Proc.\ Roy.\ Soc.\ London Ser.\ A}}
\def\RevModPhys{\journal{Rev.\ Mod.\ Phys. }}
\def\Science{\journal{Science}}
\def\SovPhysJETP{\journal{Sov.\ Phys.\ JETP}}
\def\SovPhysJETPLett{\journal{Sov.\ Phys.\ JETP Lett.}}
\def\SovJNuclPhys{\journal{Sov.\ J.\ Nucl.\ Phys.}}
\def\SovPhysDoklady{\journal{Sov.\ Phys.\ Doklady}}
\def\ZPhys{\journal{Z.\ Phys.}}
\def\ZPhysA{\journal{Z.\ Phys.\ A}}
\def\ZPhysB{\journal{Z.\ Phys.\ B}}
\def\ZPhysC{\journal{Z.\ Phys.\ C}}

\begin{titlepage}
\begin{flushright}
CERN-TH/99-71\\
Saclay-T99/026 \\
hep-ph/9903389
\end{flushright}
\vspace*{1.2cm}
\begin{center}
\baselineskip=13pt
{\Large{\bf A Boltzmann Equation for the QCD Plasma}}
\vskip0.5cm
Jean-Paul BLAIZOT\footnote{E-mail: blaizot@spht.saclay.cea.fr}

{\it Service de Physique Th\'eorique\footnote{Laboratoire de la Direction
des
Sciences de la Mati\`ere du Commissariat \`a l'Energie
Atomique}, CE-Saclay \\ 91191 Gif-sur-Yvette, France}
\vskip0.3cm
  and 

\vskip0.3cm
Edmond IANCU\footnote{E-mail: edmond.iancu@cern.ch}

{\it Theory Division, CERN\\ CH-1211, Geneva 23, 
Switzerland}

\end{center}

\vskip 1cm
\begin{abstract} 

We present a derivation of a Boltzmann equation for the QCD plasma,
starting from the quantum field equations. The derivation is based on 
a gauge covariant gradient expansion which takes consistently into
account all possible dependences on the gauge coupling assumed to be
small. We point out a limitation of the gradient expansion arising when 
the range of the interactions becomes comparable with that of the
space-time inhomogeneities of the system. The method is first applied 
to the case of scalar electrodynamics, and then to the description of
long wavelength colour fluctuations in the QCD plasma.
In the latter case, we recover B\"odecker's effective theory
and its recent reformulation by Arnold, Son and Yaffe. 
We discuss interesting cancellations among various collision terms, 
which occur in the calculation of most transport coefficients, but not 
in that of the quasiparticle lifetime, or in that of the relaxation time 
of colour excitations.

 \end{abstract}
\vskip 1.cm

\begin{flushleft}
CERN-TH/99-71\\
March 1999\\
Published in Nuclear Physics {\bf B 557} (1999) 183-236.
\end{flushleft}
\end{titlepage}

\setcounter{equation}{0}
\section{Introduction}

At  high temperature,  non-Abelian gauge theories describe
weakly coupled plasmas whose constituents, quarks and gluons
for hot QCD,  have typical momenta $k\sim T$, where $T$ is the
temperature 
\cite{BIO96,MLB96,prept}. The plasma particles may take part in
collective excitations which develop typically on a
space-time scale $\lambda\sim 1/gT$
much larger than the mean interparticle distance $\bar r
\sim 1/T$ ($g$ is the gauge coupling, assumed to be small). Such  collective
excitations  can be described in terms  of  mean fields
carrying  appropriate quantum numbers and coupled to the plasma
particles. In this long-wavelength  limit, the plasma particles
obey simple, collisionless, kinetic equations which can be viewed as the
generalization of the Vlasov equation of ordinary plasmas 
\cite{qcd,W,BIO96}.

In this paper, we shall be  interested in specific collective excitations
involving  colour fluctuations on larger wavelengths,   $\lambda\sim
1/g^2 T$. In this situation, the effects of the collisions among the plasma
particles become as important as those  of the mean fields. The kinetic
equations obeyed by the plasma particles must therefore be generalized so as
to include the collisions terms. This is what we shall do here.  The Boltzmann
equation that we shall obtain  (see  eqs.~(\ref{dn0})---(\ref{PHII}) below)
turns out to be identical to the one proposed recently by Arnold, 
Son and Yaffe \cite{ASY97}, and yields, in leading logarithmic
accuracy, B\"odeker's effective theory for the soft ($p\sim
g^2 T$) fields \cite{Bodeker98}.
The derivation presented here, starting from the quantum field equations,  
clarifies the nature of the approximations involved, and thus fixes 
its range of applicability. 
Furthermore, it  also provides some justification for
numerous previous works using ad hoc transport
equations inspired by classical transport theory
\cite{Heinz83,Baym90,Gyulassy93,Heisel93,Heisel94,Manuel94,Heisel94a,Baym97,Manuel99}. 
It should
be emphasized that transport equations
with a similar colour structure have been
also proposed in Refs.
\cite{Selik91,Gyulassy93}, and that some of the technics that we shall be using
have been used already by many authors 
\cite{Elze86,Vasak87,Elze88,EHPRept,Mrowcz90}. However, in most of these
works, the organizing role  of the various dynamical scales which appear 
in hot QCD was not recognized, which led  to unecessarily complicated, and
sometimes inconsistent,  equations.  

In fact, ``kinetic theory'' in the way we use it here could
be regarded as a powerful tool for constructing effective theories for the
soft modes of the plasma. These soft degrees of freeedom are represented
by mean fields, while the hard ones, which are ``integrated out'' using
perturbation theory, survive as induced sources for these mean fields. The
resulting effective  theory can then be studied non-perturbatively, e.g., as a
classical theory on a lattice:  recently, this strategy has received much
attention in connection with studies of baryon number violation in the high
temperature electroweak theory
\cite{McLerran,ASY96,Hu,baryo,Bodeker98,ASY97,Moore98,Manuel99}.
Let us also recall that this method has been
first demonstrated for the collective dynamics at the scale $gT$,
where we have shown  \cite{qcd,W} that  simple,
collisionless, kinetic equations resum
an infinite number of one-loop diagrams with soft external lines
and hard loop momenta, the so-called ``hard thermal loops''
\cite{BP90,FT90}.

\begin{figure}
\protect \epsfxsize=7.cm{\centerline{\epsfbox{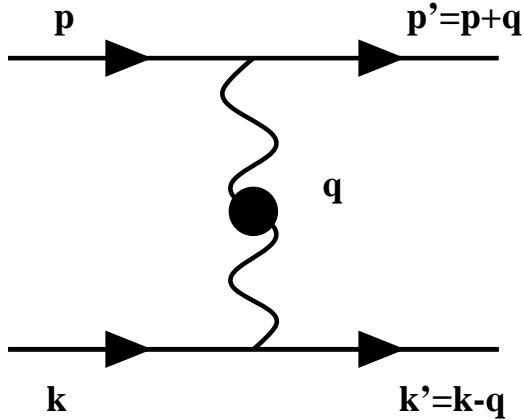}}}
         \caption{Elastic scattering in the (resummed) Born approximation.
The continuous lines refer to hard gluons (these are off-equilibrium
propagators), while the wavy line is the soft gluon exchanged in the
collision. The blob stands for HTL resummation.}
\label{Born}
\end{figure}

Let us now summarize  the main equations to be obtained below.
The collective, longwavelength  colour 
fluctuations of the hard transverse gluons
are described by a density matrix $N({\bf k},x)$ which, to the order of
interest,
 can be written in the form:
\beq\label{dn0}
N_{ab}({\bf k}, x)\,=\,N(\varepsilon_k)\delta_{ab}
 - gW_{ab}(x,{\bf v})\,({\rm d}N/{\rm d}
\varepsilon_k),\eeq
where $N(\varepsilon_k)\equiv 1/({\rm e}^{\beta \varepsilon_k}-1)$
is the Bose-Einstein thermal distribution (with $\varepsilon_k
=|{\bf k}|$), and the function 
$W(x,{\bf v})$, which parametrizes the off-equilibrium deviation,
is a colour matrix in the adjoint representation,
$W(x,{\bf v})\equiv W_a(x,{\bf v}) T^a$, which depends
upon the velocity ${\bf v}={\bf k}/\varepsilon_k$
(a unit vector), but not upon the
magnitude $k\equiv |{\bf k}|$ of the  momentum. 
The functions $W_a(x,{\bf v})$ satisfy the following transport equation:
\beq\label{W10}
(v\cdot D_x)^{ab}W_b(x,{\bf v})&=&{\bf v}\cdot{\bf E}^a(x)-
\gamma\left\{W^a(x,{\bf v})-\frac{\left\langle
\Phi({\bf v\cdot v}^\prime)W^a(x,{\bf v}^\prime)\right\rangle}
{\left\langle\Phi({\bf v\cdot v}^\prime)\right\rangle}\right\}.\eeq
In the left hand side of this equation, $v\cdot D_x$ is a gauge-covariant
drift operator (with $v^\mu\equiv (1,{\bf v})$ and
$D^\mu\equiv\del^\mu+igA^\mu$), while in the right hand
side we recognize a mean field term (${\bf v}\cdot{\bf E}^a(x)$,
with ${\bf E}^a$ the chromoelectric field) and a collision
term. The latter is proportional to the quasiparticle {\it damping rate}
$\gamma$, which appears to set 
the scale for the colour relaxation time: $\tau_{col}\sim 1/\gamma \sim
1/(g^2T\ln(1/g))$ (see below). The other notations above are as follows:
the angular brackets in the collision term denote angular average
over the directions of the unit vector ${\bf v}^\prime$ (as 
in eq.~(\ref{gamma110}) below), and the
quantity $\Phi({\bf v\cdot v}^\prime)$ is given by:
\beq\label{PHII}\Phi({\bf v\cdot v}^\prime)\equiv(2\pi)^2
\int\frac{{\rm d}^4 q}{(2\pi)^4}\,
\delta(q_0- {\bf q\cdot v})
\delta(q_0- {\bf q\cdot v}^\prime)
\Big|{}^*{\cal D}_l(q)+ ({\bf v}_t\cdot{\bf v}_t^\prime)\,
{}^*{\cal D}_t(q)\Big|^2.\,\,\,\eeq
where 
${}^*{\cal D}_{l}(q)$ and ${}^*{\cal D}_{t}(q)$ denote the resummed gluon
propagators in the electric and the magnetic channels, respectively
\cite{BIO96,MLB96}.
Up to a normalization, $\Phi({\bf v\cdot v}^\prime)$ is the total interaction
rate for two hard particles with momenta ${\bf k}$ and ${\bf p}$ (and
velocities ${\bf v}\equiv \hat{\bf k}$ and
${\bf v}^\prime \equiv \hat{\bf p}$) in the (resummed) Born approximation,
as illustrated in Fig.~\ref{Born}
(${\bf v}_t$ and ${\bf v}_t^\prime$ are the transverse projections
of the velocities with respect to the momentum ${\bf q}$ of the
exchanged gluon: e.g., $v_t^i=(\delta^{ij}-\hat q^i\hat q^j)v^j$).
The damping rate $\gamma$ is obtained
from $\Phi({\bf v\cdot v}^\prime)$ as follows:
\beq\label{gamma110}
\gamma\,=\,\frac{g^4 N_c^2 T^3}{6}\int\frac{{\rm d}\Omega'}{4\pi}
\,\Phi({\bf v\cdot v}^\prime)\,\simeq\,\frac{g^2 N_c T}{4\pi}
\ln(1/g),\eeq
and is of O$(g^2 T)$ in spite of the explicit
factor $g^4$ in front of the above integral. This is because 
the quasiparticle damping is dominated by soft
momentum transfers $q \simle gT$, which gives an enhancement factor
$\sim 1/g^2$ after the resummation of the screening effects at
the scale $gT$  \cite{BP90,Smilga90,Pisarski93,lifetime}. 

Actually, in the present approximation, $\gamma$ is even logarithmically
infrared divergent, due to the unscreened static ($q_0\to 0$) magnetic
interactions. (In writing the right hand side of eq.~(\ref{gamma110})
we have assumed an infrared cutoff $\sim g^2 T$, as it is usually
done in the literature \cite{Pisarski93,Gyulassy93,Heisel94}.)
To {\it logarithmic accuracy}, that is, by preserving only the
singular piece of the magnetic scattering element in
eq.~(\ref{PHII}) (see eq.~(\ref{singDT}) below), eqs.~(\ref{W10})--(\ref{PHII}) 
generate B\"odeker's effective theory for the soft modes
$A^\mu_a$ \cite{Bodeker98}. 

Within the same accuracy,
eq.~(\ref{W10}) can be solved to get the so-called colour conductivity
\cite{Gyulassy93,Heisel94,Bodeker98,ASY97}. 
The induced colour current is expressed in terms of $W(x,{\bf v})$ as:
\beq
{\bf j}^a(x)\,=\,m_D^2\langle {\, \bf v}\, W^a(x,{\bf v}) \rangle,
\eeq
with the Debye mass $m_D^2=g^2N_cT^2/3$.
For constant colour electric fields, we get from eq.~(\ref{W10}):
\beq
W^a({\bf v})\,=\,\frac{1}{\gamma}\, {\bf v}\cdot {\bf E}^a,
\eeq
so that 
\beq
{\bf j}^a \,=\,\frac{m_D^2}{3\gamma}\,{\bf E}^a\,\equiv\,\sigma_c {\bf E}^a,
\eeq
with the colour conductivity $\sigma_c=m_D^2/3\gamma \sim T/\ln(1/g)$.

The next section of the paper contains a derivation of the Boltzmann equation
for scalar electrodynamics (SQED). There are several reasons for this.
First, to our knowledge, this is the first consistent derivation of a
Boltzmann equation for gauge theories, starting form the quantum field
equations. Second, it serves as a preparation for the more involved non Abelian
case of  QCD which is presented in the following section. Finally, and this is
the most important, it will reveal interesting compensations which occur in
Abelian, but not in non-Abelian gauge theories. Thus, in SQED, most transport
phenomena are dominated by large angle scattering, so that the typical
relaxation times are $\tau_{tr} \sim 1/(e^4 T\ln(1/e))$,
where $e$ is the electric charge
\cite{Baym90,Baym97}. This is to be contrasted with the
quasiparticle lifetimes which are limited by small angle scatterings and 
are of order 
$\tau\sim 1/(g^2T\ln(1/g))$ (in both QED and QCD).
The same cancellations occur in most cases
for QCD as well (thus yielding, e.g., a viscous relaxation time
$\tau_{visc}\sim 1/(g^4 T\ln(1/g))$ \cite{Baym90,Heisel94a}),
except for the the relaxation of colour excitations which remains 
dominated by very soft gluon exchanges
\cite{Gyulassy93,Heisel94,Bodeker98,ASY97}. As a result, the colour relaxation
time turns out to be  of the same order as the quasiparticle lifetime, as is
evident in eq.~(\ref{W10}). This yields a colour conductivity 
$\sigma_c \sim T/\ln(1/g)$, to be contrasted with the usual, electric
conductivity\footnote{We mean here, of course, the electric conductivity 
in a QED or OCD plasma, that is, in a gauge theory without electrically 
charged vector bosons. The situation would be different in the electroweak
theory where, in the high-temperature,
symmetric, phase, the electric charge can
be efficiently randomized via small angle scatterings mediated
by the $W^\pm$-bosons
\cite{Baym97}.}: $\sigma_{el} \sim T/(e^2\ln(1/e))$ \cite{Baym97}.

The main part of the paper is section 3, which contains the derivation of the
Boltzmann equation for the QCD plasma, and a discussion of the
approximations which are needed in this derivation. 
These involve a gradient expansion of the  Dyson-Schwinger equations,
supplemented by a perturbative evaluation of the collision terms and a
linearization with respect to the off-equilibrium fluctuations. All these
approximations are commonly used in deriving kinetic equations from quantum
field theories, and they are generally seen as independent approximations (to
some extent they remain so in the case of SQED discussed in Sec. 2 below).
However, in order to fulfill the constraints imposed by a non Abelian gauge
symmetry, we shall see that it is convenient to  control all these
approximations by the same small parameter, namely the gauge coupling $g$.
Thus, for instance,  the amplitudes  of the mean fields will be 
restricted  so that 
$|A_a^\mu|
\sim  gT\,$: this guarantees  that the two terms in the soft  covariant 
derivative $D_X^\mu=\del_X^\mu+igA^\mu$ are of the same order
in $g$, $\del_X \sim gA \sim g^2 T$, so that $D_X={\rm O}(g^2T)$
can be preserved consistently in the expansion. A further difficulty that we
shall have to face is related  to the poor convergence of the  gradient
expansion when the range of the  interactions becomes comparable to the scale
of the system inhomogeneities. As we shall see this will be the main
limitation of the accuracy of the collision term. 

For completeness, we present in section 4 some diagrammatic interpretation of
the Boltzmann equation. (Previously, the connection between
Feynman graphs and the Boltzmann equation has been explored in detail
only for a scalar field theory, in Refs. \cite{Jeon93,JY96}.)
The section 5 summarizes the conclusions. 
 
\setcounter{equation}{0}
\setcounter{equation}{0}
\section{Scalar QED}

In this section, we briefly summarize the general formalism
which allows one to construct kinetic equations from the Dyson-Schwinger
equations obeyed by the non-equilibrium Green's functions
\cite{KB62,Schwinger61,Bakshi63,Keldysh64,PhysKin,Daniel83,Heinz94}.
In order to bring out the essential aspects of the formalism
while avoiding the complications specific to non Abelian gauge theories, 
we shall consider here scalar electrodynamics (SQED),
with Lagrangian:
\beq\label{Lagran}
{\cal L}&=&( D_\mu\phi) (D^\mu\phi)^*-\frac{1}{4} F_{\mu\nu}F^{\mu\nu},
\eeq
where $\phi$ is  a complex scalar field, $A_\mu$ is the photon field,
$D_\mu\equiv \del_\mu+ieA_\mu$
is a covariant derivative, and $F_{\mu\nu}$ the field strength tensor,
$F_{\mu\nu}=\del_\mu A_\nu-\del_\nu A_\mu$.

The systems that we consider are assumed to be
 initially in thermal equilibrium,
and described by the  density operator $\rho={1\over Z}\exp\left\{ -\beta H\right\}
$ where $H$ is the Hamiltonian corresponding to (\ref{Lagran}), and $Z$ is
the partition function. At some time $t_0$, a time-dependent external
perturbation (an electromagnetic current $j_\mu(x)$) starts acting on the
system, so that the Hamiltonian becomes:
\beq\label{Hj}
H_j(t)\,=\,H\,+\,\int {\rm d}^3{x}\,j(t,{\bf x})\cdot A({\bf x})\,,
\eeq
where $j\cdot A= j_\mu\, A^\mu$. The density operator at time $t$ is
given by:
\beq\label{rhoj}
{\rho}(t)\,=\, U(t,t_0)\,\rho  \,U(t_0,t),
\eeq
where  $U(t,t_0)$, the evolution operator,   satisfies:
\beq\label{eqUj}
i\del_t \,U(t,t_0)\,=\,H_j(t)U(t,t_0),\qquad\qquad
U(t_0,t_0)\,=\,1.\eeq 
In the presence of the perturbation, the gauge field $A^\mu$ develops an
expectation value:
\beq\label{avOj} {\rm Tr}\Bigl(\rho(t) A^\mu\Bigr)\,=\,{\rm Tr}\Bigl(
 \rho\,A^\mu(t)\Bigr)
\,=\,{\rm Tr}\left\{\frac{{\rm e}^{-\beta H}}{Z}\,
A^\mu(t)\right\}\equiv\,\langle A^\mu(t)\rangle\,,\eeq
with 
\beq\label{Oj}
A^\mu(t)\equiv  U^{-1}(t,t_0)\,A^\mu\,U(t,t_0)\,=\,
 U(t_0,t)\,A^\mu\,U(t,t_0),\eeq
where we have used the fact that $U(t_0,t)=U^{-1}(t,t_0)$.
More generally, we shall be interested in various
$n$-point functions, and in
particular in 2-point functions for which we shall derive equations of motion in
the next subsection. For instance, the 
time-ordered 2-point function of the charged
scalar field is given by:
\beq\label{EX}
G(t_1,t_2)&=&\langle {\rm T}\,\phi(t_1)\phi^\dagger(t_2)\rangle
\equiv {\rm Tr}\left\{\frac{{\rm e}^{-\beta H}}{Z}\,
{\rm T}\,\phi(t_1)\phi^\dagger(t_2)\right\},\ \nonumber \\
  &=& \theta(t_1-t_2) G^>(t_1,t_2)
+\theta(t_2-t_1) G^<(t_1,t_2)\eeq
where  $\phi(t) =U(t_0,t)\,\phi\,U(t,t_0)$, and the functions $G^>$ and
$G^<$ are defined by:
\beq\label{G>G<}
G^>(x,y)\,\equiv\,\langle
\phi(x)\phi^\dagger(y)\rangle,\qquad
G^<(x,y)\,\equiv\,\langle
\phi^\dagger(y)\phi(x)\rangle.\eeq
(To lighten the notation, the spatial coordinates have not been indicated
in eq.~(\ref{EX}).)
The functions $G^>$ and $G^<$  can be used to construct the 
retarded $(G_R)$ and
advanced $(G_A)$ propagators, which will also be needed:
\beq\label{A}
G_R(x,y)&\equiv &
i\theta(x_0-y_0)\Bigl[G^>(x,y)\,-\,G^<(x,y)\Bigr]\,,\nonumber\\
G_A(x,y)&\equiv & -
i\theta(y_0-x_0)\Bigl[G^>(x,y)\,-\,G^<(x,y)\Bigr]\,.\eeq 
Similar definitions hold for the photon 2-point functions. In the case of the
photon, we shall decompose the gauge field into its average value for which we
shall reserve the notation 
$A_\mu(x)$ (i.e., in the  following we identify 
$\langle A_\mu\rangle\to A_\mu$), and a fluctuating part
$a_\mu(x)$ with $\langle a_\mu\rangle =0$.
The time ordered photon propagator is then  given by:
\beq\label{PHD}
{\cal D}_{\mu\nu}(x,y)=\langle {\rm T}a_\mu(x) a_\nu(y)\rangle.
\eeq
The 2-point functions introduced above satisfy boundary conditions
which follow from their definitions (cf. eq.~(\ref{EX})).
For instance:
\beq\label{KMSSIG} 
G^<(t_0,z)\,=\, G^> (t_0-i\beta,z),
\eeq
and similarly for the photon 2-point functions and for the various
self-energies to be introduced later. Furthermore, these functions
have hermiticity properties which will be useful below.
Specifically, all the ``bigger'' ($>$) and ``lesser'' ($<$)
2-point functions are hermitian: for instance,
$(G^>(x,y))^*=G^>(y,x)$ and 
$({\cal D}^>_{\mu\nu}(x,y))^*={\cal D}^>_{\nu\mu}(y,x)$.
This, together with the definitions (\ref{A}), imply
$(G_R(x,y))^*=G_A(y,x)$ and
$({\cal D}^{\mu\nu}_R(x,y))^*={\cal D}^{\nu\mu}_A(y,x)$,
together with similar properties for the various self-energies.

In thermal equilibrium, the system is homogeneous 
(e.g., $G^<_{eq}(x,y)\equiv G^<_{eq}(x-y)$), and it is convenient
to go to momentum space. Then, the boundary condition (\ref{KMSSIG})
translates into the so-called KMS condition \cite{MLB96} :
\beq\label{KMSk}
G^>_{eq}(k)&=&{\rm e}^{\beta k_0}G^<_{eq}(k),\eeq
which implies the following structure for the equilibrium
2-point functions:
\beq\label{G><k}
G^>_{eq}(k)=\rho(k)\Bigl[1+N(k_0)\Bigr],\,\,\qquad\,\,
G^<_{eq}(k)=\rho(k)\,N(k_0)\,,\eeq
with $N(k_0)\equiv 1/({\rm e}^{\beta k_0}-1)$ and
the {spectral density} $\rho(k)\equiv G^>_{eq}(k)-G^<_{eq}(k)$.
In particular, for free, massless particles:
 \beq\label{G0}
G^<_0(k)\equiv \rho_0(k)N(k_0),\qquad 
G^>_0(k)\equiv \rho_0(k)[1+N(k_0)],\eeq
with $\rho_0(k)=2\pi \epsilon(k_0)\delta(k^2)$.

\subsection{Equations of motion for Green's functions}

\begin{figure}
\protect \epsfxsize=16.cm{\centerline{\epsfbox{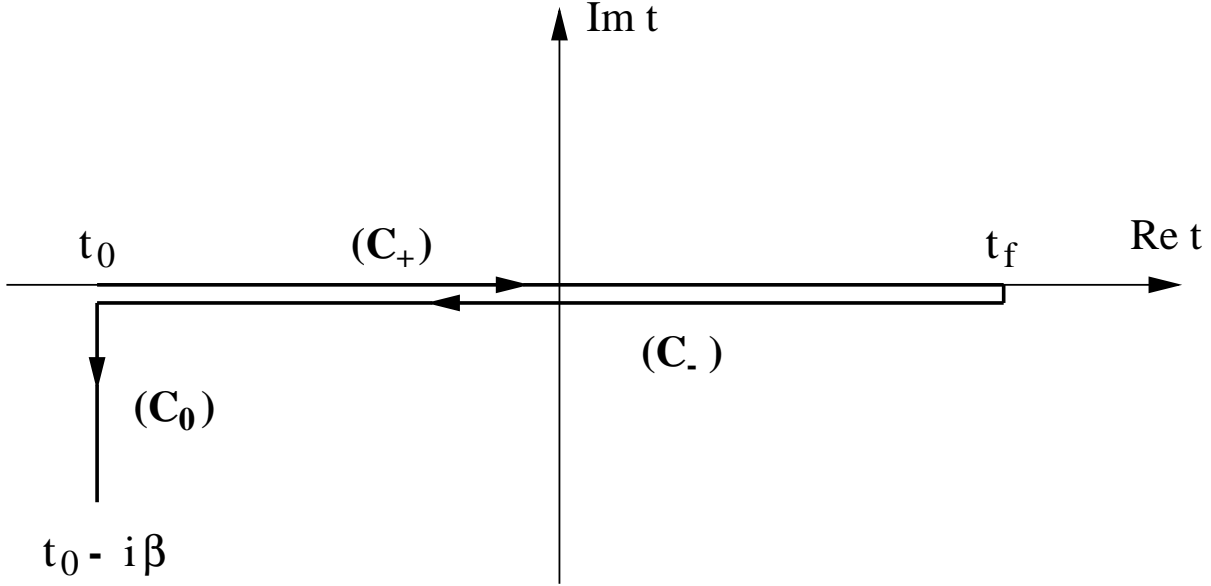}}}
         \caption{Complex-time contour for the evaluation of
the thermal expectation values: $C=C_+\cup C_-\cup C_0$.}
\label{CONT}
\end{figure}
In order to obtain the equations of motion for the 2-point functions, it is
convenient to extend their definition by allowing the time variables to take
complex values.  More specifically, we introduce, in the complex 
time plane,  the oriented contour   depicted in Fig.~\ref{CONT}. This
may be seen as the juxtaposition of three pieces:
$C=C_+\cup C_-\cup C_0$. We call $z$ the (complex) time
variable along the contour, and reserve the notation $t$ for
real times. On $C_+$, $z=t$ takes all the
real values between $t_0$ to $t_f$. On  $C_-$, we set $z=t-i\eta$
($\eta\to 0_+$) and $t$ runs backward from
$t_f$ to $t_0$. Finally, on $C_0$, $z=t_0-i\tau$, with $0<\tau\le \beta$. 
We define    a contour $\theta$-function $\theta_C$:
$\theta_C(z_1,z_2)=1$ if $z_1$ is further than $z_2$ 
along the contour (we then write $z_1 \succ z_2$),
while $\theta_C(z_1,z_2)=0$ if the opposite situation holds
($z_1 \prec z_2$).
 We can formalize this by introducing a real parameter $u$ which
is continously increasing along the contour; then, the contour $C$
is specified by a function $z(u)$, and $\theta_C(z_1,z_2)=
\theta(u_1-u_2)$. We shall later need also a contour delta function, which we 
define by:
\beq
\delta_C(z_1,z_2)\equiv \left(\frac {\del z}{\del u}\right)^{-1}
\delta(u_1-u_2).\eeq 
The definition of the propagators is then extended in a natural way. For instance,
the contour-ordered propagator of the scalar field becomes:
\beq\label{EXCONT}
G(z_1,z_2)\,\equiv\,\langle {\rm T}_C\,\phi(z_1)\phi^\dagger(z_2)\rangle
\equiv {\rm Tr}\left\{\frac{{\rm e}^{-\beta H}}{Z}\,
{\rm T}_C\,\phi(z_1)\phi^\dagger(z_2)\right\},\eeq
where ${\rm T}_C$ orders the operators on its right,  from right to left  in
increasing order of the arguments $u_i$. For time arguments
$t_1$,
$t_2$ on $C_+$, the contour propagator (\ref{EXCONT}) reduces to the 
time-ordered
propagator (\ref{EX}). For $t_1\in C_-$ and $t_2\in C_+$, 
we have $G(t_1-i\eta,t_2)\,=\,G^>(t_1,t_2)$,
while for  $t_1\in C_+$ and $t_2\in C_-$, we have
$G(t_1,t_2-i\eta)\,=\,G^<(t_1,t_2)$. 

With these definitions in hand, most of
the formal manipulations familiar in equilibrium field theory can be extended to
the case of non equilibrium. 
This is convenient for the derivation of the equations
of motion for the $n$-point functions to which we now turn.

The mean field equation is: 
\beq\label{av1}
\del_\nu F^{\mu\nu}(x) =j^\mu(x) +j^\mu_{ind}(x),
\eeq
with the induced current
\beq
j^\mu_{ind}(x)=-ie\Bigl\langle \left( D^\mu
\phi(x)\right)^\dagger\phi(x)\,-\phi^\dagger(x)\left( D^\mu
\phi(x)\right)\Bigr\rangle. 
\eeq
In this expression, $D_\mu= \del_\mu+ig(A_\mu+a_\mu)$. However, in line with
the approximations below, we can ignore the contribution of the
quantum field $a_\mu$ in the expression of the induced current. This amounts to
neglect the contribution of a connected 3-point function. We can then write:
\beq
j^\mu_{ind}(x)=ie \left(
D^\mu_x-\left(D^\mu_y\right)^\dagger\right)
\left.G^<(x,y)\right|_{x=y},
\eeq
where now, and for the rest of this section,  $D_\mu= \del_\mu+igA_\mu$, with
$A_\mu$ the average gauge potential.

In order to calculate the induced current, we need  the 2-point
function $G^<(x,y)$. An equation of motion for this function can  be obtained from
the equation of motion for the time ordered propagator: 
\beq\label{D1}
-D^2_x \,G(x,y)\,-\,i\int_C {\rm d}^4z\,\Sigma(x,z) \,G(z,y)\,=
\,i\delta_C(x,y),\eeq
where $\delta_C(x_0,y_0)$ is the contour delta function, and $\Sigma$ 
the scalar self-energy. The latter admits  the following
decomposition,  similar to that of $G$, eq.~(\ref{EX}):
\beq\label{Sig<>}
\Sigma(x,y)\,=\,-i\hat\Sigma(x)\delta_C(x,y)\,+\,
\theta_C(x_0,y_0) \Sigma^>(x,y)
+\theta_C(y_0,x_0) \Sigma^<(x,y).\eeq
We have separated out a possible singular piece
$\hat\Sigma$ (e.g., the standard tadpole diagram which generates a 
temperature-dependent mass correction \cite{Kraemmer94}).
The non-singular components $\Sigma^>$ and $\Sigma^<$
obey a boundary condition similar to (\ref{KMSSIG}).
In particular, in equilibrium,
$\Sigma^>_{eq}(k)={\rm e}^{\beta k_0}\Sigma^<_{eq}(k)$.

The equations of motion in real-time for the mean field and the
2-point functions are obtained by letting the external time variables
 $x_0$ and
$y_0$  take values on the real-time pieces of this contour, $C_+$ 
and $C_-$. For $x_0\in C_+$, the  mean field equation is formally
the same as in eq.~(\ref{av1}).
Consider now eq.~(\ref{D1}):
by choosing $x_0\in C_+$ and $y_0\in C_-$,
and by using the decompositions (\ref{EX})
and (\ref{Sig<>}),
we obtain, after some manipulations, an equation for $G^<(x,y)$:
\beq\label{eq1}
\Bigl(D^2_x+\hat\Sigma(x)\Bigr)
G^<(x,y) \,=\,-\int {\rm d}^4 z\,
\Bigl[\Sigma_R(x,z) \,G^<(z,y)
\,+\,\Sigma^<(x,z)\,G_A(z,y)\Bigr],\nonumber\\ \eeq 
together with a similar
equation where the differential operator is acting on $y$:
\beq\label{eq2}
\Bigl((D^\dagger_y)^2+\hat\Sigma(y)\Bigr)
G^<(x,y)\,=\,-\int  {\rm d}^4 z\,
\Bigl[G^<(x,z)\,\Sigma_A(z,y) 
\,+\,G_R(x,z)\,\Sigma^<(z,y)\Bigr].\nonumber\\ \eeq 
In these equations,  $D^2=D^\mu D_\mu$,
$D_\mu^\dagger=\del_\mu-ie A_\mu$, and  we have used the  definitions (\ref{A}) for
the retarded and advanced
 Green's functions, 
together with similar definitions for $\Sigma_R$ and $\Sigma_A$. One can also
obtain an equation  satisfied by $G_R(x,y)$:
\beq\label{eqGR}
\Bigl(D^2_x+\hat\Sigma(x)\Bigr)
G_R(x,y)\,+\,\int {\rm d}^4z\,\Sigma_R(x,z) \,G_R(z,y) \,
=\,\delta^{(4)}(x-y)\,.\eeq
Note that, while the  Green's functions $G^>$ and the $G^<$ and the
corresponding self-energies are coupled by
eqs.~(\ref{eq1})--(\ref{eq1}), the retarded Green's function
$G_R$ is determined by the retarded self-energy $\Sigma_R$ alone.

The above equations must be supplemented
with some approximation scheme in which, for instance,
the self-energy $\Sigma$ is
expressed in terms of the propagator $G$. Below, we shall use
perturbation theory for this purpose.
We shall refer to the above equations as the Kadanoff-Baym equations. They
were first obtained in the framework
of non-relativistic many-body theory \cite{KB62}. Note that,
 in these equations,
any explicit reference to the initial conditions and to
the KMS condition has disappeared.
These only enter as boundary conditions to be satisfied 
by the various Green's functions in the remote past.
The same set of equations has been derived 
by Keldysh  \cite{Keldysh64} to describe non-equilibrium evolutions of quantum
systems (see also \cite{Schwinger61,Bakshi63,PhysKin,Daniel83}).

\subsection{Gauge covariant Wigner transforms}

For  slowly varying off-equilibrium
perturbations, the Kadanoff-Baym equations can be transformed into
kinetic equations, as we now  explain. In thermal
equilibrium, the system is homogenous, and the two-point functions depend
only on the relative coordinates
$s^\mu=x^\mu-y^\mu$. The thermal
particles have typical energies and momenta $k\sim T$. It follows that the
2-point functions are peaked around
$s^\mu=0$, their range of variation being fixed by the thermal
wavelength $\lambda_T=1/k\sim 1/T$.
In what follows, we  shall be interested in off-equilibrium deviations
 which are slowly varying in space and time, over
a typical scale $\lambda \gg \lambda_T$. 

 In order to take advantage of the assumed separation of scales between
hard degrees of freedom (the plasma particles), and the soft degrees of
freedom (the collective excitations at scale $\lambda \gg \lambda_T$), it 
is convenient to introduce relative and central coordinates,
 \beq\label{Rel}
 s^\mu\equiv x^\mu-y^\mu,\qquad\qquad
X^\mu\equiv{x^\mu+y^\mu\over2}\,,\eeq
and to use the {\it Wigner transforms} of the 2-point functions. These are
defined as Fourier transforms with respect to the
relative coordinates $s^\mu$. For instance, the Wigner transform
of $G^<(x,y)$ is:
\beq\label{G<WIG}
  G^<(k,X)\equiv\int {\rm d}^4s
\,{\rm e}^{ik\cdot s} \,G^<\left(X+{s\over 2},X-{s\over 2}\right), \eeq
and we shall use similar definitions for the other 2-point functions.
Note that we  shall use the same symbols for the 2-point functions and their
Wigner transforms, considering that the different functions can be 
recognized from their arguments. 

The hermiticity properties of the 2-point functions
discussed after eq.~(\ref{KMSSIG})
imply similar properties for the corresponding Wigner functions.
For instance, from $(G^>(x,y))^*=G^>(y,x)$ we deduce that
$G^<(k,X)$ is a real function, $(G^<(k,X))^*=G^<(k,X)$, and similarly
for $G^>(k,X)$. Also, $(G_A(k,X))^*=G_R(k,X)$. Similar properties
hold for the photon 2-point functions and for the various
self-energies.

In gauge theories, the physical interpretation of the Wigner functions as
phase space densities is complicated by the lack of gauge covariance of the
2-point functions. To remedy this, we shall  define new, gauge invariant,
functions, whose construction may be motivated by  considering 
the conserved electromagnetic current:
\beq\label{currentSQED}
j^\mu(x)=ie \left(
D^\mu_x-\left(D^\mu_y\right)^\dagger\right)
\left.G^<(x,y)\right|_{x=y},
\eeq
where $G^<(x,y)=\langle \phi^\dagger(y)\phi(x)\rangle$ 
 is not gauge invariant. It is easy to define a 
corresponding gauge invariant  function  by multiplying it by a
parallel transporter, or ``Wilson line'',
\beq\label{U0}
U(x,y)={\rm e}^{-ie\int_\gamma {\rm d}z^\mu A_\mu(z)},
\eeq
where the path $\gamma$ 
joining $y$ to $x$ is a priori arbitrary. Thus, for instance,
\beq\label{acuteG0}
\acute G^<(x,y)\equiv \langle
\phi^\dagger(y) U(y,x) \phi(x)\rangle=  U(y,x)G^<(x,y)
\eeq
is manifestly gauge invariant. The conserved current may then be expressed in
terms of this gauge invariant function:
\beq\label{currentSQED2}
j^\mu(x)&=& \left( \del^\mu_x-\del^\mu_y \right)
\left. \acute G^<(x,y)\right|_{x=y}\nonumber\\
 &=& 2i \del_s^\mu \left.\acute G^<(s,X)\right|_{s=0}\, .
\eeq
To see this, note that $G^<(x,y)=U(x,y)\acute
G^<(x,y)$, and that $D^\mu_x
\left.U(x,y)\right|_{x=y}= 0$. Note also that the expression
for $j^\mu(x)$ is independent of the path joining $x$ and
$y$, since only an infinitesimal path is needed.

For definitness, we shall in fact 
choose $\gamma$ to be the {\it straight line} 
joining $x$ and $y$. This choice is physically motivated since, as we shall see
later, the hard particles preserve straight line trajectories
in the presence of the soft mean fields (at least, to leading order
in $e$). Moreover, as shown in Refs. \cite{Elze86,Vasak87,EHPRept}, 
such a path allows one
to interpret the covariantization procedure as the replacement
of the canonical momentum by the kinetic one (see eq.~(\ref{GGAC})
below). This being said, most of our results below will be
independent of the exact form of $\gamma$ (see, however, the discussion
after eq.~(\ref{P0})). Indeed, we 
shall mostly need the parallel transporter $U(x,y)$ in situations where the
end points $x$ and $y$ are close to each other ($|s|\simle 1/T$, with
$s\equiv x-y$), so that the variation of the field can be neglected along
the path. This is a good approximation provided $\gamma$ never goes too 
far away from $x$ and $y$, that is, provided $|z-X|={\rm O}(1/T)$
(with $X\equiv(x+y)/2$) for any point $z$ on $\gamma$.
For any such a path we can write:
\beq\label{Uapprox}
U\left(x,y\right)\,\approx\,
 {\rm e}^{\,-ies\cdot A(X)}\,,
\eeq
up to terms which involve, at least, one soft derivative $\del_X A^\mu$
(and which do depend upon the path).

Starting from $\acute G^<(x,y)$, we construct the gauge
invariant Wigner function: 
\beq\label{GGAC}
\acute G^<(k,X) &\equiv& \int {\rm d}^4 s \, {\rm e}^{ik\cdot s}\,
U\left(X-\frac{s}{2},X+\frac{s}{2}\right)
\, G^<\left(X+\frac{s}{2},X-\frac{s}{2}\right)\nonumber\\
&\approx& \int {\rm d}^4 s\,{\rm e}^{\,is\cdot (k+eA(X))} \,
G^<\left(X+\frac{s}{2},X-\frac{s}{2}\right)
\nonumber\\&=& G^<(p=k+eA(X),X).
\eeq
This formula shows that, as alluded to before, the gauge invariant
Wigner function may be  obtained from the ordinary one by the simple
replacement of the {\it canonical } momentum
$p^\mu$ by the {\it kinetic } momentum
$k^\mu=p^\mu-e A^\mu(X)$.  Returning to the current, we see that it  takes
the form:
\beq\label{JAB}
j^\mu(x)&=& 2e \int \frac{{\rm d}^4p}{(2\pi)^4}\,\left( p^\mu-e
A^\mu(X)\right) G^<(p,X)\,
\nonumber\\
 &=& 2e \int \frac{{\rm d}^4k}{(2\pi)^4}\, k^\mu\,\acute
G^<(k,X).\eeq
These two expressions for the current may be seen as the analogs of
eqs.~(\ref{currentSQED}) and (\ref{currentSQED2}). 

\subsection{Gradient expansion and kinetic equations}

For slowly varying disturbances,  taking place over a scale
$\lambda \gg \lambda_T$, we expect the $s^\mu$ dependence of
the 2-point functions to be close to that in equilibrium.
Thus, typically, $k\sim\del_s \sim T$, while $\del_X \sim 1/\lambda \ll T$.
The general equations of motion written down
in Sec. 2.1 can then be simplified with the help of
a {\it gradient expansion}, using $k$ and $X$ as most
convenient variables. 

The starting point of the gradient expansion is the equation 
obtained by taking the difference  of the Kadanoff-Baym equations
(\ref{eq1}) and (\ref{eq2}). For further reference, we shall call it
the {\it difference equation}. We then define:
\beq\label{XiAB}
\Xi(x,y)\equiv D^2_x -\, (D^\dagger_y)^2,\eeq
where:
\beq\label{d^2}
D_x^2&=&\del_x^2+2ieA(x)\cdot\del_x +ie(\del\cdot A(x))-e^2A^2(x),
\nonumber\\
(D^\dagger_y)^2&=&\del_y^2
-2ieA(y)\cdot\del_y -ie(\del\cdot A(y))
-e^2A^2(y).\eeq
By replacing the coordinates
$x^\mu$ and $y^\mu$ by $s^\mu$ and
$X^\mu$ (cf. eq.~(\ref{Rel})), and rewriting the derivatives as:
\beq\label{DERIV}
\del_x=\del_s+\half\del_X,\qquad\,\, \del_y=-\del_s+\half\del_X
\qquad\,\, \del_x^2-\del_y^2=2\del_s\cdot\del_X,\eeq   
we perform a gradient expansion in $\Xi$,  with $\del_s\sim T$ and
 $\del_X \sim 1/\lambda \ll T$, 
and preserve all the terms involving at most one soft
derivative $\del_X$. For instance,
$$A_\mu(X+s/2)\approx A_\mu(X)+ (1/2)(s\cdot \del_X) A_\mu(X).$$
A straightforward calculation yields then:
\beq\label{DIFFAB}
\Xi(s,X)\approx 2\,\del_s\cdot\del_X  
+2ie A_\mu(X)\del^\mu_X 
+2ie (s\cdot\del_XA_\mu) \del^\mu_s 
+2ie(\del_X\cdot A) 
-e^2 (s\cdot\del_XA^2)+\cdots,\nonumber \\ \eeq
where the dots stand for terms which involve at least two soft derivatives
$\del_X$.

Before taking the Wigner transform, we make the difference equation covariant
by multiplying both  sides by the parallel transport $U(y,x)$
(cf. eq.~(\ref{acuteG0})). For the
left hand side, we use the expansion (\ref{DIFFAB}) of  $\Xi(s,X)$, together
with eq.~(\ref{Uapprox}) to obtain:
\beq\label{LHS0}
U(y,x)\left(D^2_x-(D^\dagger_y)^2\right)\left(U(x,y)\acute
G^<(x,y)\right)\approx 2\Bigl(\del_s\cdot\del_X+ies^\mu
F_{\mu\nu}(X)\del_s^\nu\Bigr)\acute G^<(s,X).
\eeq
The right hand side of the difference equation
involves convolutions of the form:
\beq
C(x,y)\equiv\int{\rm d}^4 z\,\Sigma(x,z) \,G(z,y).
\eeq
Upon multiplication by $U(y,x)$, this becomes the gauge invariant quantity:
\beq\label{C0}
\acute C(x,y)&=&U(y,x)\int{\rm d}^4 z \,U(x,z)\acute\Sigma(x,z)U(z,y)\acute
G(z,y)\nonumber\\ 
 &=& \int{\rm d}^4 z\,  P(x,y,z)\, \acute\Sigma(x,z)\acute
G(z,y),
\eeq
where we have set $\acute C(x,y)\equiv U(y,x)C(x,y)$, and $P(x,y,z)$
denotes the following plaquette:
\beq\label{P0}
P(x,y,z)\equiv U(y,x) U(x,z) U(z,y).
\eeq
In line with the approximations in eq.~(\ref{LHS0}), we need the gradient
expansion of eqs.~(\ref{C0}) and (\ref{P0}) up to terms involving
one soft derivative of the background field. In each of the parallel
transporters, we choose the path $\gamma$ 
to be the straight line (cf. the discussion before eq.~(\ref{Uapprox})).
Then, the plaquette (\ref{P0}) can be 
easily expanded around the point $X=(x+y)/2$ to yield:
\beq\label{P1}
P(x,y,z)\approx \exp\left\{ -\,\frac{ie}{4} s^\mu F_{\mu\nu}(X)
\delta^\nu\right\}\,\approx\,1\, -\,\frac{ie}{4} s^\mu F_{\mu\nu}(X)
\delta^\nu,
\eeq
where $s\equiv x-y$, $X\equiv (x+y)/2$, $\delta\equiv 2(z-X)$.

We are now in position to take the  Wigner  transform. The only delicate
step concerns the transformation of  $\acute C(x,y)$, which is given by:
\beq\label{AOB}
\acute C(k,X)\approx \acute\Sigma(k,X)\acute G(k,X)+\,
\frac{i}{2}\,\Bigl\{\acute \Sigma,\, \acute
G\Bigr\}_{P.B.}-\frac{i}{2}
eF_{\mu\nu}(X)\left(\del^\mu_k\acute\Sigma\right)\left(\del_k^\nu
\acute G\right)\,+\,...\,,\eeq where
$\{A,B\}_{P.B.}$ denotes a Poisson bracket:
\beq\label{poisson}
\Bigl\{A,\,{ B}\Bigr\}_{P.B.}\equiv \del_k A\cdot\del_X{ B}
\,-\,\del_X A\cdot\del_k{ B}\,.\eeq
The third term in the r.h.s. of eq.~(\ref{AOB}), involving
$F_{\mu\nu}(X)$, comes from the plaquette (\ref{P1}) and is
therefore sensitive to the choice of $\gamma$. To the accuracy
where this term is important, we expect the gauge-invariant Wigner functions
$\acute\Sigma(k,X)$ and $\acute G(k,X)$ to be path-dependent as well.
However, such path-dependent terms will disappear in the final form
of the Boltzmann equation that we shall obtain (see eq.~(\ref{BOLO})
below).

By using eqs.~(\ref{LHS0}) and (\ref{AOB}), the
difference equation finally becomes:
\beq\label{KBSQED1}
2(k\cdot \del_X-e k\cdot F\cdot \del_k) \acute
G^<&+&(\del_X^\mu \hat\Sigma)\del_\mu^k \acute G^<
-\left\{ {\rm Re} \acute\Sigma_R,\acute G^<\right\}_{P.B.}
-\left\{  \acute\Sigma^< ,{\rm Re}\acute G_R\right\}_{P.B.}\nonumber\\
&+& eF^{\mu\nu}\left( (\del_\mu^k 
\acute\Sigma^<)(\del_\nu^k {\rm Re}
\acute G_R)+(\del_\mu^k {\rm Re}
\acute\Sigma_R)(\del_\nu^k
\acute G^<)\right)\nonumber\\ &=&
-\left( \acute G^>\acute \Sigma^< -\acute \Sigma^> \acute G^<
\right)
\eeq
In deriving the equation above, we have used 
the following relations:
\beq\label{NEQAG}
\acute G_R(k,X)- \acute G_A(k,X)&=&i\Bigl(\acute
G^>(k,X) - \acute G^<(k,X)\Bigr)\,\equiv\,
i\rho(k,X),\nonumber\\
\acute \Sigma_R(k,X)- \acute \Sigma_A(k,X)&=&i\Bigl(\acute \Sigma^>(k,X)
-\acute\Sigma^<(k,X)\Bigr)\,\equiv\,-i\Gamma(k,X),\eeq
which follow, e.g., from the definitions (\ref{A}) for $G_R$ and $G_A$
after multiplying with $U(y,x)$ and taking the Wigner transform.
Note that the right hand sides of eqs.~(\ref{NEQAG}) define two
new Wigner functions, $\rho(k,X)$ and $\Gamma(k,X)$, which
are real quantities (cf. the discussion after eq.~(\ref{G<WIG}))
and can be seen as off-equilibrium generalizations
of the corresponding spectral densities in equilibrium
(recall eq.~(\ref{G><k})).
In terms of these functions we have, for instance:
\beq\label{RETWIG}
\acute G_R(k,X)=\int_{-\infty}^\infty \frac{{\rm d}k_0^\prime}{2\pi}\,
\frac{\rho(k_0^\prime,{\bf k}, X)}{k_0^\prime-k_0-i\eta}\,,\qquad
\acute\Sigma_R(k,X)=-\int_{-\infty}^\infty \frac{{\rm d}k_0^\prime}{2\pi}\,
\frac{{\Gamma}(k_0^\prime,{\bf k}, X)}{k_0^\prime-k_0-i\eta}\,.\eeq
Eq.~(\ref{KBSQED1}) also involves:
\beq
2{\rm Re}\,\acute G_R\,=\,\acute G_R+\acute G_A,\qquad
2{\rm Re}\,\acute\Sigma_R(k,X)\,=\,
\acute \Sigma_R+\acute\Sigma_A.\,\,
\eeq

Further  manipulations allow us to put eq.~(\ref{KBSQED1}) in the form:
\beq\label{KBSQED2}
(2k -\del_k {\rm Re}\acute\Sigma)\cdot({\rm d}_X\,\acute G^<)
&+&(\del_X {\rm Re}
\acute \Sigma)\cdot (\del_k \acute G^<)
\nonumber\\
 &-& (\del_k\acute\Sigma^<)\cdot({\rm d}_X\,{\rm Re}\acute G_R)
+(\del_X\acute\Sigma^<)\cdot (\del_k {\rm Re}\acute
G_R ) \nonumber\\  &=&  -\left( \acute G^>\acute \Sigma^< -\acute
\Sigma^>\acute G^<\right),
\eeq
where ${\rm Re}\,\acute \Sigma \equiv {\rm Re}\,\acute 
\Sigma_R+\hat\Sigma$, and 
${\rm d}_X^\mu\equiv \del_X^\mu-e
F^{\mu\nu}(X)\del_\nu^k$. 
It is interesting to note that 
the corresponding equation for a scalar field theory (like
$\lambda \phi^4$) can be obtained by simply 
replacing  ${\rm d}_X^\mu$ by $\del_X^\mu$ in the
above equation \cite{prept}.

In equilibrium, both sides of eq.~(\ref{KBSQED2}) are identically zero.
This is obvious for the terms in the l.h.s.,
which involve the soft derivative $\del_X$ or mean field insertions,
and can be easily
verified for the terms in the r.h.s. by using the KMS conditions
for $G_{eq}$ and $\Sigma_{eq}$ 
(cf. eq.~(\ref{KMSk})). Thus, eq.~(\ref{KBSQED2})
is a transport equation which describes the space-time evolution
of long-wavelength fluctuations in the average
density of the charged particles. It holds to leading order 
in the gradient expansion (that is, up to terms
involving at least two powers of the soft derivative),
and to all orders in the interaction coupling strength.

To conclude this section, note that, 
within the previous approximations (that is, up to terms
involving at least two soft derivatives), the retarded
propagator $\acute G_R(k,X)$ satisfies an equation
which is formally identical to that it obeys in equilibrium:
\beq\label{Gretoff}
\Bigl(k^2-\hat\Sigma(X) - \acute\Sigma_R(k,X)\Bigr)
\,\acute G_R(k,X)\,=\,-1\,.\eeq
In order to obtain this equation, start
with eq.~(\ref{eqGR}) for $G_R(x,y)$ together with its conjugate
equation where the differential operator acts on $y$; then, consider
the {\it sum} of these two equations, and perform a gauge-invariant
gradient expansion as above. In this expansion, all the terms involving 
one soft derivative $\del_X$ cancel, and the same holds also for the
terms involving the soft mean field.
From eqs.~(\ref{RETWIG}) and (\ref{Gretoff}), we deduce
an expression for the off-equilibrium spectral density:
\beq\label{AOE}
\rho (k,X)\,=\,2\,{\rm Im}\,\acute G_R(k,X)
\,=\,\frac{\Gamma(k,X)}{\Bigl(k^2
- {\rm Re}\,\acute\Sigma(k,X)\Bigr)^2\,+\,
\Bigl(\Gamma(k,X)/2\Bigr)^2}\,,\eeq
which will be useful in discussing the quasiparticle approximation below.

\subsection{Mean-field and quasiparticle approximations}

In order to make progress with eq.~(\ref{KBSQED2}) further approximations
are needed. In particular, we shall use below
perturbation theory to express
the self-energies  $\acute \Sigma^<$ and $\acute \Sigma^>$ in
terms of the propagators  $\acute G^<$ and $\acute G^>$.
As a first step, let us consider the mean field
approximation in which the self energies $\acute\Sigma$
are neglected altogether. The
equation (\ref{KBSQED2}) reduces then to 
$(k\cdot {\rm d}_X)\acute G^<(k,X)=0$, or, more explicitly:
\beq\label{VlasovSQED}
\left( k\cdot \del_X -ek_\mu F^{\mu\nu}(X) \del_\nu^k\right) \acute
G^<(k,X)=0.
\eeq
This equation describes the motion of independent particles in the mean
field $F_{\mu\nu}$. In this approximation the spectral density remains
the same as in the free theory in equilibrium,
as obvious from eq.~(\ref{AOE}): $\rho(k,X)\approx \rho_0(k)\equiv
2\pi \epsilon(k_0)\delta(k^2)N(k_0)$. Accordingly, the solution
to eq.~(\ref{VlasovSQED}) can be written in the form:
\beq\label{GN0}
\acute G^<(k,X)=2\pi\delta(k^2)\Bigl\{ \theta(k_0)
N_+({\bf k},X)+\theta(-k_0)(1+N_-(-{\bf k},X))\Bigr\},
\eeq
where the density matrices $N_\pm({\bf k},X)$ satisfy the
 Vlasov equation \cite{PhysKin}:
\beq\label{VLASOV}
\Bigl(v\cdot \del_X\,\pm\,e({\bf E + v\times B})\cdot \bfgrad_k
\Bigr)N_\pm({\bf k},X)=0,
\eeq
with $v^\mu=(1,{\bf v})$ and
${\bf v} = \hat{\bf k}$ is the velocity of the charged particle.
The density matrices $N_\pm({\bf k},X)$ may be given
the interpretation of classical
phase-space distributions for particles and antiparticles.
In terms of them, the induced current is simply:
\beq\label{JAB0}
j^\mu(x)&=& e \int \frac{{\rm d}^3k}{(2\pi)^3}\,v^\mu\,
\Bigl(N_+({\bf k},X)\,-\,N_-({\bf k},X)\Bigr).\eeq

Going beyond the mean field approximation, we need to take into
account the various effects of the self-energies.
We shall concentrate here on a commonly used approximation 
which consists in neglecting the broadening of the single-particle
states when computing the collision terms, an approximation which we
refer to as the ``quasiparticle 
approximation''. Indeed, the interaction rate $\Gamma$ in
eq.~(\ref{AOE}) is of higher order in $e$
(specifically, $\Gamma \sim (e^2\ln(1/e))T^2$, as we shall see below),
so we can use the mean field spectral density,
$\rho (k,X)\approx \rho_0(k)$, to estimate the collision terms.
At the same time, we shall ignore 
the self-energy terms in the l.h.s. of eq.~(\ref{KBSQED2}). 
That this is consistent can be verified by power counting (we shall do this
explicitly for the QCD case, in Sec. 3.5); it is also physically
motivated from the fact that the role of these terms is to account
for the difference between $\rho (k,X)$ and $\rho_0(k)$ in the
transport equation for $\acute G^<(k,X)$ \cite{prept}.
Thus, in the quasiparticle approximation, the Wigner functions
$\acute G^<(k,X)$ and $\acute G^>(k,X)$ preserve the same on-shell
structure as in the mean field approximation, as displayed
in eq.~(\ref{GN0}).

We thus end up with the following kinetic equation:
\beq\label{BOLO}
2\left( k\cdot \del_X -ek_\mu F^{\mu\nu}(X) \del_\nu^k\right) \acute
G^<(k,X)=-\left( \acute G^>\acute \Sigma^< -\acute
\Sigma^>\acute G^<\right),
\eeq
where, in line with the weak coupling expansion, we choose
the self energies $\acute\Sigma^<$ and $\acute\Sigma^>$ so as to
reproduce the one-photon-exchange scattering in Fig.~\ref{Born}
(Born approximation). As we shall see in the next section,
this generates a collision term of the standard Boltzmann form.
Note also that eq.~(\ref{BOLO}) in independent upon the choice of the
path $\gamma$ in eq.~(\ref{U0}); indeed, the terms
which were explicitly path-dependent in eq.~(\ref{KBSQED2}) have 
disappeared in the approximations leading to eq.~(\ref{BOLO}). Moreover,
we shall verify shortly that, to the order of interest, the self-energies
$\acute\Sigma^<$ and $\acute\Sigma^>$ are path-independent as well.

In computing transport coefficients like
viscosities or electric conductivity (see, e.g.,
Refs. \cite{Baym90,Heisel94a,Baym97,JY96,BP91}), or the quasiparticle
damping rate \cite{Heisel93,prept}, it is only necessary to consider
small off-equilibrium deviations, so that the linearized version of 
eq.~(\ref{BOLO}) can be used. We then write, e.g.,
$\acute G^<\equiv G^<_{eq}+\delta\acute G^<$ and
$\acute \Sigma^<\equiv \Sigma^<_{eq}+\delta\acute \Sigma^<$ 
(with $\delta\acute G \ll G_{eq}$, $\delta\acute\Sigma \ll \Sigma_{eq}$),
and linearize the collision term with
respect to the small fluctuations $\delta\acute G$ and
$\delta\acute \Sigma\,$:
\beq\label{LIN0}
C(k,X)&\equiv&-\Bigl(\acute G^>(k,X)
\acute \Sigma^<(k,X)-\acute\Sigma^>(k,X)\acute G^<(k,X)\Bigr)
\nonumber\\ &\simeq&
-\Bigl(\Sigma^<_{eq}\,\delta\acute G^>-\Sigma^>_{eq}\,\delta\acute G^<\Bigr)
+\Bigl(\delta\acute\Sigma^>\, G_{eq}^<-
\delta\acute\Sigma^<\,G_{eq}^>\Bigr).\eeq
In the quasiparticle approximation, we further have
$G_{eq}^>\approx G_0^>$, $G_{eq}^<\approx G_0^<$ and
$\delta\acute G^< \approx \delta\acute G^>\equiv \delta\acute G$
(since $\acute G^>(k,X) - \acute G^<(k,X) =
\rho_0(k)=G^>_0(k)-G^<_0(k)$).
Then, the linearized collision term takes the form:
\beq\label{COLLIN}
C(k,X) \simeq
-\Gamma_{eq}(k)\,\delta\acute G(k,X)\,+\,
\Bigl(\delta\acute\Sigma^> \,G_0^<\,-\,
\delta\acute\Sigma^<\,G_0^>\Bigr),\eeq
where we have isolated the damping rate in equilibrium
(cf. eq.~(\ref{NEQAG})):
\beq\label{EQG}
\Gamma_{eq}(k)\,= \,\Sigma^<_{eq}(k) - \Sigma^>_{eq}(k).\eeq
Note that the quasiparticle approximation
is not a self-consistent approximation, but it is
in line with the weak coupling expansion: the collision
term generates a width which is not included in the spectral
densities which are used to estimate it; however, the neglected
terms are of higher order than those we have kept.

The most direct application of the formula above is the calculation
of the quasiparticle damping rate \cite{KB62,Heisel93,prept}. To this aim,
we consider a specific off-equilibrium deviation which is obtained 
by adding, at $t_0=0$, a particle with momentum ${\bf p}$ and energy
$p_0=\varepsilon_p\equiv |{\bf p}|$ to a plasma initially in equilibrium. 
Since, for a large system, this is a small perturbation,
we can neglect all mean field effects and assume $N({\bf p},t)$ 
to be only a function of time (here, $N({\bf p},t)\equiv N_+({\bf p},t)$;
cf. eq.~(\ref{GN0})). Moreover, for momenta ${\bf k}\ne {\bf p}$,
the distribution function does not change appreciably from the equilibrium value 
$N(\varepsilon_k)$, so that, to leading order in the external perturbation,
we can ignore the off-equilibrium fluctuations of the self-energies:
$\delta\acute\Sigma^< \approx \delta\acute\Sigma^> \approx 0$.
Then, eqs.~(\ref{GN0}), (\ref{BOLO}) and (\ref{COLLIN}) yield a very
simple equation for the fluctuation
$\delta N({\bf p},t)$ 
(with $\Gamma({\bf p})\equiv \Gamma_{eq}(p_0=\varepsilon_p,{\bf p})$) :
\beq\label{b1}
2\varepsilon_p\frac{\del}{\del t}\delta N({\bf p},t)\,=\,-\Gamma({\bf p})
\delta N({\bf p},t),
\eeq
whose solution shows exponential attenuation in time:
\beq\label{att}
\delta N({\bf p},t)\,=\,\delta N({\bf p},0)\,{\rm e}^{-2\gamma(p)t}\,.\eeq
The quasiparticle {\it damping rate} $\gamma$ is here conventionally defined as
$\gamma(p)\equiv \Gamma({\bf p})/4\varepsilon_p$ \cite{MLB96}.
(This simple picture is actually
complicated by infrared effects to be discussed
in Sec. 2.6 \cite{Smilga90,Pisarski93,lifetime}.)

\subsection{The collision terms}

\begin{figure}
\protect \epsfxsize=11.cm{\centerline{\epsfbox{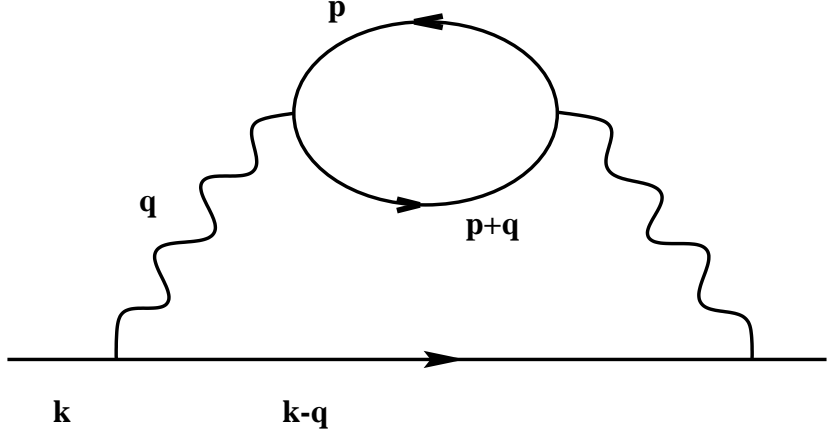}}}
         \caption{The two-loop self-energy diagram which describes collisions
in the Born approximation. The wavy lines denote free, equilibrium,
photon propagators. The other lines are off-equilibrium scalar
propagators.}
\label{S11}
\end{figure}

\begin{figure}
\protect \epsfxsize=10.cm{\centerline{\epsfbox{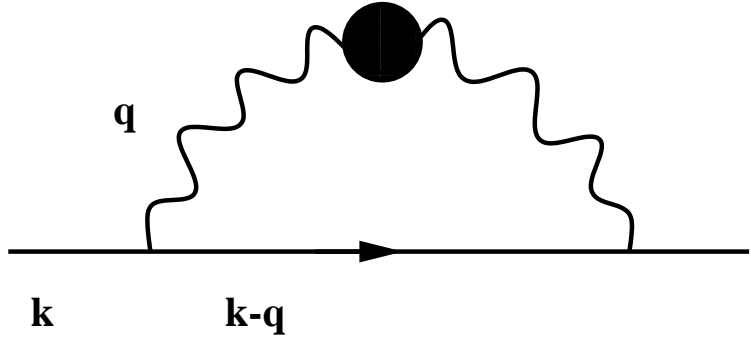}}}
         \caption{The effective one-loop self-energy which
describes collisions in the resummed Born approximation.
The blob on the photon line 
denotes the resummation of the one-loop polarization tensor (cf. 
Fig.~(\ref{PI})).}
\label{S2}
\end{figure}

\begin{figure}
\protect \epsfxsize=13.5cm{\centerline{\epsfbox{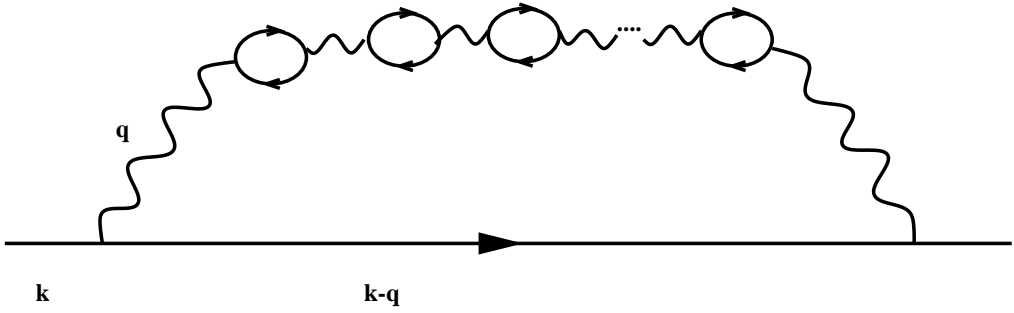}}}
         \caption{One of the multi-loop diagrams (containing $n$ bubble
insertions along the photon line) which is included in the effective 
one-loop diagram in Fig.~\ref{S2}.}
\label{Sn}
\end{figure}

We now turn to the calculation of the collisional self-energy
corresponding to Fig.~\ref{Born}. As it is well known (and will be
verified later), the corresponding transport cross section is 
dominated by relatively hard momentum transfers,
$eT\simle q \simle T$. When the photon momentum is hard,
$q\sim T$, the process in  Fig.~\ref{Born} is described by the two-loop
self-energy depicted in Fig.~\ref{S11} in which all the lines are hard,
and the scalar propagators are to
be understood as off-equilibrium propagators. The photon propagators,
on the other hand, are just free propagators in equilibrium. At
soft momenta $q\sim eT$, the relevant self-energy is given by the
effective one-loop diagram in Fig.~\ref{S2} in which both the internal 
lines denote off-equilibrium propagators; the scalar line is hard,
while the photon line
is soft and dressed by the off-equilibrium polarization tensor
in the one-loop approximation (this is denoted by a blob).
That is, the diagram in Fig.~\ref{S2} involves an infinite series
of bubble insertions along the photon line, as illustrated in 
Fig.~\ref{Sn}.

It is furthermore convenient to recognize that the diagram in Fig.~\ref{S11}
is one of the family of diagrams displayed in Fig.~\ref{Sn}; thus, we can use
the effective one-loop self-energy in Fig.~\ref{S2} to describe the collision
in Fig~\ref{Born} for {\it all} photon momenta, which we shall do in what follows.
To evaluate this diagram, we need the vertex coupling the photon to the
 scalar field in the
presence of  the classical background field
$A_\mu$. This can be read off the Lagrangian:
\beq\label{L0}
{\cal L}&=&\left(D^\dagger_\mu-iea_\mu\right) \phi^\dagger\left( D^\mu+ie
a^\mu\right)\phi\nonumber\\ &=& ie a_\mu\left[ \left(D_\mu\phi\right)^\dagger
\phi-\phi^\dagger\left(D_\mu \phi\right)\right]+e^2 a_\mu a^\mu
\phi^\dagger\phi,
\eeq
where $D_\mu= \del_\mu+ie A_\mu$. There are two relevant
vertices:
$-2ie \phi^\dagger a_\mu D^\mu \phi$ and $ -ie \phi^\dagger
\phi\del^\mu a_\mu$.
The self-energy reads then:
\beq
\Sigma(x,y)&=&e^2\biggl\{ 4\Bigl(D_x^\mu D^{\nu\,\dagger}_y
G(x,y)\Bigr){\cal D}_{\mu\nu}(x,y)+2\Bigl(D^\mu_x G(x,y)\Bigr)
\Bigl(\del^\nu_y {\cal D}_{\mu\nu}(x,y)\Bigr)
\nonumber\\&{}&\qquad +\,2\Bigl(D^{\nu\,\dagger}_y
G(x,y)\Bigr)\Bigl(\del_x^\mu {\cal D}_{\mu\nu}(x,y)\Bigr)
 +G(x,y)\Bigl(\del^x_\mu\del^\nu_y
{\cal D}^{\mu\nu}(x,y)\Bigr)\biggr\}.
\eeq
Here, ${\cal D}_{\mu\nu}(x,y)$ is the off-equilibrium photon
propagator, to be constructed shortly.
By appropriately choosing $x_0$ and $y_0$ along the contour,
we get expressions for both $\Sigma^>$ and $\Sigma^<$. For instance,
$\Sigma^>(x,y)$ will involve $G^>(x,y)$ and $D^>(x,y)$, etc.
Below, to simplify the notations,
the upper indices $>$ and $<$ will be often omitted.

We need then to evaluate the gauge invariant self energy
$\acute\Sigma(x,y)\equiv U(y,x)\Sigma(x,y)$. In doing that, we meet terms
like:
\beq
U(y,x)\,D^\mu_x\,G(x,y)=U(y,x)\,D^\mu_x\left(U(x,y)\acute G(x,y)\right).
\eeq
Performing the gradient expansion of such a term, one gets:
\beq
\lefteqn{ U(y,x)\,D^\mu_x\left(U(x,y)\acute G(x,y)\right)}\nonumber\\
&\approx & {\rm e}^{ies\cdot A(X)}\left(\del_s^\mu+ieA^\mu(X)\right)\left\{
{\rm e}^{-ies\cdot A(X)}\acute G(s,X)\right\}\nonumber\\
&\approx & \del^\mu_s \acute G(s,X).
\eeq
Note that, in the above manipulations, we have used the simple
approximation (\ref{Uapprox}) for $U(x,y)$, which makes the final result
independent of the choice the path in the Wilson line.
The same holds for all the other results in this section.

Similarly, we get
$
\del^\nu_y {\cal D}_{\mu\nu}(x,y)\approx -\del^\nu_s {\cal D}_{\mu\nu}(s,X), 
$
so that the expression \\
$D^\mu_x G(x,y)\del^\nu_y
{\cal D}_{\mu\nu}(x,y)$ becomes simply $-\del^\mu_s \acute G(s,X)\del^\nu_s
{\cal D}_{\mu\nu}(s,X)$. Proceeding in the same way for the other terms, and
performing the Wigner transform, we get:
\beq\label{SAB}\acute \Sigma(k,X)\,\approx\, e^2
 \int\frac{{\rm d}^4 q}{(2\pi)^4}\,\left(2k^\mu-q^\mu\right)
\left(2k^\nu-q^\nu\right)\acute G(k-q,X) {\cal D}_{\mu\nu}(q,X).
\eeq
This is formally the same expression as in equilibrium, except for the fact
that $\acute G(k-q,X)$ and  ${\cal D}_{\mu\nu}(q,X)$ are off-equilibrium
propagators and $k^\mu$ has to be interpreted as the kinetic
momentum. (Note that the photon propagator ${\cal D}_{\mu\nu}$ does not
need  a special treatment since it is invariant under the gauge transformations
of the background field.)

\begin{figure}
\protect \epsfxsize=13.cm{\centerline{\epsfbox{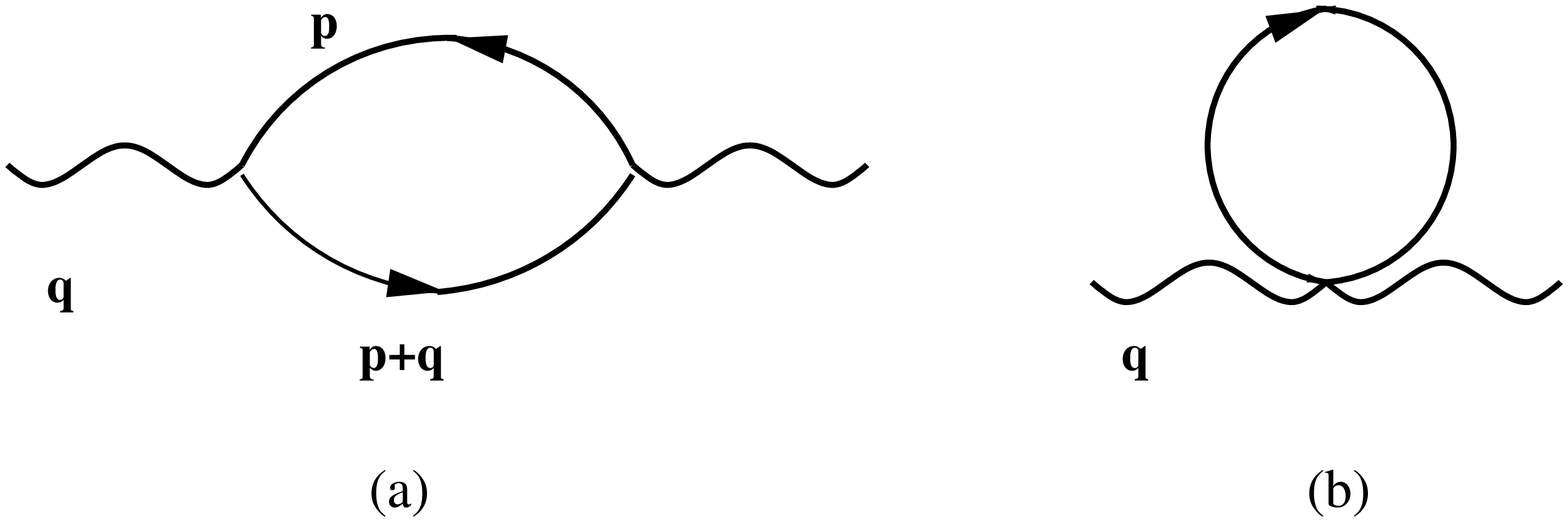}}}
         \caption{One-loop contributions to the
photon self-energy in SQED. All the internal lines are off-equilibrium
propagators.}
\label{PI}
\end{figure}

The photon propagator ${\cal D}_{\mu\nu}(x,y)$
obeys Dyson-Schwinger equations similar to
eqs.~(\ref{eq1})--(\ref{eqGR}). Specifically,
\beq\label{KBD0}
\left(g^{\mu\nu}\del^2-\del^\mu\del^\nu- \hat\Pi^{\mu\nu}\right)_x
{\cal D}^<_{\nu\rho}(x,y)
\,=\,\int {\rm d}^4z\,\Bigl(\Pi_R {\cal D}^< +
\Pi^<{\cal D}_A\Bigr)^\mu_{\,\,\,\rho}(x,y),\eeq
and similarly:
\beq\label{KBRD0}
\left(g_{\mu\nu}\del^2-\del_\mu\del_\nu- \hat\Pi_{\mu\nu}\right)_x
{\cal D}_R^{\nu\rho}(x,y)
\,-\,\int {\rm d}^4z\,\Pi^R_{\mu\nu}
 {\cal D}_R^{\nu\rho}(x,y)
\,=\,\delta_\mu^{\rho}\delta^{(4)}(x-y)\,,\eeq
where, to the order of interest, $\Pi_{\mu\nu}(x,y)$ is given
by the one-loop diagrams in Fig.~\ref{PI}.
That is, $\hat\Pi^{\mu\nu}(x)$ is the tadpole contribution
in Fig.~\ref{PI}.b, while the non-local self-energies $\Pi_R(x,y)$
and $\Pi^<(x,y)$ are determined by the graph in Fig.~\ref{PI}.a.
From the equations above, we deduce the following relation
between ${\cal D}^<$ and $\Pi^<$:
\beq\label{WRA0}
{\cal D}^<_{\mu\nu}(x,y)\,=\,-\int {\rm d}^4z_1{\rm d}^4z_2\,\,
\Bigl({\cal D}_R(x,z_1)\,
\Pi^<(z_1,z_2)\,{\cal D}_A(z_2,y)\Bigr)_{\mu\nu},\eeq
which becomes, after a gradient expansion,
\beq\label{DOEQ0}
{\cal D}^<_{\mu\nu}(q,X)\,\approx\,-\,
\Bigl({\cal D}_R(q,X)\,
\Pi^<(q,X)\,{\cal D}_A(q,X)\Bigr)_{\mu\nu},\eeq
up to corrections of O$(\del_X/q)$. Since, as we shall see shortly,
the collision terms are saturated by momenta
$q\simge eT$, the corrections to eq.~(\ref{DOEQ0})
are of higher order in $e$ provided $\del_X\simle e^2 T$.
A similar relation holds between ${\cal D}^>(q,X)$ and $\Pi^>(q,X)$.

It should be observed here that a new scale is entering the gradient
expansion. In most situations before, the soft derivative $\del_X$
appeared in combinations such as $s\cdot\del_X$ with the magnitude
of the non-locality $s^\mu$ fixed by thermal fluctuations:
$s\sim 1/T$. In eq.~(\ref{WRA0}), however, the non-localities
$x-z_1$ or $z_2-y$ are of order $1/q$ and may be interpreted as the
range of the effective interaction between the colliding particles.
Thus, the validity of the gradient expansion in this case relies
on the range of this effective interaction being small compared to
the scale of the inhomogeneities, as measured by $\del_X^{-1}$.
Now, the range of the effective interaction depends on the specific 
transport processes one is looking at. In most cases, and as a result
of cancellations to be exhibited in the next subsection, 
this range is typically of order $1/T$ (and marginally $1/eT$)
so that the gradient expansion is indeed valid to calculate
transport coefficients to leading order in $e$ already for
processes taking place on a scale $1/e^2T$.

To construct the photon self-energy out of equilibrium, we use
the interaction vertices from eq.~(\ref{L0}) and obtain,
to the order of interest,
\beq\label{PTP}
\hat\Pi_{\mu\nu}(X)=-2g_{\mu\nu}e^2
\int\frac{{\rm d}^4 p}{(2\pi)^4}\, \acute G^<(p,X),
\eeq
and (compare to eq.~(\ref{SAB})) :
\beq\label{P<0}
\Pi_{\mu\nu}^>(q,X)\,\approx\,e^2\int\frac{{\rm d}^4 p}{(2\pi)^4}\, 
\left(2p_\mu+q_\mu\right)
\left(2p_\nu+q_\nu\right)\acute G^>(p+q,X) \acute G^<(p,X),
\eeq
together with a similar expression for $\Pi^<(q,X)$ 
which involves $\acute G^<(p+q,X)$ and $\acute G^>(p,X)$.
These expressions are gauge invariant, as expected.
The tadpole piece (\ref{PTP}) enters the calculation of the retarded
propagator ${\cal D}_{R}(q,X)$,
which is related to the self-energy $\hat\Pi(X)
+ \Pi_R(q,X)$
by the same equation as in equilibrium (cf. eq.~(\ref{Gretoff})).

\begin{figure}
\protect \epsfxsize=13.5cm{\centerline{\epsfbox{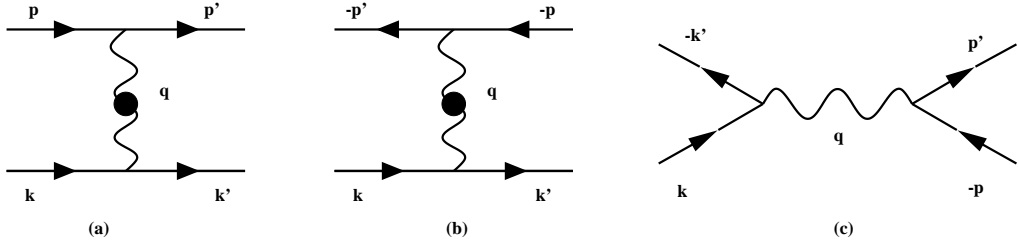}}}
         \caption{Scattering processes described by the collision
term in eq.~(\ref{COLAB}) : (a) particle-particle scattering; 
(b) particle-antiparticle scattering ($t$-channel);
(c) particle-antiparticle annihilation ($s$-channel).
In Fig. (c), the virtual photon is always hard, so it needs no resummation.}
\label{PROC}
\end{figure}

By collecting the previous results, we
finally obtain the following collision term:
\beq\label{COLAB} \lefteqn{
C(k,X)\,=\,-\int\frac{{\rm d}^4p}{(2\pi)^4} \int\frac{ {\rm d}^4 q}{(2\pi)^4}\,
\left| {\cal M}_{pk\to p'k'}\right|^2
\qquad\qquad}\nonumber\\ & & \times\left\{
\acute G^<(k,X) \acute G^<(p,X)\acute G^>(k',X)\acute G^>(p',X)-\acute
G^>(k,X) \acute G^>(p,X)\acute G^<(k',X)\acute G^<(p',X)\right\},\nonumber\\
\eeq
where $p'=p+q$, $k'=k-q$, and ${\cal M}_{pk\to p'k'}$
is the scattering matrix element:
\beq\label{MAB}
\left| {\cal M}_{pk\to p'k'}\right|^2 =e^4 (k+k')^\mu (p+p')^\nu
(k+k')^\alpha (p+p')^\beta [{\cal D}_R(q,X)]_{\mu\nu}
[{\cal D}_A(q,X)]_{\alpha\beta}\,.
\eeq
This collision term has the standard Boltzmann structure,
with a gain term and a loss term.
To be in line with the previous approximations, 
this must be evaluated with the Wigner functions
$\acute G^<$ and $\acute G^>$ in the quasiparticle
approximation, i.e., $\acute G^<(k,X)$ has the on-shell structure
exhibited in eq.~(\ref{GN0}), while $\acute G^>(k,X)$ reads similarly :
\beq\label{GN1}
\acute G^>(k,X)=2\pi\delta(k^2)\Bigl\{ \theta(k_0)
(1+N_+({\bf k},X))+\theta(-k_0)N_-(-{\bf k},X)\Bigr\}.
\eeq
Thus, the four energy variables $k_0,\,p_0,\,k^\prime_0$ and
$p^\prime_0$ in eq.~(\ref{COLAB})
are always on shell (e.g., $|k_0|=\varepsilon_k\equiv |{\bf k}|$),
but they can be either positive ($k_0=\varepsilon_k$) or negative
($k_0=-\varepsilon_k$), corresponding to particles and 
antiparticles, respectively. Choose $k_0>0$ for definitness;
then (\ref{COLAB}) describes the $t$-channel particle-particle scattering
depicted as Fig.~\ref{Born} (when all the energy variables are positive),
but also the $t$-channel particle-antiparticle scattering (when 
$k^\prime_0$ is positive, but $p_0$ and $p^\prime_0$ are both negative),
and the particle-antiparticle anihilation (or $s$-channel 
scattering: $p_0$ and $k^\prime_0$ negative, and $p^\prime_0$ positive).
These various processes are illustrated in Fig.~\ref{PROC}.

\subsection{Quasiparticle lifetimes vs. relaxation times}

At the end of Sec. 2.4, we have seen
that the collision term (\ref{COLLIN}) yields the quasiparticle lifetime 
$\tau \sim 1/\gamma$, 
which is dominated by soft momentum transfers, $q\simle eT$, 
and is typically $\tau \sim 1/(e^2 T\ln(1/e))$ \cite{Pisarski93,lifetime}.
For transport phenomena, however, it is well known
\cite{MLB96,Baym90,Heisel94a,Baym97} that the Abelian collision
term (\ref{COLAB}) is saturated by relatively large momentum transfers
$eT\simle q \simle T$, and the typical relaxation time
for off-equilibrium perturbations is $\tau_{tr} \sim 1/(e^4 T\ln(1/e))$.
The fact that $\tau_{tr} \gg \tau$ is due to
specific infrared cancellations in the collision term, that
we shall discuss now. Note that in QCD the colour algebra prohibits similar 
cancellations in the calculation of colour relaxation processes,
as we shall see in Secs. 3.8 and 4 below.

Consider then the linearized version of the collision term,
as given by eq.~(\ref{COLLIN}).
It is convenient to define (with $\beta = 1/T$) :
\beq\label{GNW}
\delta\acute G(k,X)\,\equiv \,\rho_0(k)\delta N(k,X)&\equiv&
-\rho_0(k) W(k,X)\,\frac{{\rm d}N}{{\rm d}k_0}\nonumber\\
&=&\beta \rho_0(k) W(k,X)N(k_0)[1+N(k_0)]\,\eeq
where (cf. eq.~(\ref{GN1})) :
\beq\label{DELG0}
\delta N(k,X)\,=\,\theta(k_0)
\delta N_+({\bf k},X))-\theta(-k_0)\delta N_-(-{\bf k},X).
\eeq
The function $W(k,X)$ describes the local distorsion of the
momentum distribution, as may be seen from the following equation:
\beq 
N(k,X)\,\equiv\,N(k_0)\,-\, W(k,X)\,\frac{{\rm d}N}{{\rm d}k_0}
\,\approx\,N(k_0-W(k,X)).\eeq
In terms of these new functions, the linearized collision term
takes a particularly simple form,
\beq\label{LINCOL}
C(k,X)&\approx&
-\beta \rho_0(k)\int {\rm d}{\cal T}\,|{\cal M}|^2\,N(k_0)N(p_0)
[1+N(k^\prime_0)][1+N(p^\prime_0)]\nonumber\\&{}&\qquad\,\,\,\,\times
\Bigl\{W(k,X)+W(p,X)-W(k^\prime,X)-W(p^\prime,X)\Bigr\},\eeq
with the following notation for the phase-space integral:
\beq\label{PHI}
\int {\rm d}{\cal T}\,\equiv\,
\int\frac{{\rm d}^4p}{(2\pi)^4} \int\frac{ {\rm d}^4 q}{(2\pi)^4}\,
\rho_0(p)\rho_0(p+q)\rho_0(k-q).\eeq
The matrix element $|{\cal M}|^2$ in eq.~(\ref{LINCOL}) is
to be computed with the {\it equilibrium}
retarded and advanced photon propagators ${\cal D}_{R,\,A}(q)$
(cf. eq.~(\ref{MAB})).

Following eq.~(\ref{COLLIN}), we identify the damping rate
$\Gamma(k)\equiv \Gamma_{eq}(k)$
as the coefficient of $\delta\acute G(k,X)$ in
the r.h.s. of eq.~(\ref{LINCOL}), that is, as
the term involving the fluctuation $W(k,X)\,$:
\beq\label{DC1}
 C_1(k,X)&=&
-\beta \rho_0(k)W(k,X)\int {\rm d}{\cal T}\,|{\cal M}|^2\,N(k_0)N(p_0)
[1+N(k^\prime_0)][1+N(p^\prime_0)]\nonumber\\
&\equiv&-\,\Gamma(k)\delta\acute G(k,X).\eeq
One can verify that within the present approximation the
above expression of $\Gamma$ satisfies indeed eq.~(\ref{EQG}).

Moreover, in eq.~(\ref{DC1}), 
$\Gamma(k)$ must be evaluated on the tree-level
mass-shell (i.e., at $|k_0|=\varepsilon_k$), since it is multiplied by the
on-shell fluctuation $\delta\acute G(k,X)$. 
This determines the quasiparticle damping rate,
$\gamma \equiv (1/4k)\Gamma(k_0=k)$,
which is, however, well known to be infrared divergent in
the present approximation \cite{BP90,Smilga90,Pisarski93,lifetime}.
Specifically, the leading contribution to $\gamma$ comes from
soft momenta exchange $q\simle eT$ in the $t$-channel collisions
in Figs.~\ref{PROC}.a and b.
To evaluate this contribution, we can neglect $q_0$ next to $p_0$
and $k_0$ in the thermal distributions in eq.~(\ref{DC1}), and get:
\beq\label{GTF10}
\Gamma(k)\simeq\,\int {\rm d}{\cal T}\,|{\cal M}|^2\,N(p_0)[1+N(p_0)].\eeq
To the order of interest,
we need the resummed photon propagator in the
``hard thermal loop'' approximation \cite{BIO96,MLB96},
to be denoted as ${}^*{\cal D}_R^{\mu\nu}(q)$. This yields
 (for $q\ll k,\,p$):
\beq\label{COLM10}
|{\cal M}|^2\,\simeq\,16e^4\varepsilon_k^2\varepsilon_p^2\,
 \Big|{}^*{\cal D}_l(q)
+ ({\bf \hat q \times v})\cdot ({\bf \hat q \times v}^\prime)\,
{}^*{\cal D}_t(q)\Big|^2.\eeq
where ${}^*{\cal D}_l$ and ${}^*{\cal D}_t$ are the longitudinal 
(or electric) and the transverse (or magnetic) components of the
retarded propagator, with the following IR behaviour
(below, $m_D$ is the Debye mass, $m_D^2=e^2 T^2/3$) \cite{BIO96,MLB96} :
 \beq\label{DSTAT}
{}^*{\cal D}_l(q_0\to 0,q)\,\simeq\,\frac{- 1}{q^2 + m_{D}^2}\,,
\qquad {}^*{\cal D}_t(q_0\ll q)\simeq\,\frac{1}
{q^2-i\,(\pi q_0/4q)\,m_D^2}\,.\eeq
Because of Debye screening, the electric contribution to the damping rate
$\gamma_l$ is finite and of order $e^4 T^3/m_D^2 = {\rm O}(e^2T)$.
In the magnetic sector,  the
dynamical ($q_0\ne 0$) screening \cite{BIO96,MLB96}
is not enough to make finite $\gamma_t$, which remains logarithmically
divergent (see Ref. \cite{lifetime} for more details):
\beq\label{G2LR}
\gamma_t &\simeq& \frac{e^4 T^3}{24}\,
\int{\rm d}q  \int_{-q}^q\frac{{\rm d}q_0}{2\pi}
\,\frac{1}{q^4 + (\pi m_D^2 q_0/4q)^2} \nonumber\\
&\simeq&
\frac{e^2T}{4\pi}\,\int_{\mu}^{m_D}\frac{{\rm d}q}{q}\,=\,
\frac{e^2T}{4\pi}\,\ln\frac{m_D}{\mu}\,,\eeq
where $\mu$ is an IR cutoff
and we have retained only the dominant,
logarithmically divergent, contribution.
The remaining IR divergence in eq.~(\ref{G2LR}) 
is associated to the unscreened static magnetic interactions.
Since the latter have an infinite range, one may worry that the
gradient expansion may become invalid in the calculation of the damping
rate (cf. the remark after eq.~(\ref{DOEQ0})). Recall, however,
that in the calculation of $\gamma$, the particles with which
the quasiparticle interacts are in equilibrium, and constitute therefore a
uniform background (cf. the discussion at the end of Sec. 2.4.)
Thus, the question of the relative sizes of the range
of the interaction and that of the space-time inhomogeneities is not an
issue here. Rather, the IR divergence in eq.~(\ref{G2LR}) 
is an artifact of the perturbative expansion and can be eliminated
by a specific resummation \cite{lifetime}
which goes beyond the approximations
performed in deriving the Boltzmann equation
(see however Ref. \cite{BDV98}).

The IR problem
of the damping rate does not show up in the calculation of the
transport coefficients, because the
IR contribution to $\Gamma$, the first term in the r.h.s.
of eq.~(\ref{LINCOL}), is actually compensated by a similar 
contribution to the third term, involving $W(k',X)\,$:
indeed, for soft $q$, $W(k',X)\equiv W(k-q,X)
\approx W(k,X)$, so that the first and third terms in 
eq.~(\ref{LINCOL}) cancel each other. 
As we shall see in Sec. 4, this can be understood as a cancellation
between self-energy and vertex corrections in ordinary
Feynman graphs. A similar cancellation occurs between the other
two terms in eq.~(\ref{LINCOL}), namely $W(p,X)$ and $W(p',X)$.
Thus, in order to see the leading IR ($q\ll T$) behaviour
of the full integrand in eq.~(\ref{LINCOL}), one has to expand
$W(k',X)$ and $W(p',X)$ to higher orders in $q$. This 
generates extra factors of $q$ which remove the most severe
IR divergences in the collision integral. As a result, the typical
rate involved in the calculation of the transport coefficients
is $\Gamma_{tr}\,\sim\, e^4T\ln(1/e)$, where the logarithm
originates from screening effects at the scale $eT$.

Of course, the simple arguments above are only good enough to provide
an order-of-magnitude estimate for the transport relaxation times.
In order to compute transport coefficients, one has to solve
the Boltzmann equation (\ref{BOLO}) with the linearized collision term
(\ref{LINCOL}), which is generally complicated. Explicit
solutions can be found, e.g., by using specific
Ans\"atze for the unknown function $W(k,X)$, or by variational methods.
Some calculations of this kind can be found in Refs.
\cite{Baym90,Heisel94a,Baym97,Jeon93,JY96}.

\setcounter{equation}{0}

\section{Boltzmann equation for hot QCD}
\setcounter{equation}{0}

We now come to the case of the high temperature
Yang-Mills plasma. As mentioned in the Introduction, we are interested
in the regime of  {\it ultrasoft} colour
excitations propagating on  a typical scale $\lambda
\sim 1/g^2T$. (More precisely, the spatial gradients of the fields $A^\mu_a$
are of order $g^2 T$, but their time derivatives can be even softer, i.e., of
order $ g^4T$.) The relevant response function is the induced colour
current, which we shall eventually express in the form (with $v^\mu =(1,{\bf
k}/k)$):
\beq\label{jb}
j^a_\mu(X)=2g\int\frac{{\rm d}^3k}{(2\pi)^3}\,v_\mu
\,{\rm Tr}\,\Bigl(T^a\delta N({\bf k},X)\Bigr).\eeq
where $\delta N^{ab}({\bf k},X)$ is a  density matrix  in colour space. 
The overall factor 2 stands for the two transverse polarizations. 

The density matrix $\delta N^{ab}({\bf k},X)$ is a functional of the
average fields $A^\mu_a$ and must transform covariantly under the gauge
transformations of the latter. That is, under the gauge transformation
($A_\mu=A_\mu^a T^a$):
\beq\label{GT}
A^\mu(X)\longrightarrow  h(X)\left(A^\mu(X)- \frac{i}{g}\,
\del^\mu\right) h^\dagger(X),
\eeq
where  
$h(x)=\exp(i\theta^a(x)T^a)$, 
we must have:
\beq
\delta N_{ab}({\bf k},X) \longrightarrow h_{a\bar a}(X)
\delta N_{\bar a\bar b}({\bf k},X) h^\dagger_{\bar b b}(X).\eeq
Indeed, this ensures that $j^a_\mu(X)$ transforms as a colour vector:
$j^\mu_a \to h_{ab} j^\mu_b$, or, in matrix notations,
\beq j^\mu(X)\equiv j^\mu_a T^a \longrightarrow
h(X) j^\mu(X) h^\dagger(X). \eeq
(This should be contrasted with the Abelian case, where both
the current and the distribution function are gauge invariant.)

The covariance of the density matrix $\delta N_{ab}({\bf k},X)$ should result
from a corresponding property of the off-equilibrium gluon propagator
$G^{\mu\nu}_{ab}(x,y)=\langle {\rm T}a^\mu_a(x)a^\nu_b(y)\rangle$  from which it
originates. However this propagator depends not only upon the choice of a
gauge for the average field $A^\mu_a$, but also on the gauge-fixing condition
for the fluctuating field $a^\mu_a$. 
With a generic gauge fixing,  $G^{\mu\nu}_{ab}(x,y)$ transforms in a
complicated way under the gauge transformations of $A^\mu_a$.
The situation becomes simpler when one uses the so-called
``background field gauge'' to be introduced in the next
subsection \cite{Witt67,Abbott81}. Then the gauge fixing term is covariant
under the gauge transformations of the average field $A^\mu_a$,
and  the gluon propagator $G(x,y)$ can be  turned into a
 covariant quantity by attaching Wilson lines in $x$ and $y$.
We shall then be able to maintain explicit
gauge symmetry with respect to the background field at each
step of our calculation. 

\subsection{The background field gauge}

In this method, one splits the gauge field into a
classical background field $A^a_\mu$, to be later identified with the
 average field, and a fluctuating quantum field $a_\mu^a$. The generating
functional of Green's functions is written as:
\beq\label{Zbk}
Z[j;A]=\int {\cal D}a{\cal D}\bar\zeta {\cal D}\zeta\,
\,{\rm e}^{iS_{FP}[a,\zeta,\bar\zeta;A] + i\int_C {\rm d}^4x 
j_\mu^b a_b^\mu},\eeq
with the  Fadeev-Popov action:
\beq \label{SFP}
S_{FP}[a,\zeta,\bar\zeta;A]=\int_C{\rm d}^4 x \biggl\{-\,\frac{1}{4}
\Bigl(F_{\mu\nu}^a
[A+a]\Bigr)^2+ \frac{1}{2\lambda}\Bigl(D_i[A] a^i\Bigr)^2 
+\bar\zeta^a\Bigl(D_i[A] D^i[A+a]\Bigr)_{ab}\zeta^b
\biggr\},\nonumber\\\eeq
where $D_\mu[A+a]=\del_\mu + ig(A_\mu + a_\mu)$ is the covariant derivative
for the total field $A_\mu + a_\mu$, and $F_{\mu\nu}^a[A+a]$ is the 
respective field strength tensor.  Furthermore, the gauge-fixing term
$(1/2\lambda)(D_i[A] a^i)^2$, which is of the Coulomb type,
is manifestly covariant with respect to the gauge transformations
of the background gauge field $A_\mu$. Accordingly, the exponential
in eq.~(\ref{SFP}) is invariant with respect to the following
transformations  (with matrix notations: $h(x)=\exp(i\theta^a(x)T^a)$, 
 $a_\mu=a_\mu^b T^b$, $\zeta =\zeta^a T^a$, etc.):
\beq\label{GT1}
A_\mu\,\to\,h(A_\mu -({i}/{g})\del_\mu)h^\dagger,
\qquad \,\,\,\,
j_\mu\,\to\, h j_\mu h^\dagger,\nonumber\\
a_\mu \,\to\, h a_\mu h^\dagger,\qquad\,\,
\zeta \,\to \,h \zeta h^\dagger,\qquad\,\,
\bar\zeta \,\to \,h^\dagger \bar\zeta h.\eeq
(Note the {\it homogeneous} transformations of the
quantum gauge fields $(a_\mu)$ and ghost fields $(\zeta,\,
\bar\zeta)$ in the equations
above.) Because of this symmetry, the generating functional $Z[j;A]$ is
invariant under the normal gauge transformations of its arguments, given by 
the first line of eq.~(\ref{GT1}). Then, the gluon Green's functions, 
derived from $Z[j;A]$  by differentiation with respect to $j_\mu^a$, are
gauge covariant under the same transformations.

The physical Green's functions are obtained by identifying 
the total average field to the
background field. This implies:
\beq\label{AVA}
\langle a_\mu^b(x)\rangle\,\equiv\,\frac{\delta \ln Z[j;A]}
{i\delta j^\mu_b(x)}\,=\,0\eeq 
which determines a functional relation between
the external current and the average field; we write this
as $j=j[A]$.
Then, the  2-point function  is obtained as:
\beq\label{DBK}
G_{\mu\nu}^{ab}(x,y)\,\equiv\,\langle{\rm T}_C 
\,a_\mu^a(x)a_\nu^b(y)\rangle
\,=\,-\frac{\delta^2 \ln Z[j;A]}
{\delta j^\mu_a(x)\delta j^\nu_b(y)}
\biggr|_{j[A]}\,.\eeq
Under the gauge transformations (\ref{GT1}) of $A^\mu$, it transforms
covariantly:
\beq\label{DTr}
G_{ab}^{\mu\nu}(x,y)&\to&  h_{a\bar a}(x) \,G_{\bar a\bar b}
^{\mu\nu}(x,y)\, h^\dagger_{\bar b b}(y).\eeq
The ghost propagator,
\beq\label{Delta}
\Delta^{ab}(x,y)\,\equiv\,
\langle{\rm T}_C\,\zeta^a(x)\bar\zeta^b(y)\rangle,\eeq
has the same transformation property.
Similar covariance properties hold for the higher point Green's
functions, and for the various self-energies. Note that, in practice, we shall
never have to solve the implicit eq.~(\ref{AVA}) for $j[A]$, since we shall be
able to impose the condition
$\langle a_\mu(x)\rangle =0$ directly
on the equations of motion for the Green's
functions.

In deriving the Boltzmann equation
satisfied by 
$\delta N({\bf k},X)$, it will be convenient to use the
Coulomb gauge,  which offers the most direct 
description of the physical degrees of freedom: in this gauge,
the (hard) propagating modes are entirely contained in the 
transverse components of the spatial gluon propagator $G_{ij}(x,y)$,
so that the density matrix $\delta N({\bf k},X)$
is simply the gauge-covariant Wigner transform
of $G_{ij}(x,y)$ (see below).
(In other gauges --- like the ``covariant'' ones with
gauge-fixing term $(1/2\lambda)(D_\mu[A] a^\mu)^2$ --- the physical,
transverse  degrees of freedom are mixed in all the components of
the gluon propagator $G_{\mu\nu}$. In this case, the 
density matrix $\delta N({\bf k},X)$ involves a linear combination
of the Wigner functions of the gluons and the ghosts, and it is only this
particular combination  which is gauge-fixing independent \cite{qcd}.
The intermediate calculations are cumbersome, and the explicit proof
of the gauge-fixing independence is quite non-trivial already at the mean
field approximation --- or ``hard thermal loop''  --- level
\cite{BP90,qcd}.)

In what follows we shall mostly
use the strict Coulomb gauge condition, namely:
\beq\label{COUL}
D_i[A]\,a^i\,=\,0.\eeq
 In this gauge, all the non-equilibrium
Green's functions are transverse, that is:
\beq\label{COVTR}
D^i_x[A] \,G_{i\nu}(x,y)\,=\,0\,,\eeq
and similarly for the higher point functions. The only non-trivial
components of the free retarded gluon propagator are:
\beq
G_{00}^{(0)}(k)\,=\,-\,\frac{1}{{\bf k}^2},\qquad
G_{ij}^{(0)}(k)\,=\,-\,\frac{\delta_{ij}-\hat k_i\hat k_j}{
k_0^2-{\bf k}^2}.\eeq
That is,  the electric gluon is static, and the same
is also true for the Coulomb ghost: $\Delta^{(0)}(k)=1/{\bf k}^2$.
Accordingly (with $G^<_0(k)$ and $G^>_0(k)$ as defined in eq.~(\ref{G0})),
\beq\label{W0}
G^{< \,(0)}_{ij}(k)=(\delta_{ij}-\hat k_i\hat k_j)\,G^<_0(k),\qquad
G^{>\,(0)}_{ij}(k)=(\delta_{ij}-\hat k_i\hat k_j)\,G^>_0(k),\eeq
while all the other components are zero.

\subsection{Equations of motion}

The equations of motion for the average field $A^\mu_a$
read:
\beq\label{avA1}
 (D^\nu F_{\nu\mu})^a(x)&=&j_\mu^a(x).
\eeq 
Here and in what follows, $D_\mu$ or $F_{\mu\nu}$  denote
the covariant derivative or the field strength tensor
 associated to the background field $A_\mu^a$. The induced colour current
$j_\mu^a(x)$ involves the off-equilibrium 2-point functions\footnote{There
is also a contribution to the current 
from the gluon 3-point function which, however, starts
at two-hard-loop level and is thus negligible for
what follows \cite{prept}.} for gluons and ghosts:
\beq\label{JIND}
j^{\mu}_{a}(x)\,=\,i\,g\,{\rm Tr}\, T^a\left\{
\Gamma^{\mu\rho\lambda\nu}  
 D^x_\lambda\,G_{\rho\nu}^<(x,y)\,+\,
 \Delta^<(x,y)\,(D_y^\mu)^\dagger
\right\}\bigg |_{y= x}.\eeq
We have used here the notation:
\beq\label{gamma}
\Gamma^{\mu\nu\rho\lambda}\equiv\,2g^{\mu\nu}
g^{\rho\lambda}-g^{\mu\rho}g^{\nu\lambda}-g^{\mu\lambda}g^{\nu\rho}.\eeq
Furthermore, $D^\dagger[A]=\buildchar{\del}{\leftarrow}{}
-igA^aT^a$, and the derivative $\buildchar{\del}{\leftarrow}{}$ acts
on the function on its left.

The 
Kadanoff-Baym equations for the gluon 2-point functions read
(cf. Sec. 2.1) :
\beq\label{KBYM1}
\left(g_\mu^{\,\rho}D^2-D_\mu D^\rho+ 2igF_{\mu}^{\,\rho}
\right)_xG^<_{\rho\nu}(x,y)&=&\nonumber\\
\int {\rm d}^4z\,\Bigl\{
g_{\mu\lambda}\Sigma_{R}^{\lambda\rho}(x,z)\,G^<_{\rho\nu}(z,y) 
&+&\Sigma_{\mu\rho}^{<}(x,z)G_{A}^{\rho\lambda}
(z,y)g_{\lambda\nu}\Bigr\},\eeq
and 
\beq\label{KBYM2}
G_{\mu}^{<\rho}(x,y)\left(g_{\rho\nu}\Bigl(D^\dagger\Bigr)^2
- D_\rho^\dagger D_\nu^\dagger + 2ig F_{\rho\nu}\right)_y
&=&\nonumber\\
\int {\rm d}^4z\,\Bigl\{
g_{\mu\lambda}G_{R}^{\lambda\rho}(x,z)\,\Sigma^<_{\rho\nu}(z,y) 
&+&G_{\mu\rho}^{<}(x,z)\Sigma_{A}^{\rho\lambda}
(z,y)g_{\lambda\nu}\Bigr\},\eeq
together with the gauge fixing conditions (cf. eq.~(\ref{COVTR})):
\beq\label{TR}
D^i_xG_{i\nu}(x,y)\,=\,0,\qquad\,\,
G_{\mu j}(x,y) D^{j \dagger}_y\,=\,0.\eeq
In deriving these equations, we have used symmetry properties like:
\beq\label{symm}
G^{>\,ab}_{\,\,\mu\nu}(x,y)\,=\,G^{<\,ba}_{\,\,\nu\mu}(y,x),\qquad
G^{\,\,\,ab}_{R\,\mu\nu}(x,y)\,=\,G^{\,\,\, ba}_{A\,\nu\mu}(y,x),\qquad
\eeq
and similarly for the self-energies.

In the following developments, we shall often omit the upperscripts
$>$ and $<$ on the 2-point functions, and indicate them only when
necessary, e.g., on the final equations.

\subsection{Gauge-covariant Wigner functions}

Let
$G_{ab}(x,y)$ denote any of the  2-point functions, and $G_{ab}(k,X)$ 
the corresponding Wigner function, defined as in eq.~(\ref{G<WIG}).
Unlike $G_{ab}(x,y)$, which is separately
gauge-covariant at $x$ and $y$ (cf. eq.~(\ref{DTr})), its Wigner transform
$G_{ab}(k,X)$ is not covariant. However, following what we did for SQED, we can 
construct the following  function (cf.
eq.~(\ref{acuteG0})):
\beq\label{COVG}
\acute G_{ab}(s,X)\,\equiv\,U_{a\bar a}\Bigl(X,X+{s\over 2}\Bigr)
\,G_{\bar a \bar b}\Bigl(X+{s\over 2},X-{s\over 2}\Bigr)\,
U_{\bar b b}\Bigl(X-{s\over 2},X\Bigr),\eeq
where $U(x,y)$ is the non-Abelian parallel transporter,
also referred to as a Wilson line ($A_\mu=A_\mu^a T^a$) :
\beq\label{PT1} U
(x,y)={\rm P}\exp\left\{ -ig\int_\gamma {\rm d}z^\mu A_\mu(z)\right\}.\eeq
As in the Abelian case, the path $\gamma$ is arbitrary
(see the discussion before eq.~(\ref{Uapprox})).
Under the gauge transformations of $A_\mu$, the Wilson
line (\ref{PT1}) transforms as (in matrix notations):
\beq\label{UTR} U(x,y) \longrightarrow h(x)\,U (x,y)\,h^\dagger(y)\,,\eeq
so that the function (\ref{COVG}) is indeed gauge-covariant at $X$
for any given $s$:
\beq\label{GSTR}
\acute G(s,X)  \longrightarrow h(X)\,\acute G (s,X)\,h^\dagger(X)\,.\eeq
Correspondingly, its Wigner transform 
$\acute G_{ab}(k,X)$ 
transforms covariantly as well: For any given $k$, $
\acute G(k,X)  \longrightarrow h(X)\,\acute G (k,X)\,h^\dagger(X)$. 

In principle, the equations of motion for $\acute G(s,X)$ follow from
the equations of motion (\ref{KBYM1})--(\ref{TR}) for $G(x,y)$ by replacing
$G(x,y)$ by (cf. eq.~(\ref{COVG})):
\beq G(x,y) \,=\,U(x,X)\,\acute G(s,X)\,U(X,y).\eeq
However, in contrast to what we did for SQED, in the non Abelian case we
have to proceed to a linearisation in order to preserve the consistency of
the expansion in powers of $g$. Recall indeed that the mean fields
$A^\mu_a$ are supposed to be weak and slowly varying, such that $ \del_X
\,\sim\, gA\, \ll\, T\,$.  (The ultrasoft covariant derivative is of the order
$D_X = {\rm O}(g^2 T)$, but the  simplifications we are refering to hold
already when $D_X = {\rm O}(g T)$ \cite{qcd,BIO96,prept}.) For such soft
background fields the function $\acute G(s,X)$ remains
strongly peaked at $s=0$, and vanishes  when 
$s \simge 1/T$. Over such a short scale, the mean field $A_\mu$ does not
vary significantly. Furthermore, for $s\simle 1/T$,  $g s\cdot A\ll 1$
since
$gA_\mu \ll T$. We can then expand the Wilson lines in
eq.~(\ref{COVG}) in powers of $g$ and get, to leading non-trivial order:
\beq\label{approxPT}
U_{ab}(x,y)\,\simeq\, \delta_{ab}\,-\,ig\Bigl(s \cdot A_{ab}(X)\Bigr).\eeq
This should be compared to eq.~(\ref{Uapprox}) in SQED: both expressions
hold to leading order in an expansion in soft gradients, but in the
non-Abelian expression (\ref{approxPT}) we have also performed an
expansion in powers of the gauge field. In what follows, we will never
need to go beyond the simple approximation (\ref{approxPT}).

Similarly, we shall see that the off-equilibrium fluctuations
$\delta G\equiv G - G_{eq}$ are perturbatively small: $\delta G \sim (D_X/T)
G_{eq} \sim g^2 G_{eq}$. Thus, by writing:
\beq G\equiv G_{eq}+ \delta G,\qquad\qquad
\acute G\equiv G_{eq} +\delta\acute G,\eeq
in eq.~(\ref{COVG}), and recalling that $G_{eq}^{ab}=\delta^{ab}G_{eq}$, 
we can easily obtain the following relation between
$\delta\acute G$ and $\delta G$, valid to leading order in $g$:
\beq\label{COV}
\delta\acute G (s,X)\,\simeq\, \delta G (x,y) \,+\,
ig\Bigl(s \cdot A(X)\Bigr)G_{eq}(s),\eeq
or, equivalently:
\beq\label{delG}
\delta\acute G(k,X)\simeq\delta G(k,X)+g(A(X)\cdot\del_k)G_{eq}(k).\eeq 
Note that both terms in the r.h.s. of eq.~(\ref{COV}) or (\ref{delG})
are of the same order, namely of O$\Bigl((D_X/T) G_{eq}\Bigr)$.
On the other hand, the terms which have been neglected in going
from eq.~(\ref{COVG}) to eq.~(\ref{COV}) are down by, at least,
one more power of $D_X/T$.

Consider now a term like $D^\mu_x G(x,y)$ which appears in
eqs.~(\ref{KBYM1})--(\ref{TR}). Clearly, such a term
transforms in the same way as $G(x,y)$, so it can be treated in a 
similar way (cf. eq.~(\ref{COVG})). Then, we can write:
\beq\label{COVDEL}
\delta\Bigl(D^\mu_x G(x,y)\Bigr)\equiv
D^\mu_x G(x,y) -\del^\mu_s G_{eq}(s)\,\simeq\,\del^\mu_s
\delta\acute G(s,X) \,-\,ig\Bigl(s \cdot A(X)\Bigr)\del^\mu_s
G_{eq}(s),\,\,\eeq
which parallels eq.~(\ref{COV}). In particular, since the equilibrium
gluon Wigner function is transverse, $\del_i G^{i\nu}_{eq} = 0$, 
eqs.~(\ref{TR}) and (\ref{COVDEL}) show
that the {\it gauge-covariant} Wigner function
is transverse as well:
\beq\label{TRAC}
\del^i_s \acute G_{i\nu}(s,X)\,=\,0,\qquad\,\,{\rm or}\qquad
k^i \acute G_{i\nu}(k,X)\,=\,0.\eeq

Finally, we have to express the induced current (\ref{JIND})
in terms of the gauge-covariant Wigner functions.
Since it vanishes in equilibrium,  it involves only
the off-equilibrium deviations of the Wigner functions
of the gluons and the ghosts.
We have:
\beq\label{jg}
j_{\mu}^{a}(X)\,=\,g\int\frac{{\rm d}^4k}{(2\pi)^4}\,{\rm Tr}
\,T^a \Bigl\{- k_\mu \delta\acute G_{\nu}^{<\,\nu}(k,X)
+ \delta\acute G_{\mu\nu}^<(k,X)k^\nu
 -k_\mu \delta\acute \Delta^<(k,X) \Bigr\},\eeq
where the following property has been used (cf. eq.~(\ref{COVDEL})):
\beq D^\mu_x G(x,y)\Big |_{y=x}\,=\,\del^\mu_s 
\acute G(s,X)\Big |_{s=0}.\eeq
Like (\ref{JIND}), eq.~(\ref{jg}) holds in an arbitrary gauge.
In Coulomb's gauge it can be
further simplified: as we shall see in the next section,
only the transverse fluctuations $\delta\acute G_{ij}(k,X)
\equiv (\delta_{ij}-\hat k_i\hat k_j)\delta \acute G(k,X)$
matter for the calculation of $j^\mu$, so that:
\beq\label{JFIN}
j^{\mu}_{a}(X)\,=\,2g\int\frac{{\rm d}^4k}{(2\pi)^4}\,k^\mu {\rm Tr}
\Bigl\{T^a \delta \acute G^<(k,X)\Bigr\}.\eeq

\subsection{The non-Abelian Vlasov equation}

In this section, we shall study eqs.~(\ref{KBYM1})--(\ref{TR}) 
in the limit where the all the terms involving self-energies
can be neglected. As in the case of SQED, this amounts to a mean field
approximation in which the hard gluons are allowed to  scatter on the average
colour fields $A^\mu_a$, but not among themselves. The resulting equations are:
\beq\label{MFEQ}
\left(g_\mu^{\,\,\rho}D^2-D_\mu D^\rho+ 2igF_{\mu}^{\,\,\rho}
\right)_xG_{\rho\nu}(x,y)&=&0,\nonumber\\
G_{\mu}^{\,\,\rho}(x,y)\left(g_{\rho\nu}\bigl(D^\dagger\bigr)^2
- D_\rho^\dagger D_\nu^\dagger + 2ig F_{\rho\nu}\right)_y
&=&0,\eeq
where $G$ denotes either one of the functions $G^>$ or $G^<$. The outcome of
the present subsection is the Vlasov equation for the gluon
density matrix. The derivation is not new \cite{qcd}, 
except for the use of the Coulomb
gauge. However, since this involves manipulations  which will be essential for
the evaluation of the collision terms, we present it in detail.

The equations (\ref{MFEQ})
 involve hidden powers of $g$,  associated with the soft 
inhomogeneities ($\del_X \sim g^2T$) and with the amplitudes of the
mean fields ($A \sim gT$ and $gF_{\mu\nu} \sim g^4T^2$).
The purpose of the covariant gradient expansion is precisely to isolate
all the terms of leading order in $g$. (Actually, all the
manipulations in this subsection apply already for inhomogeneities
at the scale $gT$, when $\del_X \sim gA \sim gT$ 
and $gF_{\mu\nu} \sim g^2T^2$ \cite{qcd}.)

As in Sec. 2.3, we start by considering the difference of the
two equations (\ref{MFEQ}). Let us look at
the first term in the l.h.s. of this difference
equation,  which we denote as:
\beq\label{Xi}
\Xi(x,y)\equiv D^2_x G(x,y)\,-\,G(x,y) (D^\dagger_y)^2,\eeq
where $D_x^2$ and $(D^\dagger_y)^2$ are given by eq.~(\ref{d^2}), except
that the derivatives in $(D^\dagger_y)^2$ are now understood
to act on their left. (Minkowski indices are omitted to simplify
the notations; they will be reestablished when needed.)
Proceeding as in  Sec. 2.3, and paying attention to the colour
algebra, we obtain:
\beq\label{DIFF1}
\Xi(s,X)&=& 2\del_s\cdot\del_X G +2ig\Bigl[A_\mu(X),\del^\mu_s G\Bigr]
+ig\Bigl\{A_\mu(X),\del^\mu_X G\Bigr\}
+ig\Bigl\{(s\cdot\del_X)A_\mu,\del^\mu_s G\Bigr\}\nonumber\\
&{}&+ig\Bigl\{(\del_X\cdot A), G\Bigr\} - g^2\left[A^2(X),G\right]
-\frac{g^2}{2}\left\{(s\cdot\del_X)A^2,G\right\}\,+\,...\,,\eeq
where the right parantheses 
(the braces) denote commutators (anticommutators) of colour matrices,
and the dots stand for terms which involve at least two soft derivatives
$\del_X$.

 At this point, we use the fact that $A\sim gT$
and $\delta G\equiv G-G_{eq}\,\sim\, g^2G_{eq}$ (as will be verified
a posteriori), with $G_{eq}\approx G^{(0)}$ in the mean field
approximation. To leading order in $g$, eq.~(\ref{DIFF1})
then simplifies to:
\beq\label{DIFF2}
\Xi(s,X)\approx 2(\del_s\cdot\del_X) \delta G +2ig\left[A_\mu,
\del^\mu_s \delta G\right]
+2ig(s\cdot\del_X)A_\mu\,(\del^\mu_s G^{(0)})
+2ig (\del_X \cdot A) G^{(0)}\,,\nonumber\\\eeq
where all the terms are of order $g^4 T^2G_0$. Taking now the Wigner transform,
we get:
\beq \label{DIFF3}
\Xi(k,X)\,\approx\,2\Bigl[k\cdot D_X,\, \delta  G(k,X)\Bigr] +
2 gk^\mu \Bigl(\del_X^\nu A_\mu(X)\Bigr)\del_\nu G^{(0)}(k),\eeq
where $G(k,X)$ is the ordinary Wigner transform of $G(x,y)$,
defined as in eq.~(\ref{G<WIG}). This can be rewritten in a
gauge-covariant form by replacing $\delta G=
\delta\acute G-g(A\cdot\del_k)G^{(0)}$ (cf. eq.~(\ref{delG})):
\beq \label{KIN}
\Xi_{\mu\nu}(k,X)\,\approx \,2\Bigl[k\cdot D_X, \delta\acute
G_{\mu\nu}(k,X)\Bigr] -
 2 gk^\alpha F_{\alpha\beta}(X) \,\del^\beta G_{\mu\nu}^{(0)}(k),\eeq
where the Minkowski indiced have been reintroduced.

We return now to eqs.~(\ref{MFEQ}). Since we
 are mainly interested in the transverse gluon Wigner function
$\delta \acute G_{ij}(k,X)$, let us focus on the 
components $\mu=i$ and $\nu=j$: 
\beq\label{MFIJ}
D_x^2 G_{ij} - D^x_iD^x_0G_{0j}+ 2igF_{i}^{\,\rho}(x)
G_{\rho j}&=&0,\nonumber\\
G_{ij}\Bigl(D_y^\dagger\Bigr)^2 - G_{i0} 
D_{0\,y}^\dagger D_{j\,y}^\dagger + 2igG_{i\rho}F^{\rho}_{\,\nu}(y)
&=&0.\eeq
(In writing these equations, we have also used the
gauge-fixing constraint (\ref{TR}) to simplify some terms.)
When taking the difference of these equations, we first meet
(cf. eq.~(\ref{KIN})):
\beq\label{GIJ1}
D_x^2 G_{ij} -G_{ij}\Bigl(D_y^\dagger\Bigr)^2\,
\longrightarrow \,
2\Bigl[k\cdot D_X, \delta\acute G_{ij}\Bigr]
 - 2 gk^\alpha F_{\alpha\beta}(X) \,\del^\beta G^{(0)}_{ij}(k).\eeq
Note the following identity, which will be useful later:
\beq\label{ID1}\lefteqn{ 
k^\alpha F_{\alpha\beta}\del^\beta G^{(0)}_{ij}(k)\equiv
k^\alpha F_{\alpha\beta}\del^\beta [
(\delta_{ij}-\hat k_i\hat k_j)G_0(k)]}\nonumber\\
& &=(\delta_{ij}-\hat k_i\hat k_j)k^\alpha F_{\alpha\beta}\del^\beta
G_0 -k^\alpha F_{\alpha l}\,\frac{k_i\delta_{jl}
+k_j\delta_{il}-2\hat k_i\hat k_j k_l}
{{\bf k}^2}\,G_0.\,\,\,\eeq
The  terms involving $G_{i0}$ and $G_{0j}$ vanish in equilibrium, and remain
small out of equilibrium, but nevertheless their conribution to
eqs.~(\ref{MFIJ}) is non-negligible: Indeed, we shall verify shortly that
$G_{i0}
\sim (D_X^2/T^2)G_0$, which is one order higher than
the transverse fluctuations $\delta G_{ij}\sim (D_X/T)G_0$.
However, the hard derivatives multiplying
$G_{i0}$ and $G_{0j}$ in  eqs.~(\ref{MFIJ}) do {\it not} cancel
in the difference of the two equations, in contrast to what happens
with the spatial components
$\delta G_{ij}$. Specifically, $\del_x^2-\del_y^2=2\del_s\cdot\del_X
\sim TD_X$, while $\del^x_i\del^x_0 \sim \del^s_i\del^s_0
\sim T^2$. Therefore, the difference:
\beq  D^x_iD^x_0G_{0j}- G_{i0} D_{0\,y}^\dagger D_{j\,y}^\dagger
\,\approx\, \del^s_0(\del^s_iG_{0j}-\del^s_jG_{i0})\,\sim\,
 (D_X)^2 G_0,\eeq
is of the same order as, e.g., $(\del_s\cdot \del_X) G_{ij}$.
We thus have to evaluate these terms properly, which we shall do
later, with the following results:
\beq\label{GI0}
G_{0j}(k,X)\,\approx\,  2ig F_{0l}\,\frac{\delta_{lj}-\hat k_l\hat k_j}
{{\bf k}^2}\,G_0(k),\qquad
G_{i0}^{ab}(k,X)=G_{0i}^{ba}(-k,X).\eeq
(The second equality above follows from the symmetry property 
(\ref{symm}).) The
corresponding contribution to the kinetic equation for
$\delta \acute G_{ij}(k,X)$ reads then:
\beq\label{GIJ2}
-\Bigl(D^x_iD^x_0G_{0j}- G_{i0} D_{0\,y}^\dagger D_{j\,y}^\dagger\Bigr)
\,\longrightarrow \, -2gk^0F_{0l}\,\frac{k_i\delta_{jl}
+k_j\delta_{il}-2\hat k_i\hat k_j k_l}
{{\bf k}^2}\,G_0(k).\eeq
Finally, in the last terms in  eqs.~(\ref{MFIJ}) --- the terms
involving the field strength tensor --- we can replace
 $F_{i}^{\,\rho}(x)\approx F_{i}^{\,\rho}(X)$
and $G_{\rho j}(k,X) \approx G_{\rho j}^{(0)}(k)=\delta_{\rho l}
(\delta_{lj}-\hat k_l\hat k_j)G_0$, to get:
\beq\label{GIJ3}
-2ig\Bigl(F_{il}(X) G_{lj}^{(0)}(s) - G_{il}^{(0)}(s)
F_{lj}(X)\Bigr)\,\longrightarrow \, -2g G_0(k)\Bigl(
F_{il}\hat k_l\hat k_j + F_{jl}\hat k_l\hat k_i\Bigr).\eeq 
By using the identity (\ref{ID1}), it is easy to recognize
the role of the two contributions in eqs.~(\ref{GIJ2}) and (\ref{GIJ3}):
this is to cancel the non-transverse piece in the
r.h.s. of eq.~(\ref{GIJ1}). Finally,
$\delta\acute G_{ij}(k,X)$ satisfies the following kinetic equation: 
 \beq\label{GVLAS}
\Bigl[k\cdot D_X,\,\delta\acute G_{ij}(k,X)\Bigr]
\,-\,(\delta_{ij}-\hat k_i\hat k_j)
gk^\alpha  F_{\alpha\beta}(X)\del^\beta G_0(k)\,=\,0.\eeq
It is transverse, as anticipated:
\beq\label{GIJ}
\delta\acute G_{ij}(k,X)\,\equiv\,
(\delta_{ij}-\hat k_i\hat k_j)\delta\acute G(k,X),\eeq
with the new function $\delta\acute G(k,X)$ satisfying:
 \beq\label{VLAS}
\Bigl[k\cdot D_X,\,\delta\acute G(k,X)\Bigr]
\,=\,g\,k^\alpha  F_{\alpha\beta}(X)\del^\beta G_0(k).\eeq
Since $k\sim T$, $D_X \sim g^2T$ and $g F_{\alpha\beta}\sim 
(D_X)^2 \sim g^4T^2$,
eq.~(\ref{VLAS}) implies  $\delta\acute G \sim (D_X/T)G_0
\sim g^2G_0$, as anticipated.

Eq.~(\ref{GVLAS}) is the main result of this subsection.
In order to complete its proof, we still have to justify
eq.~(\ref{GI0}) for $G_{0j}$. To this aim, we shall consider
the first eq.~(\ref{MFEQ}) with $\mu=0$ and $\nu=j$. This 
reads
\beq\label{eqGI}{\bf D}_x^2 G_{0j}\,+\,2ig F_{0l}(x)G_{lj}(x,y)\,=\,0.\eeq
To the order of interest, ${\bf D}_x^2 \approx \bfgrad_s^2$,
$F_{0l}(x)\approx F_{0l}(X)$
and $G_{lj}(k,X) \approx G_{lj}^{(0)}(k)$. Then, eq.~(\ref{GI0})
is just the Wigner transform of eq.~(\ref{eqGI}).

To conclude this section, let us remark that $\delta\acute G(k,X)$ is of the
form (compare to eq.~(\ref{GN0}) in SQED):
\beq\label{GN}
\delta\acute G_{ab}(k,X)&=&\rho_0(k)\delta N_{ab}(k,X)\nonumber\\
&\equiv&2\pi\delta(k^2)\Bigl\{\theta(k_0)\delta N_{ab}({\bf k},X)
\,+\,\theta(-k_0)\delta N_{ba}(-{\bf k},X)\Bigr\},\eeq
where the structure of the second line follows from the first
symmetry property (\ref{symm}), and the density matrix $\delta N_{ab}({\bf
k},X)$
 satisfies the equation  \cite{qcd}
(with $v^\mu =(1,{\bf k}/k)$):
\beq\label{N}
\left[ v\cdot D_x,\,\delta N({{\bf k}},x)\right]=-\, g\,
{\bf v}\cdot{\bf E}(x)\frac{{\rm d}N}{{\rm d}k},\eeq
which may be seen as the non-Abelian generalization
of the Vlasov equation.
Note also that eqs.~(\ref{VLAS}) and (\ref{GN}) hold for both
 $\delta\acute G^<$ and  $\delta\acute G^>$, which are equal
in the mean field approximation:
\beq\label{G<>}
\delta\acute G^<(k,X)\,\approx\,
\delta\acute G^>(k,X)\,\equiv\,\delta\acute G(k,X)
\,=\,\rho_0(k)\delta N(k,X).\eeq
This results from the fact that the spectral density 
 $\rho(k,X)=G^>(k,X)-G^<(k,X)$ is not modified
in the present approximation: $\rho(k,X)\approx 
\rho_0(k) \equiv 2\pi \epsilon(k_0)\delta(k^2)$.

\subsection{Collision terms in QCD}

As we have seen in the previous section, the colour background field 
$A^\mu_a$ induces a fluctuation $\delta\acute G_{ij} \sim (D_X/T)G_0$ in the
 Wigner function of the hard transverse gluons.
For $D_X\sim g^2T$, this fluctuation is  of order $g^2 G_0$, and the various
terms in eq.~(\ref{GVLAS}) are all of order $D_X^2 G_0 \sim g^4 T^2 G_0$.
In this case,  the collision terms cannot be neglected  and must be 
added in the r.h.s. of eq.~(\ref{GVLAS}).

In order to compute these terms, we consider, as usual, the
difference of the self-energy terms in the r.h.s. of eqs.~(\ref{KBYM1})
and (\ref{KBYM2}). These involve convolutions of self-energies and
propagators which yield, after a Wigner transform,
\beq\label{PBQCD}
\int {\rm d}^4z\,G(x,z)\,\Sigma(z,y) 
\,\longrightarrow G(k,X) \Sigma(k,X)\,+\,
\frac{i}{2}\,\Bigl\{G,\, \Sigma\Bigr\}_{PB}\,+\,...\,,\eeq
up to terms involving, at least, two soft derivatives.
The Poisson bracket $\{G,\, \Sigma\}_{PB}$ is defined as
in eq.~(\ref{poisson}).
To simplify writing, we have left aside the Minkowski indices;
these will be added on the final equations.
Note also that $\Sigma$ and $G$ are
colour matrices, so their ordering is important.

Collecting  all the terms without soft derivatives in the
r.h.s. of eqs.~(\ref{KBYM1}) and (\ref{KBYM2}), we obtain:
\beq\label{COL0}
C(k,X)&\equiv&i\Bigl(\Sigma_R G^< - G^< \Sigma_A
+\Sigma^< G_A - G_R \Sigma^<\Bigr)\nonumber\\
&=&-\,\frac{1}{2}\Bigl(\{G^>,\Sigma^<\} - \{\Sigma^>,G^<\}\Bigr)
-i[{\rm Re} \Sigma_R, G^<] +i[{\rm Re} G_R,\Sigma^<],\eeq
where the various parantheses stand for colour commutators
or anticommutators.
In writing the second line above, we have also used the relations
(\ref{NEQAG}).

We now proceed to some approximations. Recall first that both the soft
gradients, and the amplitudes of the  background
fields and of the fluctuations  $\delta G$ or $\delta\Sigma$,
are controlled by powers of $g$. Writing for instance 
 $G^>(k,X) \equiv G_{eq}^>(k) + \delta G^>(k,X)$ and
$\Sigma^<(k,X) \equiv \Sigma_{eq}^<(k) + \delta \Sigma^<(k,X)$,
we  have $\delta G^> \sim g^2 G_{eq}^>$, and similarly
$\delta \Sigma^< \sim g^2\Sigma_{eq}^<$ (see the next section for the latter
estimate). Thus, to leading order in $g$, we can linearize $C(k,X)$,
eq.~(\ref{COL0}), with respect to the off-equilibrium fluctuations.
Since the equilibrium two-point functions are diagonal in colour
(e.g.,  $G_{eq}^{ab}=\delta^{ab}G_{eq}$),
the two commutator terms in eq.~(\ref{COL0}) simply vanish,
while the anticommutator terms yield:
\beq\label{COL}
C(k,X)\,\simeq\,-\,\Bigl(G_{eq}^>\delta\Sigma^< \,+\,
\delta G^>\Sigma^<_{eq}\Bigr)\,+\,\Bigl(\delta\Sigma^>G_{eq}^<
\,+\,\Sigma^>_{eq}\delta G^<\Bigr).\eeq
Each of the terms in the above equation is of order $g^2 G_{eq}
\Sigma_{eq}$. At this order, the Poisson bracket in 
eq.~(\ref{PBQCD}) can be neglected. Indeed: 
\beq\label{pb}
\Bigl\{G,\, \Sigma\Bigr\}_{PB}\,\equiv\,
\del_k G \cdot \del_X \Sigma \,-\,\del_X G \cdot\del_k \Sigma
\,=\,\del_k G \cdot \del_X \delta\Sigma \,-\,\del_X \delta G 
\cdot \del_k \Sigma,\eeq
where we have used, e.g., $\del_X \Sigma(k,X)= \del_X \delta\Sigma(k,X)$.
With $\delta \Sigma \sim g^2\Sigma_{eq}$ and a similar estimate for
$\delta G$, each of the two terms above is $\sim g^2(\del_X/T)
G_{eq} \Sigma_{eq} \,\sim\,g^4 G_{eq} \Sigma_{eq}$.

Thus,  at the order 
of interest, the only relevant collision terms are those displayed
in eq.~(\ref{COL}).
This corresponds to the {\it quasiparticle approximation}
introduced in Sec. 2.4. Indeed, it can be verified that with the
Poisson brackets  excluded,  the hard gluon spectral density $\rho(k,X)$
satisfies the same equation as in the mean field approximation
\cite{prept}, so that $\rho(k,X)\approx 
\rho_0(k) \equiv 2\pi \epsilon(k_0)\delta(k^2)$. This has the
consequence discussed at the end of the previous section, namely:
\beq\label{GQP}
\delta\acute G^<(k,X)\,\approx\,
\delta\acute G^>(k,X)\,\equiv\,\delta\acute G(k,X)
\,=\,2\pi\delta(k^2)\epsilon(k_0)\delta N(k,X),\eeq
and the density matrix $\delta N(k,X)$ has the
structure displayed in eq.~(\ref{GN}).
Here, however, $\delta N(k,X)$ will be shown to satisfy a Boltzmann-like
equation, with the collision terms in eq.~(\ref{COL}).
In the same approximation, the equilibrium 2-point functions
$G_{eq}^>$ and $G_{eq}^<$ coincide with the free respective functions,
as given in eqs.~(\ref{W0}) and (\ref{G0}).

We end this section by completing the following two tasks:
({\it i}) First, we shall rewrite the collision terms (\ref{COL}) in a manifestly
gauge-covariant way. ({\it ii}) Then, we shall specify the tensor structure
of the collision terms in Minkowski space.

For point ({\it i}), it is enough to replace the non-covariant
fluctuations $\delta \Sigma$ and $\delta G$ in eq.~(\ref{COL}) by the
corresponding gauge-covariant expressions
$\delta\acute \Sigma$ and $\delta\acute G$ (cf. eq.~(\ref{delG})):
\beq\label{SIGG}
\delta G(k,X)&=&\delta\acute G(k,X)-g(A(X)\cdot\del_k)G_{eq}(k),
\nonumber\\
\delta \Sigma(k,X)&=&\delta\acute
\Sigma(k,X)-g(A(X)\cdot\del_k)\Sigma_{eq}(k).\eeq
 This yields:
\beq\label{COLC}
C(k,X)&=&-\,\Bigl(G_{eq}^>\,\delta\acute\Sigma^< \,+\,
\delta\acute G^>\Sigma^<_{eq}\Bigr)\,+\,\Bigl(\delta\acute
\Sigma^>G_{eq}^< \,+\,\Sigma^>_{eq}\,\delta\acute G^<\Bigr),\eeq
which turns out to be the same expression as above, eq.~(\ref{COL}), except
for the replacement of ordinary by gauge-covariant Wigner functions: 
The corrective terms in eq.~(\ref{SIGG}) do not contribute to 
$C(k,X)$ since they are proportional
to the collision term in equilibrium, which is zero:
\beq g(A(X)\cdot\del_k)\Bigl(G_{eq}^>\,\Sigma^<_{eq}
\,-\,G_{eq}^<\,\Sigma^>_{eq}\Bigr)\,=\,0.\eeq
And, actually, eq.~(\ref{COLC}) is formally the same as
in SQED (cf. eq.~(\ref{LIN0})).

Concerning point ({\it ii}), recall that the equilibrium 2-point functions
$G^{< \,\mu\nu}_{eq}$ and $G^{>\,\mu\nu}_{eq}$ --- which coincide here with
the corresponding tree-level functions; cf. eq.~(\ref{W0}) ---
have only spatial, and transverse, components. These will 
in turn project the
tranverse components of the self-energy fluctuations 
$\delta\acute \Sigma^>_{ij}$ and $\delta\acute \Sigma^<_{ij}$
in the collision term (\ref{COLC}). Accordingly, the kinetic equation
for $\delta\acute G_{ij}(k,X)$,  which reads (cf. eq.~(\ref{GVLAS})) :
\beq\label{GVLAS1}
2\Bigl[k\cdot D_X,\,\delta\acute G_{ij}(k,X)\Bigr]
\,-\,2(\delta_{ij}-\hat k_i\hat k_j)
gk^\alpha  F_{\alpha\beta}(X)\del^\beta G_0(k)\,=\,C_{ij}(k,X),\eeq
admits a transverse solution:
\beq \delta\acute G_{ij}(k,X) \equiv
(\delta_{ij}-\hat k_i\hat k_j)\delta\acute G(k,X)\,,\eeq
as in the mean field approximation.
Defining transverse projections in the usual way, e.g.,
\beq \label{ST}
\delta\acute\Sigma_T^>(k,X) \equiv (1/2)
(\delta_{ij}-\hat k_i\hat k_j)
\delta\acute\Sigma_{ij}^>(k,X),\eeq
we are finally led to the following kinetic equation for
$\delta\acute G(k,X)\,$:
 \beq\label{bolt}
2\Bigl[k\cdot D_X,\,\delta\acute G(k,X)\Bigr]
\,-\,2gk^\alpha  F_{\alpha\beta}(X)\del^\beta G_0(k)\,=\,
C(k,X),\eeq
with the collision term:
\beq\label{COLL}
C(k,X)&=&-\Gamma_T(k)\delta\acute G(k,X)\,+\,
\Bigl(\delta\acute\Sigma_T^> G_0^<\,-\,
\delta\acute\Sigma^<_TG_0^>\Bigr).\eeq
We have recognized here the equilibrium damping rate for 
the transverse gluons (cf. eq.~(\ref{EQG})):
\beq\label{Gamma}
\Gamma_T(k) \,=\,(\Sigma^<_{eq}(k)\,-\,\Sigma^>_{eq}(k))_T .\eeq
(Note that, in what follows, the subscript $T$ on
transverse quantities will be often omitted.)
Eqs.~(\ref{bolt})--(\ref{COLL}) are manifestly covariant
under the gauge transformations of the background field.

\subsection{The hard gluon self-energy out of equilibrium}

In this subsection, we use perturbation theory to
compute the transverse gluon self-energy
$\delta\acute\Sigma(k,X)$
to the order of interest. Specifically, we shall find that
$\delta\acute\Sigma \sim g^2 \Sigma_{eq}\sim g^4 T^2$,
so that the collision terms in the r.h.s. of eq.~(\ref{bolt}) are
of the same order as the drift and mean field terms in the l.h.s.

We start with the ordinary (i.e.,
non-gauge-covariant) self-energy $\Sigma(x,y)$ out of equilibrium.
As in SQED, the leading-order collision term corresponds to
scattering via one gluon exchange, as illustrated in Fig.~\ref{Born}.
However, as stated in the Introduction, we are mostly interested here
in colour relaxation, for which the relevant collisions are dominated by
soft momentum transfers, $g^2 T\simle q\simle gT$ (cf.
Sec. 3.8 below). Accordingly, the virtual
gluon in Fig.~\ref{Born} is always soft
and the self-energy which describes this collision is the effective
one-loop diagram depicted in Fig.~{\ref{S2}. This is formally
the same diagram as in SQED, except that, now, the continuous line in 
Fig.~{\ref{S2} refers to a hard transverse gluon and the wavy line
to a soft virtual one (which can be longitudinal or transverse).
The bubble on the wavy line denotes, as usual, 
the resummation of one-loop polarization tensor in the
propagator of the soft gluon. 

In thermal equilibrium, the hard line in 
Fig.~{\ref{S2} is a free propagator, while the soft
one is the HTL-resummed propagator, as introduced in Sec. 2.6.
(In the HTL approximation, the gluon and photon propagators are
formally the same up to the replacement of the Abelian
Debye mass $m_D^2=e^2T^2/3$ by the non-Abelian one, $m_D^2=
g^2  N_c T^2/3$ \cite{BIO96,MLB96}.)
Then, the self-energy in Fig.~{\ref{S2} yields a contribution of
O$(g^2 T^2)$ to the thermal interaction rate $\Gamma(k)$, eq.~(\ref{Gamma}).
Thus, the first collision term in eq.~(\ref{COLL})
can be estimated as:
\beq\label{est1}
\Gamma(k)\delta\acute G(k,X)\,\sim\,g^2 T^2
\delta\acute G\,\sim\,g^4 T^2 G_0.\eeq
Consider now the other terms in  eq.~(\ref{COLL}), which
involve the off-equilibrium self-energy $\delta\Sigma$.
A typical term contributing to  $\delta\Sigma$ is obtained
by replacing the equilibrium propagator $G_0$ in the hard
line in Fig.~{\ref{S2} by the respective off-equilibrium
fluctuation $\delta G$. Thus,   
\beq \label{est2}
\delta\Sigma\,\sim\,(\delta G/G_0) \Sigma_{eq}
\,\sim\,g^2 \Sigma_{eq}
\,\sim\,g^4 T^2,\eeq
which contributes to the collision terms at the same order as the
damping rate in eq.~(\ref{est1}).
Moreover, the off-equilibrium effects enter also 
the soft gluon line in Fig.~{\ref{S2}, via the polarization 
tensor. This will be computed 
in the next section, where we shall see that the net effect 
is also of order $g^4$, as in eq.~(\ref{est2}).

Let us turn now to the explicit evaluation of $\Sigma$. To this aim,
we need the three-gluon vertex between two hard gluons and a soft
one in the presence of the background field. This can be read on 
the Yang-Mills action, in the following way: Split the fluctuating
gluon field (that we have originally denoted as $a^\mu_b(x)$;
cf. eq.~(\ref{Zbk})) into soft and hard components,
and use different notations for the two. That is, replace\footnote{Note
that we preserve the notation $A^\mu_a(x)$ for the colour
background field.}
$a^\mu_b(x)\longrightarrow a^\mu_b(x) + {\cal A}^\mu_b(x)$,
where the new field $a^\mu$ is hard (it carries momenta $k\sim T$),
while the field ${\cal A}^\mu$ is soft (with typical momenta
$g^2 T \ll q \ll T$). There is no difficulty with this separation
(e.g., no problems with gauge symmetry) since the collision terms
will be saturated by soft momenta $q\simle gT$.
The Yang-Mills piece of the action reads then:
\beq S_{YM}&=&-\int_C{\rm d}^4 x \,\frac{1}{4}
\Bigl(F_{\mu\nu}^a [A+a+{\cal A}]\Bigr)^2\nonumber\\
 &=&\frac{1}{2}\int_C{\rm d}^4 x \,a^\mu_a\left\{
g_{\mu\nu}D^2 - D_\mu D_\nu + 2ig F_{\mu\nu}
\right\}_{ab} a^\nu_b\,+\,\dots\,\eeq
where $D_\mu\equiv D_\mu[A+{\cal A}] = D_\mu[A] + ig{\cal A}_\mu$,
and the dots stands for terms of cubic or quartic order in $a^\mu$,
which are unimportant here.
We still have to isolate the trilinear couplings $a^\mu a^\nu
{\cal A}^\rho$ from the equation above. After some algebra
and integration by parts, the relevant interaction piece of
the action is obtained as:
\beq\label{SINT}
S_I\,=\,ig\int_C{\rm d}^4 x \,a^\mu_a\left\{
g_{\mu\nu}({\cal A}\cdot D)\,-\,\frac{1}{2}({\cal A}_\mu D_\nu +
{\cal A}_\nu D_\mu) + 2[D_\mu,{\cal A}_\nu]\right\}_{ab}a^\nu_b,\eeq
where now $D_\mu \equiv D_\mu[A]=\del_\mu + igA_\mu$ is the covariant
derivative defined by the background field alone, and all the fluctuating
fields are explicit.

Only the first two terms in eq.~(\ref{SINT}) will be important. In these terms,
the covariant derivatives act on the hard fields $a^\mu$ and give rise to
vertices with hard momenta. The third term, on the other hand,
involves a covariant derivative acting on the soft field ${\cal A}^\mu$,
and is subleading (by a factor of $q/k \simle g$).
In what follows, we shall ignore this term, and focus on the
self-energy built out of the first two terms in $S_I$. We write:
\beq\label{S}
S_I\,=\,\frac{ig}{2}\,\Gamma_{\mu\nu\rho\lambda}
\int_C{\rm d}^4 x \,a^\mu {\cal A}^\rho D^\lambda[A] a^\nu,\eeq
with an implicit trace over the colour indices (the symbol
$\Gamma_{\mu\nu\rho\lambda}$ has been defined in eq.~(\ref{gamma})).
With the three-particle vertex above, it is a straightforward
exercise to construct the self-energy displayed in Fig.~{\ref{S2}.
This reads:
\beq\label{S1}
\Sigma^{ab}_{\mu\nu}(x,y)=-\,4\times 
\frac{g^2}{4}\,\Gamma_{\mu\alpha\gamma\rho}
\Gamma_{\nu\beta\delta\lambda} \,(T^a)_{cd}\, (T^b)_{\bar c\bar d}
\,{\cal D}^{\gamma\delta}_{d\bar d}(x,y)\Bigl(D^\rho_x
G^{\alpha\beta}(x,y) D^{\lambda\,\dagger}_y
\Bigr)_{c\bar c}, \eeq
where $G^{\alpha\beta}(x,y)$ is the hard gluon propagator and 
${\cal D}^{\gamma\delta}(x,y)$ the
soft gluon propagator:
\beq {\cal D}^{\gamma\delta}_{ab}(x,y)
\,\equiv\,\langle{\rm T}_C{\cal A}^\gamma_a(x)
{\cal A}^\delta_b(y)\rangle.\eeq
Finally, the factor 4 takes into account the fact that, strictly
speaking, there are three other terms similar to the one above,
which yield the same contribution to the order of 
interest (see below). 

 By choosing the time variables
in eq.~(\ref{S1}) on opposite sides of the contour,
we deduce expressions for the Wigner functions
$\Sigma^>$ and $\Sigma^<$. The upperscripts $>$ and
$<$ will be often omitted, to simplify writing.

Consider first the equilibrium limit of eq.~(\ref{S1}),
where $A^\mu=0$ and the internal propagators
are unit matrices in colour: By using $(T^a)_{cd}\, (T^b)_{cd}
= -N_c\delta^{ab}$ and going to momentum space, we obtain
$(\Sigma_{eq})_{ab}^{\mu\nu}(k)=
\delta_{ab}\Sigma_{eq}^{\mu\nu}(k)$ with :
\beq\label{SEQ}
\Sigma^{eq}_{\mu\nu}(k)\,=\,g^2N_c\,\Gamma_{\mu\alpha\gamma\rho}
\Gamma_{\nu\beta\delta\lambda} \int\frac{{\rm d}^4 q}{(2\pi)^4}
\,(k-q)^\rho(k-q)^\lambda\,{\cal D}_{eq}^{\gamma\delta}(q)\,
G^{\alpha\beta}_{eq}(k-q).\eeq 
In the construction of the collision terms, we shall need only the difference
$\Sigma^<-\Sigma^>$ (cf. eqs.~(\ref{COLL}) and (\ref{Gamma})),
and the resulting integral will be  dominated by soft
momenta. With this in mind, we shall neglect $q$ next to $k\sim T$
in the vertices in eq.~(\ref{SEQ}). Furthermore, to the same
order, $G_{eq}\approx G_0$ (which has only spatial and transverse
components; cf. eq.~(\ref{W0})) and ${\cal D}_{eq}
\approx \,{}^*{\cal D}$ (which is the HTL-resummed gluon propagator;
cf. Sec. 2.6). We finally get the following estimate for the 
transverse gluon self-energy in equilibrium:
\beq\label{SEQT}
\Sigma_{eq}^<(k)\equiv (1/2)
(\delta_{ij}-\hat k_i\hat k_j)\Sigma_{ij}^{eq\,<}
\,=\,4g^2N_c k^\rho k^\lambda
\int\frac{{\rm d}^4 q}{(2\pi)^4}\,\,
{}^*{\cal D}^<_{\rho\lambda}\,(q)G_0^<(k-q),\eeq 
together with a similar expression for $\Sigma_{eq}^>(k)$. 
The following identity has been useful in performing the Minkowski
algebra (with 
${\cal P}_{ij}(\hat{\bf k})\equiv\delta_{ij}-\hat k_i\hat k_j$):
\beq (1/2){\cal P}_{ij}(\hat{\bf k})\,\Gamma_{il\gamma\rho}k^\rho\,
{\cal P}_{lm}(\hat{\bf k})\,\Gamma_{jm\delta\lambda}k^\lambda
\,=\,4k_\gamma k_\delta.\eeq

We now return to the general expression in eq.~(\ref{S1})
and evaluate the off-equilibrium fluctuation $\delta\Sigma$.
Since we consider only small deviations away from equilibrium,
we can linearize this expression, as we have already done
for the collision terms in Sec. 3.5.
We thus get, keeping explicit only the colour indices:
\beq
\delta\Sigma^{ab}(x,y)&=&(T^a)_{cd}\, (T^b)_{\bar c\bar d}
\,\Bigl\{\delta {\cal D}_{d\bar d}(x,y)\Bigl(\del^\rho_x
\del^\lambda_y G_0(x-y)\Bigr)\delta_{c\bar c}\nonumber\\
&{}&\qquad\qquad\qquad +\,
\delta_{d\bar d}{}^*{\cal D}(x-y)\,\delta\Bigl(D^\rho_x
G(x,y) D^{\lambda\,\dagger}_y
\Bigr)_{c\bar c}\Bigr\}.\eeq
From this, we shall construct the gauge-covariant self-energy
$\delta\acute\Sigma^{ab}(s,X)$, as explained in Sec. 3.3.
First, we replace the non-covariant fluctuations in the internal
propagators $\delta {\cal D}$ and $\delta G$ in terms of the
corresponding gauge-covariant fluctuations
$\delta \acute{\cal D}$ and $\delta \acute G$ 
(cf. eq.~(\ref{COV}) and (\ref{COVDEL})):
\beq\label{COVS}
\delta{\cal D} (x,y)&=&\delta\acute{\cal D}(s,X) \,-\,
ig\Bigl(s \cdot A(X)\Bigr)\,{}^*{\cal D}(s),\nonumber\\
\delta\Bigl(D^\rho_x
G(x,y) D^{\lambda\,\dagger}_y\Bigr)&=&
-\del^\rho_s\del^\lambda_s \delta\acute G (s,X)
\,+\,ig\Bigl(s \cdot A(X)\Bigr)
\del^\rho_s\del^\lambda_s G_0.\eeq
Then, we define the covariant self-energy as in eq.~(\ref{SIGG}):
\beq
\delta\acute \Sigma(s,X)\,=\,\delta
\Sigma(x,y)\,+\,ig\Bigl(s \cdot A(X)\Bigr)\Sigma_{eq}(s),\eeq
with $\Sigma_{eq}$ of eq.~(\ref{SEQ}).
The result of these operations has the rather simple form:
\beq
\delta\acute\Sigma^{ab}(s,X)=(T^aT^b)_{cd}
\,\Bigl\{\delta\acute{\cal D}_{cd}(s,X)\del^\rho_s\del^\lambda_s G_0(s)
\,+\,{}^*{\cal D}(s)
\del^\rho_s\del^\lambda_s \delta\acute G_{cd}(s,X)\Bigr\}
\eeq
so that,  after a Wigner transform:
\beq\lefteqn{
\delta\acute\Sigma^{ab}(k,X)= -(T^aT^b)_{cd}
\int\frac{{\rm d}^4 q}{(2\pi)^4}
(k-q)^\rho(k-q)^\lambda}\nonumber\\
&{}&\times
\Bigl\{\delta\acute{\cal D}_{cd}(q,X)G_0(k-q)
+{}^*{\cal D}(q) \delta\acute G_{cd}(k-q,X)\Bigr\}.\eeq
By putting back the factors of $g^2$ and the Minkowski indices,
we finally obtain the following expression
for the gauge-invariant self-energy fluctuation
$\delta\acute\Sigma^{ab}_{\mu\nu}\,$:
\beq\label{DELS}
\delta\acute\Sigma^{ab}_{\mu\nu}(k,X)&=&
g^2\,\Gamma_{\mu\alpha\gamma\rho}
\Gamma_{\nu\beta\delta\lambda}\, (T^aT^b)_{cd}
\int\frac{{\rm d}^4 q}{(2\pi)^4}
\,(k-q)^\rho(k-q)^\lambda\,\nonumber\\
&{}&\qquad\qquad\times\Bigl\{\delta\acute{\cal D}_{cd}^{\gamma\delta}
(q,X)G_0^{\alpha\beta}(k-q)
+{}^*{\cal D}^{\gamma\delta}(q) \delta\acute G_{cd}^{\alpha\beta}
(k-q,X)\Bigr\}.\nonumber\\\eeq
This expression  is  very close to the 
corresponding expression in equilibrium: eqs.~(\ref{SEQ}) and 
(\ref{DELS}) involve the same momentum-dependent vertices,
and the equilibrium propagators of eq.~(\ref{SEQ}) have been
simply replaced in eq~(\ref{DELS}) by the (linearized version) of the
respective off-equilibrium propagators. 
The only significant difference is the colour structure,
which is trivial in equilibrium.

As before, we can neglect the soft momentum $q$ in the vertices
of eq.~(\ref{DELS}), and take the transverse projection of
this expression, to obtain:
\beq\label{DELST}
\delta\acute\Sigma^{ab}_T(k,X)
&=&4g^2k_\rho k_\lambda\, (T^aT^b)_{cd}
\int\frac{{\rm d}^4 q}{(2\pi)^4}
\Bigl\{{}^*{\cal D}^{\rho\lambda}
(q) \delta\acute G_{cd}(k-q,X)+\delta\acute{\cal D}_{cd}^{\rho\lambda}
(q,X)G_0(k-q)\Bigr\},\nonumber\\
&{}&\eeq
which is the non-equilibrium generalization of eq.~(\ref{SEQT}).

To conclude this section, let us return to a previous remark
according to which the complete self-energy $\Sigma$ should
involve three other terms in addition to the one displayed in
eq.~(\ref{S1}). In these terms, the covariant derivatives act 
differently on the two internal propagators, which then
results in modification of the momentum-dependent vertices.
For instance, we meet terms like 
$D^\rho_x D^\lambda_y\Bigl(G(x,y){\cal D}(x,y)\Bigr)$ which,
after covariantization and Wigner transform, yield the same result
as in eq.~(\ref{DELS}), except for the replacement $(k-q)^\rho
(k-q)^\lambda\longrightarrow k^\rho k^\lambda$ in the vertices.
However, this difference is not important here since we neglect
$q$ next to $k$ in the vertices. The same holds for the other two terms,
so that the total contribution is, indeed, four times the contribution
of the term displayed in eq.~(\ref{S1}). Hence the factor
4 in eq.~(\ref{S1}).

\subsection{The off-equilibrium propagator of the soft gluon} 

The above expression for $\delta\acute\Sigma(k,X)$ involves the
off-equilibrium propagator of the soft gluon
$\delta\acute{\cal D}^{\rho\lambda}(q,X)$, to which we turn now.
The relevant
off-equilibrium effects are encoded in the soft gluon
polarization tensor, which we   denote by 
$\Pi_{\mu\nu}^{ab}(x,y)$. In equilibrium, this reduces to the
corresponding hard thermal loop $\delta^{ab}\Pi_{\mu\nu}^{eq}(x-y)$.
Thus, the calculation below provides a generalization
of the HTL polarization tensor out of equilibrium.

The Kadanoff-Baym equations for the soft gluon propagator
${\cal D}_{\mu\nu}(x,y)$ are formally identical to those for
$G_{\mu\nu}(x,y)$, i.e., eqs.~(\ref{KBYM1}) and (\ref{KBYM2}).
For instance:
\beq\label{KBD}
\left(g_\mu^{\,\,\rho}D^2-D_\mu D^\rho+ 2igF_{\mu}^{\,\,\rho}
\right)_x{\cal D}^<_{\rho\nu}
\,-\,\int {\rm d}^4z\,
g_{\mu\lambda}\,\Pi_R^{\lambda\rho}\,{\cal D}^<_{\rho\nu}
\,=\,\int {\rm d}^4z\,g_{\lambda\nu}\,\Pi_{\mu\rho}^{<}\,
{\cal D}_{A}^{\rho\lambda}.\eeq
We shall also need the retarded  propagator
${\cal D}_R^{\mu\nu}(x,y)$, which obeys (cf. eq.(~\ref{eqGR})):
\beq\label{eqDR}
\left(g^\mu_{\,\,\rho}D^2-D^\mu D_\rho+ 2igF^{\mu}_{\,\,\rho}
\right)_x{\cal D}_R^{\rho\nu}(x,y)
\,-\,\int {\rm d}^4z\,
g_{\lambda\rho}\,\Pi_R^{\mu\lambda}\,{\cal D}_R^{\rho\nu}
\,=\,g^{\mu\nu}\delta^{(4)}(x-y)\,.\eeq
A priori, the self-energy $\Pi_{\mu\nu}$ in these equations
involves both interactions with the hard fields $a^\mu_a$
and self-interactions of the fields
${\cal A}_\mu^a$. However $\Pi_{\mu\nu}$ will be dominated
by the one-loop diagrams depicted in Fig.~\ref{PI} where
the internal lines are hard.

As in the Abelian case, the two equations above imply a relation
between ${\cal D}^<$ and $\Pi^<$ (cf. eq.~(\ref{WRA0})) :
\beq\label{WRA}
{\cal D}^<_{\rho\lambda}(x,y)\,=\,-\int {\rm d}^4z_1{\rm d}^4z_2\,\,
g_{\rho\alpha}{\cal D}_R^{\alpha\mu}(x,z_1)\,
\Pi_{\mu\nu}^<(z_1,z_2)\,{\cal D}_A^{\nu\beta}(z_2,y)g_{\beta\lambda}
.\eeq
A similar relation holds in between ${\cal D}^>$ and $\Pi^>$.
In particular, in thermal equilibrium,
\beq\label{DEQ}
{}^*{\cal D}^<_{\rho\lambda}(q)\,=\,-\,
\Bigl({}^*{\cal D}_R(q)\,
\Pi_{eq}^<(q)\,{}^*{\cal D}_A(q)\Bigr)_{\rho\lambda}
,\eeq
where ${}^*{\cal D}(q)$ is the HTL-resummed propagator.

Out of equlibrium, we can compute the Wigner transform of eq.~(\ref{WRA})
to get:
\beq\label{DOEQ}
{\cal D}^<_{\rho\lambda}(q,X)\,=\,-\,
\Bigl({\cal D}_R(q,X)\,
\Pi^<(q,X)\,{\cal D}_A(q,X)\Bigr)_{\rho\lambda}.\eeq
This equation holds up to corrections of O$(\del_X/q)$.
This is an important limitation here since
$\del_X\sim g^2 T$, while $g^2 T \simle q \simle gT$. 
However, since it will turn out that the colour relaxation rate
is only logarithmically sensitive to the ultrasoft momenta $q\sim g^2T$,
it can be argued that 
the terms which have been neglected in the above gradient expansion
are suppressed by a factor of $1/\ln(1/g)$. In this sense,
eq.~(\ref{DOEQ}) is still correct to logarithmic accuracy.

The difficulty we are facing here comes
from the necessity to perform a gradient expansion in the presence
of long range interactions, and has already been alluded to
in the case of SQED in Sec. 2.5 (see the discussion
after eq.~(\ref{DOEQ0})). In principle, one could develop a more
accurate approximation scheme by allowing for a collision term
 non-local in $X$.
Specifically, in the exact equation (\ref{WRA}), we can safely
treat $x$ and $y$ as neighbouring points, with $|x-y|\sim 1/T$, since
the propagator ${\cal D}^<(x,y)$ is to enter the hard gluon
self-energy $\Sigma(x,y)$ in eq.~(\ref{S1}); we can proceed similarly
with the points $z_1$ and
$z_2\,$: $|z_1-z_2|\sim 1/T$, since the polarization tensor
$\Pi_{\mu\nu}^<(z_1,z_2)$ is dominated by the hard loop
(e.g., $p$ is a hard momentum in eq.~(\ref{DELPI}) below).
But the points $x$ and $z_1$ (or $y$ and $z_2$) are relatively
distant one from the other, since they are related by the long-range
propagator ${\cal D}_R(x,z_1)$ (respectively, by
${\cal D}_A(z_2,y)$). Thus, a more accurate gradient expansion should 
treat $X\equiv(x+y)/2$ and $X'\equiv(z_1+z_2)/2$ as distinct points
(the ``end-points'' of the virtual gluon line in Fig.~\ref{Born}),
which would then lead to a collision term which is non-local in $X'$.
However, the construction of such a non 
local collision term goes beyond our present goal, and
we shall stick to the local expression in eq.~(\ref{DOEQ}).

The gauge-covariant fluctuation $\delta\acute{\cal D}^<(q,X)$
is obtained from eq.~(\ref{DOEQ}) by first linearizing with respect
to the off-equilibrium fluctuations (e.g., $\Pi^<(q,X)\equiv
\Pi_{eq}^<(q)+\delta\Pi^<(q,X)$, with $\delta\Pi^< \sim g^2 \Pi_{eq}^<$),
and then replacing the ordinary 2-point functions with the corresponding
covariant ones. This gives:
\beq \label{DCOV}
\delta\acute{\cal D}^<(q,X)\equiv\delta{\cal D}^<(q,X)
+g(A(X)\cdot \del_q){}^*{\cal D}^<(q)\,=\,-{}^*{\cal D}_R(q)
\,\delta\acute\Pi^<(q,X){}^*{\cal D}_A(q)\,\,\dots,\,\,\eeq
where the dots stand for terms involving the off-equilibrium
deviations of the retarded, or advanced, functions, but the equilibrium
self-energy $\Pi_{eq}^<(q)$. It is easy to verify that such terms 
will eventually cancel in the collision terms (as they do in
the Abelian case: cf. the remark after eq.~(\ref{PHI})),
so we shall ignore them in what follows. 

We thus need the off-equilibrium polarization tensor 
$\delta\acute\Pi^<(q,X)$ for soft $q$. This is determined by
one-loop diagrams which look formally as in SQED
(cf. Fig.~\ref{PI}), except that, in QCD, the internal lines denote
hard transverse gluons.
The tadpole diagram in Fig.~\ref{PI}.b does not contribute to the
collisional self-energies $\Pi^>$ and $\Pi^<$, but only to the
retarded self-energy $\Pi_R$, which we know
already to be the HTL (recall that we only need this in
equilibrium; cf. eq.~(\ref{DCOV})).
Therefore, in what follows we shall focus
on the non-local\footnote{In Coulomb's gauge, there is another
tadpole coming from the diagram in Fig.~\ref{PI}.a where one of the internal 
lines is transverse, and the other one is longitudinal (and static)
\cite{BP90}. This too contributes to the retarded/advanced propagators,
but not to the collisional self-energies.}
self-energy in Fig.~\ref{PI}.a, which we evaluate
by using the three-particle vertex  $a^\mu a^\nu {\cal A}^\rho$ 
in eq.~(\ref{S}). This yields:
\beq\label{PI1}
\Pi^{ab}_{\mu\nu}(x,y)&=&
\frac{g^2}{4}\,\Gamma_{\mu\rho\alpha\beta}
\Gamma_{\nu\lambda\gamma\delta} \,(T^a)_{cd}\, (T^b)_{\bar c\bar d}
\Bigl\{\Bigl(D^\rho_x G^{\alpha\gamma}(x,y)\Bigr)_{c\bar c}
\Bigl(G^{\beta\delta}(x,y) D^{\lambda\,\dagger}_y\Bigr)_{d\bar d}
\nonumber\\&{}&\qquad\qquad\qquad\qquad
\times G^{\alpha\gamma}_{c\bar c}(x,y)
\Bigl(D^\rho_x G^{\beta\delta}(x,y) D^{\lambda\,\dagger}_y
\Bigr)_{d\bar d}\Bigr\}.\eeq
Starting with this expression, the linearization and the 
covariantization proceed along the same lines
as for the hard gluon self-energy in eq.~(\ref{S1}).
(This is legitimate since the loop integral is dominated by hard
momenta, so that $\Pi_{\mu\nu}(x,y)$ is
localized at $|x - y| \simle 1/T$.)
In this process, we use identities like
eq.~(\ref{COV}), (\ref{COVDEL}) and (\ref{COVS})
to replace ordinary Wigner functions by gauge-covariant ones,
and define the covariant polarization tensor as usual:
\beq \delta\acute \Pi(s,X)\,=\,\delta
\Pi(x,y)\,+\,ig\Bigl(s \cdot A(X)\Bigr)\Pi_{eq}(s),\eeq
with
\beq\label{PIEQ}
\Pi^{eq}_{\mu\nu}(q)\,=\,\frac{g^2N_c}{2}\,
\Gamma_{\mu\rho\alpha\beta}\Gamma_{\nu\lambda\gamma\delta}
 \int\frac{{\rm d}^4 p}{(2\pi)^4}
\,p^\rho p^\lambda\,G_0^{\alpha\gamma}(p+q)\,
G_0^{\beta\delta}(p).\eeq
Here again, we have neglected $q\simle gT$ next to $p\sim T$
in the vertices.

The final result is quite predictible: The {\it covariantized}
fluctuation $\delta\acute \Pi(q,X)$ is formally similar to the
equilibrium self-energy (\ref{PIEQ}), except for the replacement
of the equilibrium internal lines by (linearized) off-equilibrium
{\it gauge-covariant} Wigner functions. Once again, this simple result
holds only after covariantization, and implies that the internal
momenta in eq.~(\ref{DELPI}) below should be interpreted
as {\it kinetic} momenta (recall the discussion after
eq.~(\ref{DELS})). Specifically:
\beq\label{DELPI}
\delta\acute\Pi^{ab}_{\mu\nu}(q,X)&=&
\frac{g^2}{2}\,\Gamma_{\mu\rho\alpha\beta}
\Gamma_{\nu\lambda\gamma\delta}\, (T^aT^b)_{cd}
\int\frac{{\rm d}^4 p}{(2\pi)^4}\,p^\rho p^\lambda\nonumber\\
&{}&\qquad\times
\Bigl\{G_0^{\alpha\gamma}(p+q)\,\delta\acute G^{\delta\beta}_{dc}
(p,X)\,+\,\delta\acute G^{\alpha\gamma}_{cd}(p+q,X)\,
G_0^{\delta\beta}(p)\Bigr\}.\eeq
At this point, we remember that the hard gluon transverse
functions (in or out of equilibrium) are purely spatial and
transverse, so that the above equation can be further simplified to:
\beq\label{PIT}
(\delta\acute\Pi^>)^{ab}_{\mu\nu}(q,X)&=&4g^2\, (T^aT^b)_{cd}
\int\frac{{\rm d}^4 p}{(2\pi)^4}\,p^\mu p^\nu\nonumber\\
&{}&\qquad\times\Bigl\{
G_0^>(p+q)\,\delta\acute G^<_{dc}(p,X)
\,+\,\delta\acute G^>_{cd}(p+q,X)\,G_0^<(p)\Bigr\},\eeq
where we have reestablished the upperscripts $>$ and $<$.

\subsection{The Boltzmann equation for colour}
 
We are now in position to explicitly compute the collision terms in
eq.~(\ref{COLL}). We write $C= C_1+C_2$, with
\beq\label{COLL1}
C_1\equiv -\Gamma(k)\delta\acute G(k,X),\qquad
C_2\equiv \delta\acute\Sigma^>(k,X)G_0^<(k)-
\delta\acute\Sigma^<(k,X)G_0^>(k).\eeq
The first piece $C_1$ involves the (equilibrium) damping rate
for hard transverse gluons $\Gamma(k)$, which can be computed from
eqs.~(\ref{SEQT}) and (\ref{DEQ}) above:
\beq\label{GT2}
\Gamma(k)&=&4N_cg^2\int\frac{{\rm d}^4 q}{(2\pi)^4}\,
k^\rho k^\lambda \Bigl({}^*{\cal D}^>_{\rho\lambda}(q)
\,G_0^>(k-q)-{}^*{\cal D}^<_{\rho\lambda}(q)
\,G_0^<(k-q)\Bigr)\nonumber\\
&=&-4N_cg^2\int\frac{{\rm d}^4 q}{(2\pi)^4}\,\Bigl[k_\rho
{}^*{\cal D}^{\rho\mu}_R(q)\,{}^*{\cal D}^{\nu\lambda}_A(q)
 k_\lambda\Bigr]\Bigl(\Pi^>_{\mu\nu}(q)G_0^>(k-q)-\Pi^<_{\mu\nu}(q)
G_0^<(k-q)\Bigr).\nonumber\\&{}&\eeq
Here, $\Pi_{\mu\nu}\equiv\Pi^{eq}_{\mu\nu}$ is the polarization tensor
in equilibrium, as given in eq.~(\ref{PIEQ}). By inserting it
in eq.~(\ref{GT2}), one finds (with $p'\equiv p+q$ and $k'\equiv k-q$):
\beq\label{GTF}
\Gamma(k)\,=\,N_c^2\int_{p\,q}\,|{\cal M}|^2
\Bigl\{G_0^<(p)G_0^>(p')G_0^>(k')-
G_0^>(p)G_0^<(p')G_0^<(k')\Bigr\},\eeq
where
\beq
\int_{p\,q}\,\equiv\,\int\frac{{\rm d}^4 q}{(2\pi)^4}
\int\frac{{\rm d}^4 p}{(2\pi)^4}\,,\eeq
and we have recognized the matrix element squared
for the collision depicted in Fig.~\ref{Born} :
\beq\label{COLM}
|{\cal M}|^2\,\,\equiv\,\,16g^4\Bigl[k_\rho
{}^*{\cal D}^{\rho\mu}_R(q)p_\mu\Bigr]\Bigl[
 p_\nu{}^*{\cal D}^{\nu\lambda}_A(q) k_\lambda\Bigr].\eeq
The second piece $C_2$ in eq.~(\ref{COLL1}) can be similarly computed
by using eqs.~(\ref{DELST}), (\ref{DCOV}) and (\ref{PIT}). One gets:
\beq\label{C2}
C_2&=&C_{21}\,+\,C_{22}\,+\,C_{23}\nonumber\\
C_{21}^{ab}&\equiv&-N_c (T^aT^b)_{cd}\int_{p\,q}\,|{\cal M}|^2
\Bigl\{G_0^<(p)G_0^>(p')G_0^<(k)-
G_0^>(p)G_0^<(p')G_0^>(k)\Bigr\}
\delta\acute G_{cd}(k',X),\nonumber\\
C_{22}^{ab}&\equiv&-(T^aT^b)_{c\bar c}(T^cT^{\bar c})_{d\bar d}
\int_{p\,q}\,|{\cal M}|^2 \Bigl\{G_0^<(k)G_0^>(k')G_0^>(p')-
G_0^>(k)G_0^<(k')G_0^<(p')\Bigr\}
\delta\acute G_{\bar d d}(p,X),\nonumber\\
C_{23}^{ab}&\equiv&-(T^aT^b)_{c\bar c}(T^cT^{\bar c})_{d\bar d}
\int_{p\,q}\,|{\cal M}|^2 \Bigl\{G_0^<(k)G_0^>(k')G_0^<(p)-
G_0^>(k)G_0^<(k')G_0^>(p)\Bigr\}
\delta\acute G_{d\bar d}(p',X).\nonumber\\&{}&
\eeq
In writing these equations, we have used the fact that 
$\delta\acute G^< \simeq \delta\acute G^>\equiv\delta\acute G$
in the present approximation (cf. eq.~(\ref{GQP})).
The piece $C_{21}$ comes
from the first term in the r.h.s. of eq.~(\ref{DELST}), which describes
fluctuations in the hard propagator inside $\Sigma(k,X)$ (the lower line
in Fig.~\ref{S2}). The other two pieces, $C_{22}$ and $C_{23}$, come from
the second term in eq.~(\ref{DELST}) and describe fluctuations in the
soft (upper) line in Fig.~\ref{S2}. Clearly, these three terms $C_{21}$
$C_{22}$ and $C_{23}$ are associated with fluctuations along the
external lines ``to be summed over'' in Fig.~\ref{Born} --- namely, the lines
with momenta $p$, $p'$ and $k'$ ---, as opposed to $C_1$ which
describes fluctuations along the incoming line with momentum $k$.

We shall verify in a moment that the phase-space integrals in
eqs.~(\ref{GTF}) and (\ref{C2}) are indeed dominated by soft exchanged 
momenta $q$, which justifies our previous approximations. This allows
us to make some further simplifications, as follows:
Recall first that all the Wigner functions in these equations are
distributions  with support on the tree-level
mass-shell (cf. eq.~(\ref{GQP})). E.g.,
\beq
G^<_0(k')=\rho_0(k-q)N(k_0-q_0),\qquad 
\delta\acute G(p',X)=\rho_0(p+q)\delta N(p+q,X),\eeq
with $\rho_0(k)=2\pi \epsilon(k_0)\delta(k^2)$.
In such expressions, we cannot neglect $q$ next to $k$ (or $p$)
within the mass-shell $\delta$-functions, but we can still do that 
in the occupation numbers:
\beq
G^<_0(k')\approx \rho_0(k-q)N(k_0),\qquad 
\delta\acute G(p',X)\approx \rho_0(p+q)\delta N(p,X).\eeq
This yields, e.g., (compare to the Abelian expression in
eq.~(\ref{GTF10})) :
\beq\label{GTF1}
\Gamma(k)&\simeq&N_c^2\int {\rm d}{\cal T}\,|{\cal M}|^2
\Bigl\{N(p_0)[1+N(p_0)][1+N(k_0)] - [1+N(p_0)]N(p_0)N(k_0)
\Bigr\}\nonumber\\
&=&N^2_c\int {\rm d}{\cal T}\,|{\cal M}|^2\,N(p_0)[1+N(p_0)],\eeq
and the phase-space measure $\int {\rm d}{\cal T}$ has been defined
in eq.~(\ref{PHI}). Similarly, 
\beq\label{C21}
C_{21}^{ab}&\simeq& \rho_0(k)N_c (T^aT^b)_{cd}\delta N_{cd}(k,X)
\int {\rm d}{\cal T}\,|{\cal M}|^2\,N(p_0)[1+N(p_0)],\eeq
which is of the same form as the damping rate contribution
$C_1\equiv -\Gamma(k)\delta\acute G(k,X)$
(cf. eq.~(\ref{GTF1})), and can be combined with the latter to yield
(in matrix notations):
\beq\label{C121}
C_1+C_{21}\,=\,-\rho_0(k)\,\frac{N_c}{2}\,[T^a,[T^a,\delta N(k,X)]]
\int {\rm d}{\cal T}\,|{\cal M}|^2\,N(p_0)[1+N(p_0)].\eeq
The remaining two terms $C_{22}$ and $C_{23}$ can be similarly
simplified to yield:
\beq\label{C223}
C_{22}^{ab}+C_{23}^{ab}&\simeq&\rho_0(k)
(T^aT^b)_{c\bar c}(T^cT^{\bar c})_{d\bar d}
\int {\rm d}{\cal T}\,|{\cal M}|^2\,N(k_0)[1+N(k_0)]
\Bigl(\delta N_{d\bar d}-\delta N_{\bar d d}
\Bigr)(p,X)\nonumber\\ &=&\rho_0(k)\,\frac{N_c}{2}\,(T^e)_{ab}
\int {\rm d}{\cal T}\,|{\cal M}|^2\,N(k_0)[1+N(k_0)]\,
{\rm Tr}\,\Bigl(T^e\delta N(p,X)\Bigr),\eeq
where the second line follows after some elementary colour algebra.

Thus, unlike the Abelian case in Sec. 2.6, where the IR contributions
to the four terms in eq.~(\ref{LINCOL}) cancel each other,
here we do not have a complete
cancellation because of the non-trivial colour structure.
By putting together eqs.~(\ref{C121}) and (\ref{C223}), 
we finally derive the following expression for the non-Abelian collision
term:
\beq\label{COLF}
C[\delta N]&=&-\rho_0(k)\,\frac{N_c}{2}
\int {\rm d}{\cal T}\,|{\cal M}|^2\Bigl\{N(p_0)[1+N(p_0)]\,
[T^a,[T^a,\delta N(k,X)]]\,-\,\nonumber\\
&{}&\qquad\qquad\qquad\qquad-\,N(k_0)[1+N(k_0)]\,T^a\,
{\rm Tr}\,(T^a\delta N(p,X))\Bigr\}.\eeq
Together, eqs.~(\ref{bolt}) and (\ref{COLF}) determine a Boltzmann
equation which describes colour relaxation in hot QCD. By also using
eq.~(\ref{GN}), this can be rewritten as
an equation for the density matrix $\delta N_{ab}({\bf k},X)$.
To this aim, we need the positive-energy projection ($k_0=
\varepsilon_k \equiv |{\bf k}|$) of the collision term.
For soft $q$ and $k_0= \varepsilon_k \sim T$,
$k_0^\prime$ is positive as well:
\beq 
\rho_0(k')|_{k_0=\varepsilon_k}\,=\,
\frac{2\pi}{2\varepsilon_{k-q}}\,\delta(\varepsilon_k-q_0-
\varepsilon_{k-q})\,\simeq\,\frac{2\pi}{2\varepsilon_k}\,
\delta(q_0- {\bf q\cdot v}),\eeq
where ${\bf v}=\hat{\bf k}$ is the velocity of the incoming
particle with momentum ${\bf k}$.
Concerning $p_0$, this can be either positive, $p_0=\varepsilon_p$,
or negative, $p_0=-\varepsilon_p$, and the two situations yield
identical contributions. We thus replace
 (with ${\bf v}^\prime=\hat{\bf p}$):
\beq
\rho_0(p)\rho_0(p+q) &\to& 
2\, \left(\frac{2\pi}{2\varepsilon_p}\right)^2
\,\delta(p_0-\varepsilon_p)
\delta(q_0- {\bf q\cdot v}^\prime).\eeq
Also, for on-shell momenta $k_0=\varepsilon_k$ and
$p_0=\varepsilon_p$, the matrix element (\ref{COLM}) becomes:
\beq\label{COLM1}
|{\cal M}|^2\,=\,16g^4\varepsilon_k^2\varepsilon_p^2\,
 \Big|{}^*{\cal D}_l(q)
+ ({\bf \hat q \times v})\cdot ({\bf \hat q \times v}^\prime)\,
{}^*{\cal D}_t(q)\Big|^2,\eeq
with the same notations as in the Abelian case (cf. eq.~(\ref{COLM1})).

By also using the identity  $N(p)[1+N(p)]=-T({\rm d}N/{\rm d}p)$,
we are finally led to the following Boltzmann equation,
which is the main result in this paper:
\beq\label{B}
\left[ v\cdot D_X,\,\delta N({{\bf k}},X)\right]\,+\, g\,
{\bf v}\cdot{\bf E}(X)\frac{{\rm d}N}{{\rm d}k}\,=\,C[\delta N],\eeq
with the collision term\footnote{Note that the overall
normalizations in eqs.~(\ref{COLF}) and (\ref{COLON})  are different.}:
\beq\label{COLON}
C[\delta N]\equiv  g^4 N_c T
\int\frac{{\rm d}^3 p}{(2\pi)^3}\,\Phi({\bf v\cdot v}^\prime)
\biggl\{\frac{{\rm d}N}{{\rm d}\varepsilon_p}
\,[T^a,[T^a,\delta N({\bf k},X)]]\,
-\,\frac{{\rm d}N}{{\rm d}\varepsilon_k}\,T^a\,
{\rm Tr}\,(T^a\delta N({\bf p},X))\biggr\},
\nonumber\\{}\eeq
and the collision integral
(${\bf v}_t$ and ${\bf v}^\prime_t$ are the velocity projections
transverse to ${\bf q}$):
\beq\label{PHI2}\Phi({\bf v\cdot v}^\prime)\equiv(2\pi)^2
\int\frac{{\rm d}^4 q}{(2\pi)^4}\,
\delta(q_0- {\bf q\cdot v})
\delta(q_0- {\bf q\cdot v}^\prime)
\Big|{}^*{\cal D}_l(q)+ ({\bf v}_t\cdot{\bf v}_t^\prime)\,
{}^*{\cal D}_t(q)\Big|^2.\,\,\,\eeq
The collision terms above are identical to those
written down in Ref. \cite{ASY97} on an heuristic basis
(cf. eq.~(3.26) in Ref. \cite{ASY97}).

The above equations can be further simplified by noticing that
the corresponding solution
$\delta N({\bf k},X)$ can be written in the form:
\beq\label{dn}
\delta N({\bf k}, X)\equiv - gW(X,{\bf v})\,({\rm d}N/{\rm d}k),\eeq
where the new function $W(X,{\bf v})$ (a colour matrix) depends
upon the velocity ${\bf v}$ (a unit vector), but not  on the
magnitude $k\equiv |{\bf k}|$ of the momentum.
This function satisfies a simpler equation
(we change $X \to x$ from now on since this
is the only space-time variable left in all the equations to come):
\beq\label{W}
\left[ v\cdot D_x,\,W(x,{\bf v}) \right]
&=&{\bf v}\cdot{\bf E}(x)\,-\,m_D^2\,
\frac{g^2T}{2}\int\frac{{\rm d}\Omega'}{4\pi}
\,\Phi({\bf v\cdot v}^\prime)\nonumber\\
&{}&\qquad\qquad\qquad\times\Bigl\{[T^a,[T^a, W(x,{\bf v})]]\,-\,T^a\,
{\rm Tr}\,(T^a W(x,{\bf v}^\prime)\Bigr\},\qquad\qquad\eeq
where the angular integral runs over all the directions of the unit
vector ${\bf v}^\prime$, and the Debye mass $m_D^2$ comes out
after performing the radial integral over $p\equiv |{\bf p}|\,$:
\beq\label{MD}
m^2_D\,\equiv\,-\,\frac{g^2 N_c}{\pi^2}\int_{0}^\infty 
{\rm d}p \,p^2\,\frac{{\rm d}N}{{\rm d}p}\,=\,\frac{g^2N_c T^2}{3}\,.\eeq
The factorized structure of the colour density matrix 
in eq.~(\ref{dn}) has been first recognized
at the level of the mean field (or Vlasov) approximation 
(cf. eq.~(\ref{N})), where it has been shown to have a simple
physical interpretation \cite{W}.
It is remarkable that such a structure persists after the inclusion
of the collision terms.

Consider furthermore the colour structure of eq.~(\ref{W}).
In constructing the induced current (\ref{jb}), we only need the first
colour moment of the density matrix $W(x,{\bf v})$, namely 
$W_a(x,{\bf v})\equiv (1/N_c){\rm Tr}\,(T^a W(x,{\bf v}))$.
The equation satisfied by $W_a(x,{\bf v})$ follows from eq.~(\ref{W})
by taking the appropriate colour trace, and reads:
\beq\label{W1}
(v\cdot D_x)^{ab}W_b(x,{\bf v})&=&{\bf v}\cdot{\bf E}^a(x)-m_D^2
\frac{g^2 N_c T}{2}\int\frac{{\rm d}\Omega'}{4\pi}
\,\Phi({\bf v\cdot v}^\prime)\Bigl\{W^a(x,{\bf v})-
W^a(x,{\bf v}^\prime)\Bigr\}.\nonumber\\&{}&\eeq
This is the equation which has been announced in the Introduction
(cf. eq.~(\ref{W10})).
The collision term in this equation has a simple
physical interpretation. We may indeed identify
\beq\label{gamma1}
m_D^2\,
\frac{g^2 N_c T}{2}\int\frac{{\rm d}\Omega'}{4\pi}
\,\Phi({\bf v\cdot v}^\prime)\,=\,
\gamma\,=\,\frac{g^2 N_c T}{4\pi}\Bigl
(\ln\frac{m_D}{\mu}+{\rm O}(1)\Bigr),\eeq
which is the {\it damping rate} 
$\gamma\equiv\Gamma(k_0=k)/4k$ for a hard transverse gluon with velocity
${\bf v}$ \cite{Pisarski93,lifetime}. In the present approximation,
the gluon damping rate suffers from the same logarithmic IR
divergence as the damping rate for a charged particle in the Abelian case
(cf. Sec. 2.6). In the equation above, this has been cutoff by hand, 
by introducing an IR cutoff $\mu$ (see the r.h.s. of eq.~(\ref{gamma1})).

In previous studies of the colour conductivity or the
damping rates, it has been generally assumed that the IR cutoff 
is provided by non-perturbative
magnetic screening at the scale $g^2 T$ \cite{MLB96,prept}.
This yields a damping rate $\gamma \simeq \alpha N_cT\ln(1/g)$
(with $\alpha\equiv g^2/4\pi$), but the constant
term under the logarithm cannot be determined: indeed, this is sensitive
to the details of the magnetic screening, which remains poorly understood.
In the framework of the effective theory recently proposed by
B\"odeker \cite{Bodeker98}, $\mu$ should be rather understood
as an intermediate scale $\mu\sim g^2T\ln(1/g)$ separating the perturbative
physics at ``hard'' ($k\sim T$) and ``semi-hard''
($\mu\simle q\simle gT$) momenta from the non-perturbative physics
at ``ultrasoft'' ($q \simle g^2 T < \mu$) momenta.
Then, $\mu$ will hopefully cancel in any complete calculation,
via a matching between the perturbative and the non-perturbative
(e.g., lattice) calculations. However, our above derivation of the
collision term shows that, within the present formalism, it is not
possible to go beyond leading-log accuracy (which makes the issue
of matching superfluous): this reflects the present limits of the gradient
expansion in the presence of long range interactions
(cf. the discussion after eq.~(\ref{DOEQ})).

Within this logarithmic accuracy, the integral in eq.~(\ref{PHI2})
can be performed by replacing the matrix element for transverse 
scattering by its infrared limit as $q_0\ll q\to 0$, namely 
(see eq.~(\ref{DSTAT})) \cite{lifetime}:
\beq \label{singDT}
|{}^*{\cal D}_t(q_0,q)|^2\,\simeq\,
\frac{1} {q^4 + (\pi m_D^2 q_0/4q)^2}\,\,\,
\longrightarrow_{q\to 0}\,\,\,\frac{4}{m_D^2}\,\frac{\delta(q_0)}{q}\,.\eeq
One thus obtains:
\beq
\Phi({\bf v\cdot v}^\prime)\,\simeq\,\frac{2}{\pi^2 m_D^2}\,
\frac{({\bf v\cdot v}^\prime)^2}{\sqrt{1-({\bf v\cdot v}^\prime)^2}}\,
\ln(1/g)\,,
\eeq
where the $\ln(1/g)$ comes from (compare to eq.~(\ref{G2LR})) :
\beq \int_\mu^{m_D}\frac{{\rm d}q}{q}\,\simeq\,\ln\,\frac{gT}{\mu}\,\simeq\,
\ln\,\frac{1}{g}\,,\eeq
with $\mu\sim g^2 T\ln(1/g)$.
Then, eq.~(\ref{W1}) reduces to
(with $\gamma \simeq \alpha N_cT\ln(1/g)$) :
\beq\label{W2}
(v\cdot D_x)W(x,{\bf v})\,=\,{\bf v}\cdot{\bf E}(x)\,-\,\gamma
\left\{W(x,{\bf v})-\frac{4}{\pi}\int\frac{{\rm d}\Omega'}{4\pi}
\frac{({\bf v\cdot v}^\prime)^2}{\sqrt{1-({\bf v\cdot v}^\prime)^2}}
W(x,{\bf v}^\prime)\right\}.\,\,\eeq
This is precisely the kinetic equation
which generates B\"odeker's effective theory \cite{Bodeker98,ASY97}.

The previous equations have been obtained by working in the Coulomb
gauge for the hard field fluctuations (cf. eq.~(\ref{COUL})), but
we expect the final results --- namely, the Boltzmann equation in
 eqs.~(\ref{B})--(\ref{COLON}), or (\ref{W1}) ---
to be actually gauge-fixing independent.
Except for the collision term, this has
been explicitly proven in \cite{qcd} (see also Refs. \cite{BP90,FT90}).
Moreover, on physical grounds, the collision term should
be gauge-fixing independent as well,
since it involves only the off-equilibrium fluctuations of the
(hard) transverse gluons, together with the (gauge-independent)
matrix element squared (\ref{COLM1}).
(See also the discussion after eq.~\ref{Delta}.)

Furthermore, the previous equations are manifestly covariant under the
gauge transformations of the background fields. This ensures that
the induced current (\ref{jb}), 
which can be expressed in terms of $W_a(x,{\bf v})$ as follows:
\beq\label{j1}
j^\mu_a(x)&=&m_D^2\int\frac{{\rm d}\Omega}{4\pi}
\,v^\mu\,W_a(x,{\bf v}),\eeq
transforms indeed as a colour vector in the adjoint representation.
Moreover, this current is covariantly conserved, 
\beq\label{COVCONS}
D_\mu j^\mu\,=\,0,\eeq
as necessary for
the consistency of the mean field equations of motion (\ref{avA1}).
To verify eq.~(\ref{COVCONS}), use eq.~(\ref{W1})
together with the fact that
the collision terms vanishes after angular averaging:
\beq
\int\frac{{\rm d}\Omega}{4\pi}
\int\frac{{\rm d}\Omega'}{4\pi}
\,\Phi({\bf v\cdot v}^\prime)\Bigl\{W^a(x,{\bf v})-
W^a(x,{\bf v}^\prime)\Bigr\}\,=\,0.\eeq

The quantities $W_a(x,{\bf v})$ may be regarded as functionals
of the mean fields $A^\mu_a(x)$, as given implicitly by
the Boltzmann equation (\ref{W1}). The current (\ref{j1}) itself
acts as a generating functional for the one-particle-irreducible amplitudes
of the ultrasoft colour fields. We can formally write (see, e.g., Refs. 
\cite{BIO96,prept,qcd}) :
\beq\label{exp}
j^{a}_\mu \,=\,\Pi_{\mu\nu}A_a^\nu
+\frac{1}{2}\, \Gamma_{\mu\nu\rho}^{abc} A_b^\nu A_c^\rho+\,...
\eeq
where $\Pi_{\mu\nu}(P)$ is the polarization tensor for the ultrasoft 
($P\sim g^2T$) gluons, $\Gamma_{\mu\nu\rho}^{abc}$ is a correction
to the 3-gluon vertex, etc. For external momenta of order $g^2 T$
or less, the amplitudes in eq.~(\ref{exp}) are of the same order of
magnitude as the ``hard thermal loops'' \cite{BP90,FT90,qcd},
which they generalize by including the effects
of the collisions among the hard particles. Like the HTL's,
the above amplitudes are gauge-fixing independent, and satisfy
simple Ward identities which follow from the 
conservation law (\ref{COVCONS}) by successive differentiations with 
respect to the fields $A^\mu_a$. For instance:
\beq\label{Wtr}
p^\mu\,\Pi_{\mu\nu}(p)&=&0,\nonumber\\
p_1^\mu\Gamma_{\mu\nu\rho}(p_1,p_2,p_3)&=&\Pi_{\nu\rho}
(p_3)-\Pi_{\nu\rho}(p_2)\,.\eeq
Such identities express the fact that the effective theory
at the scale $g^2 T$, as obtained from the Boltzmann equation 
\cite{Bodeker98}, is gauge invariant.
However, unlike the HTL's --- which in terms of Feynman graphs
correspond to one-loop diagrams \cite{BP90,FT90} ---, each of
the ultrasoft amplitudes in eq.~(\ref{exp}) receives contributions
from an infinite series of multi-loop Feynman graphs, 
which all contribute at the same order in $g$.
This will be explained in the next section and, in more detail,
in a further publication \cite{USA}.

\setcounter{equation}{0}
\section{Diagrammatic interpretation of the collision terms}

In this section we provide
a diagrammatic interpretation of the collision terms. 
For the case of a scalar field theory, such an interpretation
has been worked out in much detail in Refs. \cite{Jeon93,JY96},
where the Boltzmann equation has been actually derived by
resumming appropriate classes of Feynman graphs. Here, where the
Boltzmann equation has been constructed by using different
technics, it is still interesting to understand what are the
diagrams which have been effectively resummed in this construction. 
To this aim,
it is convenient to go back to eqs.~(\ref{COLL1}), (\ref{GTF}) and
(\ref{C2}) where the off-equilibrium fluctuations (like, e.g.,
$\delta\acute G(k,X)$) are unambiguously associated with each
of the colliding fields in Fig.~\ref{Born}.

Consider first the piece
$C_1\equiv -\Gamma(k)\delta\acute G(k,X)$ which involves the
interaction rate $\Gamma(k)$ for the incoming particle with
momentum $k$ (cf. eq.~(\ref{GTF})). This can be moved into the
l.h.s. of the kinetic equation and combined with the drift term
$k\cdot D_X$ to yield:
\beq\label{B2}
\Bigl(2(k\cdot D_x)+\Gamma(k)\Bigr)\delta\acute G(k,x)\,=\,
2g(k\cdot  F\cdot \del_k) G_0(k)\,+\,C_{21}\,+\,C_{22}\,+\,C_{23},\eeq
with $C_{2i}$ as given in eqs.~(\ref{C2}). After the inclusion of
$\Gamma$, the drift operator describes also the decay of the incoming
particle. For instance, the corresponding Green's function satisfies 
(with $k_0=k$ and $\gamma = \Gamma/4k$; cf. eq.~(\ref{gamma1})) :
\beq\label{Gret}
-i\,(v\cdot D_x + 2\gamma)\,\Delta(x,y;{\bf v})=\delta^{(4)}(x-y),
\eeq
with the retarded solution (below, $t \equiv x^0-y^0$):
\beq\label{GR}
\Delta_{R}(x,y;{\bf v})&=&i\,\theta (t)\,\delta^{(3)}
\left({{\bf x}}-{{\bf y}}-{\bf v}t
\right )\,{\rm e}^{-2\gamma t}\,U(x,y),\eeq
which shows exponential attenuation in time with a rate equal to
$2\gamma$ (compare to eq.~(\ref{b1})).

The inclusion of the interaction rate $\Gamma(k)$ in the drift operator
amounts to an approximate self-energy resummation in the hard gluon propagator.
At the level of the original Kadanoff-Baym
equations (\ref{KBYM1}) and (\ref{KBYM2}), this amounts to moving
all the terms involving $G^<(x,y)$ (including the terms $\Sigma_R G^<$ 
and $\Sigma_A G^<$) into the left hand sides of these equations.
What we would like to argue now is that the other collision
terms in the r.h.s. of eq.~(\ref{B2}) can be similarly recognized
as vertex corrections.

\begin{figure}
\protect \epsfxsize=13.cm{\centerline{\epsfbox{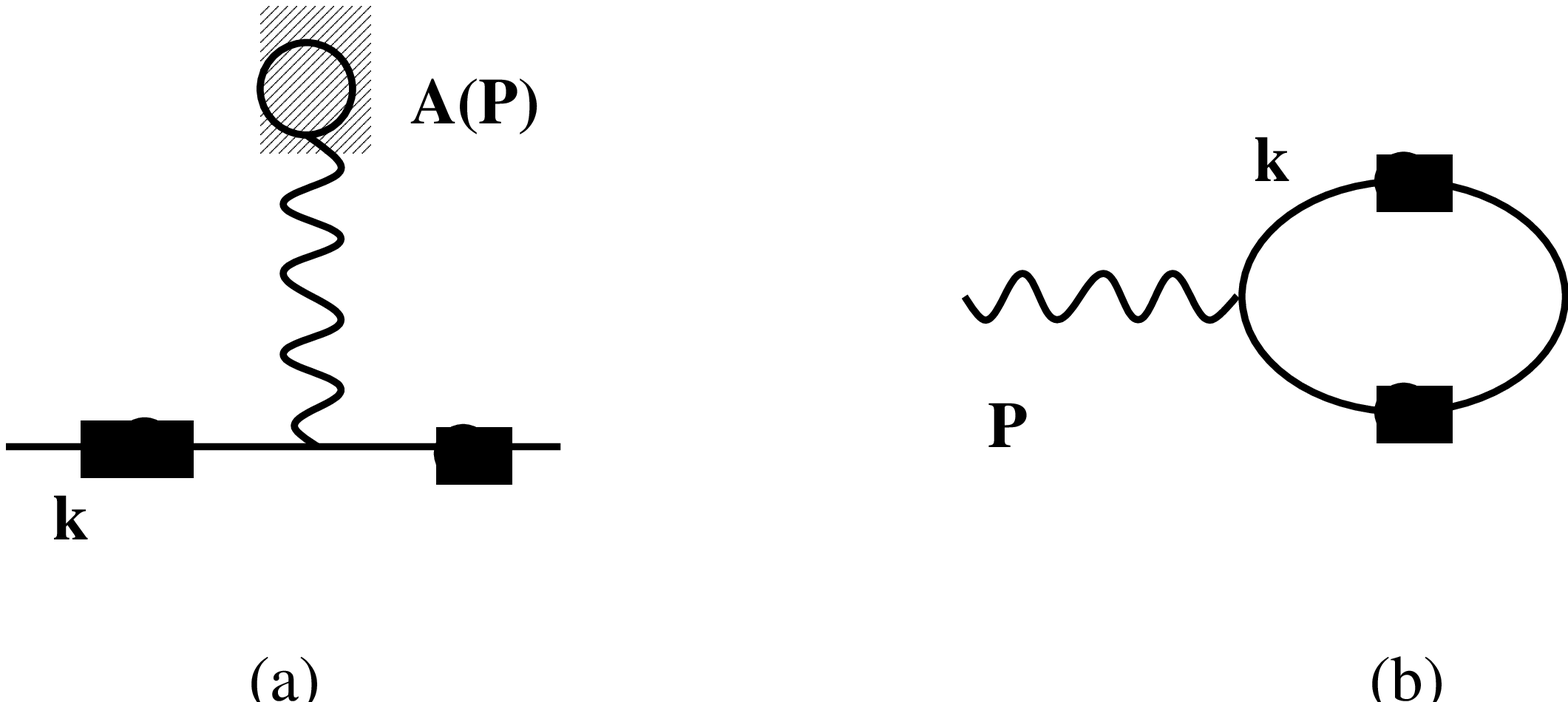}}}
         \caption{Diagrammatic representation of eq.~(\ref{FL0}):
(a) the first-order fluctuation $\delta N^{(0)}(k,P)$, as induced
by the mean field $A^\mu(P)$; (b) the corresponding contribution to
the polarization tensor $\Pi^{\mu\nu}(P)$.
The continuous line with a blob denotes the hard gluon propagator
after the inclusion of the damping rate.}
\label{N0}
\end{figure}

For more clarity, consider the response of the plasma to an arbitrarily
weak electric mean field ${\bf E}_a$. Then, non-linear effects are 
negligible and the relevant response function is the polarization 
tensor:
$j^\mu_a(P)\simeq \Pi^{\mu\nu}(P)A_\nu^a(P)$, where $P\sim g^2 T$
is the (ultrasoft) momentum carried by the background field.
In the mean field approximation, and to linear order in the electric 
field, the induced fluctuation reads (cf. eq.~(\ref{N})):
\beq\label{FL}
\delta N(k,P)\,=\,\frac{-g {\bf v\cdot E}(P)}{v\cdot P + i\eta}
\,\frac{{\rm d}N}{{\rm d}k}\,,\eeq
where the small imaginary part $i\eta$ with $\eta \to 0_+$ stands
for retarded boundary conditions. 

After adding the collision terms,
we end up with an integro-differential equation for $\delta N$
(see, e.g., eqs.~(\ref{W1}) or (\ref{B2})), whose resolution is
non-trivial already in the weak field limit. However, in order to see
what are the relevant graphs, it is sufficient to consider the
perturbative solution obtained by iterations. First, it is
straightforward to resum the equilibrium damping
rate $\Gamma$ (i.e., the collision 
term $C_1$): according to eqs.~(\ref{B2}) and (\ref{Gret}), this amounts
to including the damping rate in eq.~(\ref{FL}). This provides us
with the 0th order iteration for $\delta N$:
\beq\label{FL0}
\delta N^{(0)}(k,P)\,=\,\frac{-g {\bf v\cdot E}(P)}{v\cdot P + 2i\gamma}
\,\frac{{\rm d}N}{{\rm d}k}\,.\eeq
This can be given the diagrammatic representation in Fig.~\ref{N0}.a
where the blob on the continuous line denotes the resummation
of the damping rate in the hard gluon propagator, and the wavy line
with a bubble attached to it represents a mean field insertion.
In this approximation, the polarization tensor $\Pi^{\mu\nu}(P)$
is given be the one-loop diagram in Fig.~\ref{N0}.b.

With $\delta N^{(0)}$ as above, we can compute the first iteration
for the collision terms $C_{2i}$, $i=1,\,2,\,3$.
We shall not write down the corresponding formulae, but simply
look for their interpretation in terms of Feynman graphs.
The term $C_{21}$ in eq.~(\ref{C2}) is associated with a fluctuation
$\delta N(k',X)$ in the hard, lower, propagator in Fig.~\ref{S2}; 
its first iteration is obtained
by using the approximation (\ref{FL0}) for this fluctuation, and has
the diagrammatic representation in Fig.~\ref{1IT}.a. To the same order,
$C_{22}$ is represented in Fig.~\ref{1IT}.b, while $C_{23}$ has a similar
representation.
With these approximations for the collision terms, one can compute
the first iteration $\delta N^{(1)}(k,P)$, as well as its contribution 
to $\Pi^{\mu\nu}(P)$: the latter is shown in Figs.~\ref{N1}.a and \ref{N1}.b.
Clearly, Fig.~\ref{N1}.a is a vertex correction, which has to supplement
the self-energy resummations in Fig.~\ref{N0}.b as required by gauge symmetry.
Similarly, the diagram in Fig.~\ref{N1}.b is a different kind of
vertex correction which involves two hard loops (and one soft one).

\begin{figure}
\protect \epsfxsize=13.cm{\centerline{\epsfbox{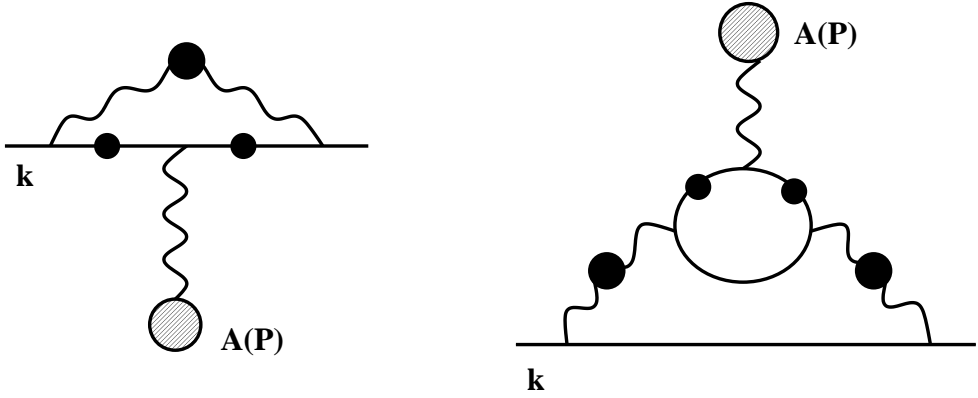}}}
         \caption{Diagrammatic representation of the collision terms
$C_{21}$ (fig. (a)) and $C_{22}$ (fig.  (b)), as computed with
the first-order fluctuation $\delta N^{(0)}(k,P)$.
The wavy line with a blob denotes the soft gluon propagator in the HTL
approximation. The other notations are as in Fig.~\ref{N0}.}
\label{1IT}
\end{figure}

It should be clear by now what are the diagrams generated by further
iterations: these are the ladder diagrams
displayed in Fig.~\ref{L} where the continuous lines are hard transverse 
gluons dressed with the
damping rate $\Gamma$, and the (internal) wavy lines are soft 
gluons which may be longitudinal or transverse and are dressed
with the hard thermal loop. Indeed, the phase-space integrals giving 
$\Gamma$ and $C_{2i}$ are individually dominated by soft $q$ momenta, 
in the range $g^2 T\simle q \simle gT$. Note also that the diagrams
in Fig.~\ref{L} may involve an arbitrary number of loops.
However, from the previous derivation of the Boltzmann equation,
we know that, for an external momentum $P\sim g^2 T$, they all
contribute at the same order in $g$, namely,
at the same order as the corresponding 
hard thermal loop \cite{BIO96,MLB96} (see also Ref. \cite{USA}).

\begin{figure}
\protect \epsfxsize=13.cm{\centerline{\epsfbox{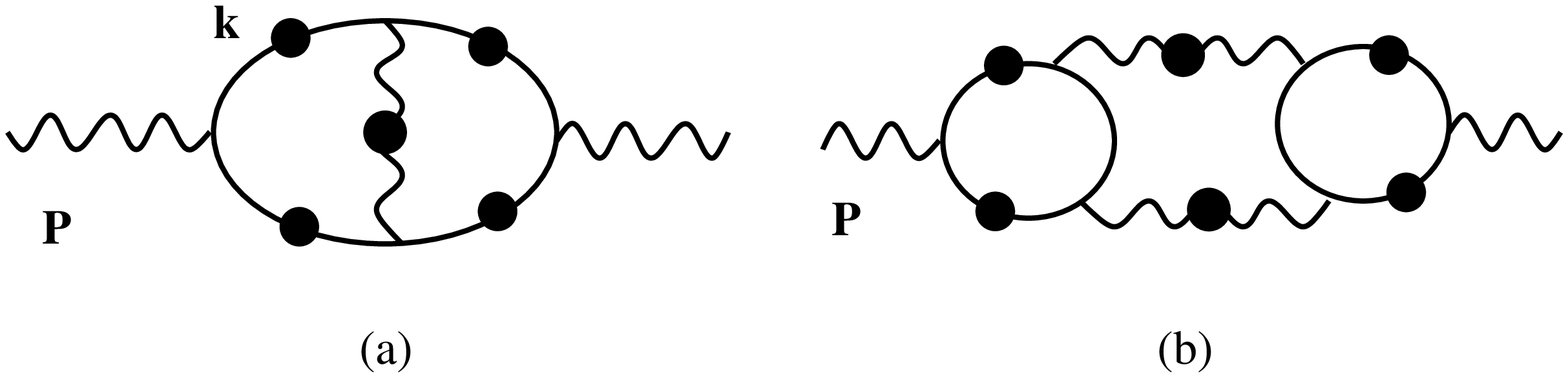}}}
         \caption{Loop corrections to the polarization
 tensor in Fig.~\ref{N0}.b,
as generated by the collision terms depicted in Fig.~\ref{1IT}.}
\label{N1}
\end{figure}

Except for the colour structure of the vertices, similar diagrams
are resummed by the Abelian Boltzmann equation as well. There too,
the damping rate $\Gamma$ of the charged particle is dominated by soft
momenta transfer, and, in fact, it coincides with the corresponding
rate for gluons up to a colour factor
 (compare eqs.~(\ref{GTF10}) and (\ref{GTF1})).
Nevertheless, we have seen in Sec. 2.6 that the Abelian collision terms 
are {\it not} dominated by 
soft momenta exchange, but  receive contributions
from all the momenta between $gT$ and $T$.
In order for this to be consistent with the diagrammatic
picture in Fig.~\ref{L}, there must be some cancellations among the
Feynman graphs, in such a way that the globl contribution of soft ($q
\simle gT$) internal photons cancels out in their sum.

The relevant cancellations have been discussed in Sec. 2.6,
and can be also read off the formulae in Sect. 3.8.
The contributions of the soft exchanged momenta $q\simle gT$
to the individual collision terms are listed in eqs.~(\ref{GTF1}), 
(\ref{C21}) and (\ref{C223}). If there was not for the colour structure,
the vertex correction in eq.~(\ref{C21}) would exactly cancel
the self-energy correction in eqs.~(\ref{GTF1}). This is what happens
in the Abelian case (cf. the discussion in Sec. 2.6)
and explains, for instance, why there is no effect
of the damping rate $\gamma \sim e^2 T\ln(1/e)$ of the charged particle
on the polarization tensor $\Pi_{\mu\nu}$ for soft photons.
Diagramatically, this 
corresponds to a cancellation of the soft photon effects in between
self-energy and vertex corrections in any of the bubbles depicted
in Fig.~\ref{L}, as it has been also verified
via the direct analysis of the Feynman graphs, by Lebedev and Smilga
\cite{Smilga90} (see also Refs. \cite{Kraemmer94,Petit98}).

\begin{figure}
\protect \epsfxsize=14.cm{\centerline{\epsfbox{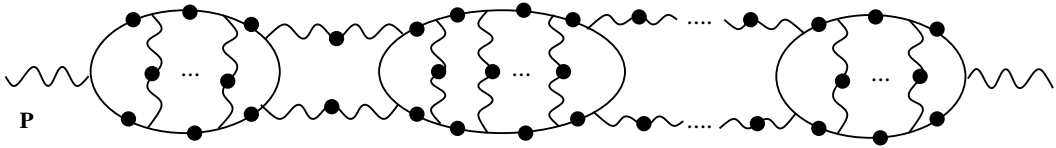}}}
         \caption{A generic ladder diagram of the type resummed in the
solution to the Boltzmann equation.}
\label{L}
\end{figure}

Similarly, in the absence of colour effects,
there would be an exact cancellation between the soft
contributions to $C_{22}$ and $C_{23}$ (cf. eq.~(\ref{C223})):
if we refer to the diagram in  Fig.~\ref{1IT}b, this cancellation
reflects the well known fact that there is no photon HTL with three
(or more) external legs \cite{BIO96,MLB96}.

In QCD, on the other hand, these cancellations are not complete, because
of the non-trivial colour structure (cf. eqs.~(\ref{C121}) and
(\ref{C223})). In fact, for $\delta N(k,X)\equiv \delta N_a T^a$,
we have (refer to eq.~(\ref{C21})) $(T^aT^b)_{cd}\delta N_{cd}
=(N_c/2)\delta N_{ab}$, so that the vertex correction in eq.~(\ref{C21})
only cancels {\it half} of the damping rate contribution in 
eq.~(\ref{GTF1}). Furthermore, since $\delta N\equiv
\delta N_a T^a$ is an antisymmetric colour matrix, the two terms
$C_{22}$ and $C_{23}$ give identical contributions which
{\it add} together (rather than subtract, as in QED) to give the final
result in eq.~(\ref{C223}). Accordingly, the polarization
tensor $\Pi_{\mu\nu}(P)$ for soft ($P\sim g^2T$) gluons {\it is}
sensitive to the hard gluon damping rate $\gamma\sim
g^2T\ln(1/g)$, as discussed at the end of the previous section.
(To our knowledge, this has been first observed by B\"odecker 
\cite{Bodeker98}.)

The previous discussion also shows that, for
{\it colourless} fluctuations in the hot QCD plasma (as involved, for
instance, in the calculation of the shear
viscosity), the pattern of the cancellations is
the same as in QED: the relevant collision terms are
dominated by relatively large ($gT\simle q \simle T$) transferred
momenta. Thus, the typical inverse relaxation time for such
fluctuations is $\tau^{-1}\sim g^4 T\ln(1/g)$, where
the logarithm comes from $\log(T/m_D)$,
that is, from the screening effects at the scale $gT$
\cite{Baym90,Heisel94a,Baym97}. 

\setcounter{equation}{0}
\section{Conclusions}

Starting from first principles, we have derived a Boltzmann
equation which describes the long wavelength colour excitations 
of a high temperature Yang-Mills plasma. Our derivation  relies
on a gauge-covariant gradient expansion of the Kadanoff-Baym equations
for the off-equilibrium dynamics of the plasma.
This expansion can be also intrepreted as an expansion in powers of $g$, 
in the sense that the plasma inhomogeneities (the ``soft'' 
gradients $\del_X$), the strength of the colour mean fields $A^\mu_a$
and the  off-equilibrium deviations of the distribution functions
are all contolled by powers of $g$. Specifically, our derivation
applies to the case where $\del_X \sim gA \sim g^2 T$, so that the
soft {\it covariant} derivatives $D_X=\del_X+igA \sim g^2T$ are
consistently preserved in the perturbative expansion. This, together
with a judicious choice of the gauge fixing (the background field gauge)
and a proper definition of gauge-covariant Wigner functions, has allowed
us to maintain explicit gauge symmetry at all steps of our construction.

In this framework, the Boltzmann equation has emerged as the quantum
transport equation at leading order in $g$. Note that our present,
leading order, perturbative expansion encompasses several approximations 
which are generally seen as independent approximations when deriving
transport equations: the gradient expansion (slowly varying disturbances),
the weak field expansion (small perturbations), the quasiparticle
approximation (well defined ``quasiparticles'', with a long
lifetime) and the (dressed) Born approximation (one-gluon exchange)
for the collision term. Note also that, in spite of being obtained
within a systematic perturbative expansion, the Boltzmann equation
actually resums an infinite series of ladder diagrams and therefore
generates a non-perturbative effective theory for the ``ultrasoft''
($\del_X\sim g^2T$) colour fields. For instance, the polarization
tensor for these fields, as obtained by solving the Boltzmann equation,
is equivalent to the sum of an infinite number of ladder diagrams
of the perturbation theory.

Remarkably, the colour structure of the collision terms
is precisely that predicted by Arnold, Son and Yaffe, on
the basis of simple heuristic arguments \cite{ASY97}
(this is also consistent
with some previous results in Refs. \cite{Selik91,Gyulassy93,Heisel94}).
Accordingly, the effective theory generated by our Boltzmann equation
is precisely that obtained by B\"odecker using
a different method \cite{Bodeker98} (see Ref. \cite{ASY97} for 
a discussion of the relation between the Boltzmann equation
and B\"odecker's theory).

It is important to emphasize the accuracy limits of the Boltzmann equation
presented here. The collision term in eq.~(\ref{W10})
is only known to logarithmic accuracy, which means that 
both its functional form and the overall coefficient $\gamma$
which fixes the time scale for colour relaxation are known
only up to corrections of relative order $1/\ln(1/g)$.
For instance, $\gamma =\alpha N_cT\Bigl(
\ln(1/g)+{\rm O}(1)\Bigr)$ where the constant terms under the
logarithm is not accurately given by the present formalism.
There are two main sources for such a limitation:
({\it i}) The gradient expansion in the presence of long range
interactions: the one-gluon exchange interaction has a typical
range $\sim 1/q$ with $g^2T \simle q \simle gT$
which is marginally the same as the inhomogeneity scale in the
problem, $\lambda\sim 1/g^2 T$. Since the collision terms are only
logarithmically sensitive to the low momenta $q\sim g^2T$, it follows
that the gradient expansion for the collision kernel (i.e., for
the exchanged gluon propagator in eq.~(\ref{WRA})) is only correct
to logarithmic accuracy. As discussed after eq.~(\ref{DOEQ}),
this limitation can be avoided, at least in principle,
by relaxing the gradient expansion so as to allow
for a collision kernel which is non-local in $X$.
({\it ii}) Still related to the long-ranged collision kernel:
the behaviour of the gluon propagator at very soft momenta 
$q\sim g^2 T$ is not correctly described by the HTL approximation,
since it is sensitive to the non-perturbative physics. 

Of course, the Boltzmann equation constructed in this
paper can also be used to study colourless fluctuations,
as involved, e.g., in the calculation of the shear viscosity
\cite{Baym90,Heisel94a,JY96}. The relevant collision terms are
displayed in eqs.~(\ref{COLL1}), (\ref{GTF}) and
(\ref{C2}), where the off-equilibrium fluctuations have now
a trivial colour structure (e.g., $\delta\acute G_{\bar d d}(p,X)
\equiv \delta_{\bar d d}\delta\acute G(p,X)$, etc.).
This yields the standard Boltzmann equation (in its linearized
version), as already used in calculations of the shear viscosity
for the quark gluon plasma \cite{Baym90,Heisel94a}.

\end{document}